%% file: Angantyr.tex
\documentclass{JHEP3}
\usepackage{amsmath}
\usepackage{axodraw4j}
\usepackage{epsfig}
\usepackage{amssymb}
\usepackage{mathrsfs}
\usepackage{mathcomp}
\usepackage{enumerate}
\usepackage{cite}
\usepackage[utf8]{inputenc}
\usepackage{xspace}
\usepackage{ifthen}
\usepackage{url}

\newcommand{\pythia}{P\protect\scalebox{0.8}{YTHIA}\xspace}
\newcommand{\dipsy}{\protect\scalebox{0.8}{DIPSY}\xspace}

\newcommand{\pytppp}{P\protect\scalebox{0.8}{YTHIA}8\xspace}

\newcommand{\sherpa}{S\protect\scalebox{0.8}{HERPA}\xspace}

\newcommand{\alice}{ALICE\xspace}

\newcommand{\angantyr}{Angantyr\xspace}
\newcommand{\fritiof}{Fritiof\xspace}
\newcommand{\etal}{\textit{et~al.}}

\providecommand{\eqref}[1]{eq.~(\ref{#1})\xspace}
\newcommand{\eq}[1]{(\ref{#1})\xspace}
\renewcommand{\eqref}[1]{eq.~(\ref{#1})\xspace}

\newcommand{\eqsref}[1]{eqs.~(\ref{#1})\xspace}

\newcommand{\tab}[1]{\ref{#1}}
\newcommand{\tabref}[1]{table~\tab{#1}}

\newcommand{\fig}[1]{\ref{#1}}
\newcommand{\figref}[1]{figure~\fig{#1}}
\newcommand{\figrefs}[1]{figures~\fig{#1}}

\newcommand{\citeref}[1]{ref.~\cite{#1}}
\newcommand{\citerefs}[1]{refs.~\cite{#1}}

\newcommand{\sect}[1]{\ref{#1}}
\newcommand{\sectref}[1]{section~\sect{#1}}
\newcommand{\sectrefs}[1]{sections~\sect{#1}}

\newcommand{\appref}[1]{appendix~\sect{#1}}

\def\text{\mathrm}
\def\eg{\emph{e.g.}}
\def\ie{\emph{i.e.}}
\def\cf{\emph{c.f.}}
\def\etc{\emph{etc.}}
\def\mrm#1{\mathrm{#1}}
\def\mbf#1{\mathbf{#1}}
\def\sub#1{\ensuremath{_{\mrm{#1}}}}

\def\subtot{\sub{tot}}

\def\subabs{\sub{abs}}

\def\subel{\sub{el}}

\def\subdtot{\sub{D}}
\def\subsdp{\sub{Dp}}
\def\subsdt{\sub{Dt}}
\def\subdd{\sub{DD}}
\def\subdiff{\sub{diff}}
\def\subinel{\sub{inel}}

\def\subwt{\sub{Wt}}
\def\subwp{\sub{Wp}}

\def\sigNN{\ensuremath{\sigma^{\scalebox{0.5}{\textit{NN}}}}}
\def\bNN{\ensuremath{b_{\mu\nu}}}

\def\pom{{\ensuremath{\mathrm{I\!P}}}}
\def\ppom{\ensuremath{\mathrm{p}\pom}}
\def\sqrts{\ensuremath{\sqrt{s}}}

\newcommand{\sqrtsNN}{\ensuremath{\sqrt{s_{\mbox{\tiny $N\!N$}}}}}

\newcommand{\pomeron}{Pomeron\xspace}

\def\llangle{\left\langle}
\def\rrangle{\right\rangle}

\def\prot{\ensuremath{\mrm{p}}}
\def\pp{\ensuremath{\mrm{pp}}}
\def\pn{\ensuremath{\mrm{pn}}}
\def\np{\ensuremath{\mrm{np}}}
\def\nn{\ensuremath{\mrm{nn}}}

\def\pd{\ensuremath{\mrm{pd}}}
\def\pA{\ensuremath{\mrm{p}A}}

\def\pPb{\ensuremath{\mrm{pPb}}}

\def\eA{\ensuremath{\mrm{e}A}}

\def\ep{\ensuremath{\mrm{ep}}}
\def\AA{\ensuremath{AA}}
\def\NN{\ensuremath{NN}}

\def\PbPb{\ensuremath{\mrm{PbPb}}}
\def\XeXe{\ensuremath{\mrm{XeXe}}}

\def\pd{\ensuremath{\mrm{pd}}}

\def\ColText(#1,#2)[#3]#4{\Text(#1,#2)[#3]{#4}}

\def\showcommentsflag{0}
\newcommand{\showcomments}{\def\showcommentsflag{1}}

\newcounter{commentcounter}%
\newcommand{\comment}[1]{\ifnum\showcommentsflag > 0%
\addtocounter{commentcounter}{1}%
{{\Red{\ensuremath{\ddagger^{\arabic{commentcounter}}}}}}%
\marginpar{\raggedright\tiny\it{{\Red{\ensuremath{\ddagger^{\arabic{commentcounter}}}}} {#1}}}
\fi%
}
\newcommand{\commentdel}[2]{\ifnum\showcommentsflag > 0%
\Red{\sout{#1}}\comment{#2}%
\fi
}
\newcommand{\commentadd}[2]{\ifnum\showcommentsflag > 0%
\comment{#2}\Red{#1}%
\else
#1
\fi
}
\newcommand{\commentchange}[3]{\ifnum\showcommentsflag > 0%
\Red{\sout{#2}}\comment{#3}\Red{#1}%
\else
#1
\fi
}
\newcommand{\nocomment}[1]{\ifnum\showcommentsflag > 0%
{\tiny\it\Red{\{#1}\}}
\fi%
}
\newcommand{\nocommentdel}[1]{\ifnum\showcommentsflag > 0%
\Red{\sout{#1}}%
\fi
}
\newcommand{\nocommentadd}[1]{\ifnum\showcommentsflag > 0%
\Red{#1}%
\else
#1
\fi
}
\newcommand{\nocommentchange}[2]{\ifnum\showcommentsflag > 0%
\Red{\sout{#2}}\Red{#1}%
\else
#1
\fi
}

\showcomments
\preprint{LU-TP 18-19\\
  MCnet-18-12\\
  arXiv:1806.10820 [hep-ph]\\
}

\title{
  The \angantyr model for Heavy-Ion Collisions in
  \pytppp
  \footnote{Work supported in part by the Swedish
    Research Council, contracts number 2016-03291, 2016-05996 and
    2017-0034, in part by the European Research Council (ERC) under
    the European Union’s Horizon 2020 research and innovation
    programme, grant agreement No 668679, and in part by the MCnetITN3
    H2020 Marie Curie Initial Training Network, contract 722104.  }}

\author{Christian Bierlich, Gösta Gustafson, Leif Lönnblad and Harsh Shah\\
  Dept.~of Astronomy and Theoretical Physics,
  Sölvegatan 14A, S-223 62 Lund, Sweden\\
  E-mail: \email{Christian.Bierlich@thep.lu.se},
  \email{Gosta.Gustafson@thep.lu.se},
  \email{Leif.Lonnblad@thep.lu.se}, and \email{Harsh.Shah@thep.lu.se}}
\abstract{We present a new model for building up complete exclusive
hadronic final states in high energy nucleus collisions. 
It is a direct extrapolation of high energy pp collisions
 (as described by \pythia), and thus bridges a large part of the existing
 gap between heavy ion and high energy physics phenomenology. 
The model is inspired by the old \fritiof model and the notion of wounded nucleons.
 Two essential features are the treatment of multi-parton interactions and
 diffractive excitation in each $NN$ sub-collision. Diffractive excitation 
is related to fluctuations in the nucleon partonic sub-structure, 
and fluctuations in both projectile and target are here included for the first time. 
The model is able to give a good description of general final-state properties such 
as multiplicity and transverse momentum distributions, both in \pA\ and \AA\ collisions.
 The model can therefore serve as a baseline for understanding the non-collective 
background to observables sensitive to collective behaviour. As \pythia does not 
include a mechanism to reproduce the collective effects seen in pp collisions, 
such effects are also not reproduced by the present version of Angantyr. 
Effects of high string density, shown to be able to reproduce \textit{e.g.} higher 
strangeness ratios and the ridge in pp, will be added in future studies.}

\begin{document}

\sloppy
\newpage
\section{Introduction}
\label{sec:intro}

At hadron collider experiments at RHIC and LHC, protons as well as
large nuclei, are collided, and the results are interpreted to obtain
better knowledge about the dynamics of the fundamental interactions at
high energies.
The strong nuclear force plays a central role, but 
the studies of proton--proton (\pp) collisions and heavy ion collisions 
respectively, are often carried out in quite different ways.

In the case of \pp\ collisions, so--called "general purpose Monte
Carlo event generators", such as \sherpa \cite{Gleisberg:2008ta},
Herwig~7 \cite{Bellm:2015jjp} and \pytppp \cite{Sjostrand:2014zea},
have been established as cornerstones in aiding our
understanding. These event generators have over the last three decades
succeeded in simultaneously simulating the dynamics of strong and
electroweak processes from very high momentum transfer scales where
perturbation theory is applicable, down to scales around
$\Lambda_{QCD}$, where one must rely on models inspired by analogies
to electrodynamics or results from lattice QCD. This has resulted in a
remarkably precise description of the majority of observations in
proton--proton collisions, which both further experimental and
theoretical developments often rely heavily upon.

In high energy heavy ion collisions, the landscape is quite
different. Here efforts are more often directed towards signals for
the formation of the Quark--Gluon Plasma (QGP), and studies of its
properties. The existence of such a phase is demonstrated in lattice
calculations and it is presumed to have existed in the hot, early
Universe. In this area event generators also exist, but are usually
more "special purpose" than "general purpose", each attempting to
describe a specific array of observations ascribed to the formation of
a QGP. Event generators generating full exclusive events also exist,
and the ones most frequently used in analyses investigating particle
production mechanisms are, arguably, EPOS-LHC \cite{Pierog:2013ria},
AMPT \cite{Lin:2004en} and HIJING \cite{Wang:1991hta}. At least for
the bulk event properties, these three generators have for many years
defined the "golden standard" for Monte Carlo comparisons to
experimental data. In \sectref{sec:other-models} we outline some of
the main similarities and differences between these models and our
own.

Several features, which in heavy ion physics are interpreted as a QGP
effect, are also observed in \pp\ collisions at the LHC, which may
indicate that the dynamics at play in these two types of collision
systems are in fact very similar. Two typical examples are enhanced
strangeness \cite{ALICE:2017jyt} and the formation of a "ridge"
\cite{Khachatryan:2016txc}. This immediately raises a challenge for
the general purpose \pp\ event generators and their underlying
models. If a QGP is indeed formed even in \pp\ collisions, then the
effects of such a formation should be included. On the other hand, if
the flow-like effects in \pp\ collisions have a different,
non-thermal, origin,
then it might be possible to capture the general features of nuclear
collisions by adding a nuclear structure "on top" of existing \pp\
models. 


In the present paper we will primarily address the second of
these possibilities, presenting a model, henceforth called
``\angantyr'', which is an extrapolation of \pp\ dynamics 
to collisions with nuclei with a minimum of adjustable 
parameters. In this way it forms a bridge between heavy ion and 
high energy hadron phenomenology. \angantyr
is a generalisation to \AA\ collisions of the model for 
\pA\ scattering in \citeref{Bierlich:2016smv}, which was
able to reproduce general features in \pA\ collisions, like
multiplicity as a function of (measured) centrality, rapidity
distributions, and to a certain degree also $p_\perp$
distributions. Like \pytppp and the 
model in \citeref{Bierlich:2016smv}, \angantyr does not include
an assumption of a hot thermalised medium. The model can therefore
 serve as a baseline for understanding
 the non-collective background to observables sensitive to
 collective behaviour.

Before discussing the generalisation to heavy ion collisions, we want
to discuss some features of \emph{high energy pp scattering}, which are 
important for this generalisation. 

First, as will be discussed in more detail below, diffractive excitation is important.
At high energies the real part of the \pp\ amplitude is small, and usually 
neglected in applications to collisions with nuclei.
Diffraction (elastic scattering and diffractive excitation) is then
the shadow of absorption into inelastic (non-diffractive)
channels. Absorption is here specified by colour exchange between
projectile and target, while diffraction corresponds to colour neutral
(Pomeron) exchange. In the Good--Walker formalism diffractive
excitation is then part of the diffractive beam, when the projectile
mass eigenstate (the proton) is a (coherent) linear combination of 
scattering eigenstates with different absorption probability. These eigenstates 
have in \citerefs{Hatta:2006hs, Avsar:2006jy} been interpreted as different parton cascades.

Secondly multiple partonic sub-collisions are very important at high energies.
Here we use the scheme from \citeref{Sjostrand:1987su}, 
as implemented in \pytppp, to describe inelastic non-diffractive events. Hard
scattering is also seen in diffractive events, and here we use the Ingelman--Schlein 
formalism \cite{Ingelman:1984ns}, which is also included in the \pytppp package.

A generalisation of the formalism for \pp\ collisions to an event generator 
for \emph{\pA\ and \AA\ collisions} will have \emph{four separate components}:

(i) It is necessary to determine nucleon positions within the colliding nuclei.
Here a number of MCs are already available to generate nucleon distributions, 
see \eg\ \citerefs{Broniowski:2007nz,
Rybczynski:2013yba,Alvioli:2009ab,Alvioli:2011sk}.

(ii) One has to calculate the number of interacting nucleons and binary
\NN\ collisions. This is generally performed using the Glauber
formalism~\cite{Glauber:1955qq, Miller:2007ri}. This formalism is
based on the eikonal approximation in impact parameter space, where
the projectile nucleon(s) are assumed to travel along straight lines
and undergo multiple sub-collisions with nucleons in the target. The
importance of including diffractive excitation was early pointed out
by Gribov~\cite{Gribov:1968jf}, but has often been neglected also in
recent applications (see \eg\ the review by Miller \etal\
\cite{Miller:2007ri})\footnote{As an example, in many analyses the \NN\
interaction has been approximated by a ``black disk model'', where
diffractive excitation of individual nucleons is completely neglected.}.
As mentioned above, diffractive excitation is a consequence of
fluctuations in the nucleon substructure. An important point is then that 
a nucleon in the projectile is fixed in the same state during its passage through
the target nucleus. (And similarly the state of a target nucleon is fixed 
through the projectile nucleus.) 

Fluctuations in the \emph{projectile} proton in \pA\ collisions was
studied by Heiselberg \etal~\cite{Heiselberg:1991is}, for estimates of
the number of individual $NN$ sub-collisions. This formalism was
further developed in several papers (see \citerefs{Blaettel:1993ah,
 Alvioli:2013vk, Alvioli:2014sba, Alvioli:2014eda} and further
references in there). It is often referred to as the
``Glauber--Gribov'' colour fluctuation model (GGCF or just GG), and is
used in several experimental analyses, \eg\ in
\citerefs{Aad:2015zza, Adam:2014qja}. 

As discussed in \citeref{Bierlich:2016smv}, taking averages over
target nucleon states is enough for calculations of cross sections and
the number of wounded nucleons in \pA\ collisions, \emph{provided}
diffractively excited nucleons are also counted as wounded nucleons.
For a generalisation to \AA\ collisions it is, however, necessary to
take into account individual fluctuations in both projectile and
target nucleons. As far as we know, \angantyr is the first model where 
this condition is satisfied.

(iii) One must estimate the contribution to the final state from
each interacting nucleon. The \angantyr model is here inspired by the
old \fritiof model for \pA\ and \AA\ collisions
\cite{Andersson:1986gw, Pi:1992ug} and the notion of ``wounded'' nucleons
\footnote{This is also the case for the HIJING model.}.
Bia\l as, Bleszy\'{n}ski, and Czy\.z \cite{Bialas:1976ed} showed that the 
production of soft particles is determined by the number of wounded 
(or participant) nucleons, rather than the number of individual \NN\
sub-collisions. (The latter was later seen to be correlated to hard processes, 
like production of high $p_\perp$ particles or vector bosons.)
In the early \fritiof model \cite{Andersson:1986gw} it was
assumed that an interacting nucleon suffers a longitudinal momentum
exchange with a distribution $\sim dQ/Q$, leading to an excited mass
$\sim dM^2/M^2$. When hadronising like a colour string this gives \emph{on
average} a triangular distribution in rapidity. This behaviour was also later
obtained by Bia\l as and Czy\.z in an analysis of dAu collisions at
RHIC~\cite{Bialas:2004su}.

The \fritiof model did not explicitly include diffractive excitation.
We note, however, that if the mass distribution
for diffractive excitation can be approximated by
$dP\propto dM^2/M^2$, then the contribution from a diffractively
excited nucleon is very similar to the contribution from an average
wounded nucleon in the \fritiof model or from the analysis in
\citeref{Bialas:2004su}. The wounded nucleons in \fritiof can
therefore effectively represent both non-diffractively and
diffractively wounded nucleons.

(iv) At high energies, the \textit{hard} partonic sub--collisions (scaling with 
\NN\ sub-collisions rather than wounded nucleons) play a very essential role.
It is therefore necessary to account for those specifically
in events with multiple \NN\ collisions, \eg\ 
when one projectile nucleon interacts with several target nucleons (or \emph{vice versa}).
In \citeref{Bierlich:2016smv} we introduced the concept of
\emph{primary} and \emph{secondary absorptive} interactions, when a
projectile nucleon is interacting absorptively with more than one
target nucleon. The corresponding \NN\ parton-level event could be
generated using the full multi-parton interaction (MPI) machinery in
\pytppp, for both absorptive and diffractive interactions. To
generate fully exclusive final states in \AA\ collisions, we then have to
calculate all sub-collisions between a nucleon $\mu$ in the projectile
and nucleon $\nu$ in the target, study the number of multiple
sub-interactions for all nucleons $\mu$ and $\nu$, and here separate
diffractive from non-diffractive (absorptive) interactions. This
process is fully described in \sectref{sec:woundet-to-fs}.

We now return to the question of QGP formation. In the current version
of \angantyr the generated partonic states are hadronised using the
string fragmentation model in \pytppp, without including any
final-state collective effects. In this way the model can be used as a
starting point for implementing and analysing new models for
collectivity. As an example we showed \cite{Bierlich:2015rha} that an
enhanced strangeness production can be expected in (high multiplicity)
\pp\ collisions, due to overlapping colour strings forming ``ropes'', in
agreement with experimental observations \cite{ALICE:2017jyt}.
Furthermore we demonstrated in
\citerefs{Bierlich:2014xba,Bierlich:2017vhg} that the enhanced
density also ought to give an outward pressure, which may explain the
observed flow-like effects in \pp\ scattering.

In the present version of the model we limit ourselves to general
features like distributions of particle density in rapidity and
$p_\perp$, postponing a discussion of flow-like effect to a coming
publication. We would like to emphasise, however, that the model can
still be used as an important tool for understanding non-flow effects
on experimental observables designed to measure flow and other
collective behaviours.

\FIGURE[ht]{
  \includegraphics[width=0.9\textwidth]{figs/eventgen-100.mps}
  \caption{\label{fig:flowchart}Flowchart showing the programmatic
    structure of \angantyr.  In order to make predictions for heavy
    ion collisions, several parts of a normal \pytppp simulation needs
    to be modified, and tuned accordingly. In the flowchart we
    illustrate how each separate part is tuned to either $e^+ e^-$,
    \ep\ or \pp\ data, while no tuning is done to heavy ion data.}
}

In \figref{fig:flowchart} we show how the structure described above is
put together and tuned in the concrete simulation. Since all parts of
the simulation; GG colour fluctuations to generate the number of
sub-collisions, the \pytppp MPI model, the parton shower and the
hadronisation model rely on a number of parameters, these parameters
need to be tuned, and a large part of this paper describes how
this procedure is carried out. We want to emphasise from the
beginning that all parts are tuned to data from collisions of smaller
systems, $e^+ e^-$, \ep\ and \pp, and no tuning is done to heavy ion
data. The results can thus be regarded as real predictions depending
only on the chosen extrapolation procedure, and not a specific choice
of parameters.

The layout of the paper follows the workflow of the generation
procedure as shown in \figref{fig:flowchart}, and implemented in
\pytppp. In \sectref{sec:nucsel} we discuss how to calculate the
number of wounded nucleons and the number of individual $NN$
sub-collisions. Here we include fluctuations both in the distribution
of nucleons in the nuclei and in the individual nucleon states, both
for nucleons in the projectile nucleus and in the target nucleus. We
note that a projectile (target) nucleon is fixed in the same
diffractive eigenstate through the passage through the target
(projectile) nucleus. If it is then not absorbed, it may end as
diffractively excited, when projected to the system of mass
eigenstates.
Then in \sectref{sec:woundet-to-fs} we discuss how to generate the
parton-level sub-events for the different kinds of sub-collisions, and
in \sectref{sec:stacking} we describe the procedure for stacking these
sub-events together into complete exclusive hadronic final-states in
\AA. In \sectref{sec:sasd} we then make a digression to discuss the
details of the generation of secondary absorptive sub-collisions,
before we present some sample results in
\sectref{sec:results}. In \sectref{sec:uncertainties} we discuss model
uncertainties, especially related to our treatment of secondary absorptive
sub-collisions. Finally we discuss differences and similarities
between our approach and other heavy ion event generators in
\sectref{sec:other-models}, before presenting some conclusions and an
outlook in \sectref{sec:conclusion-outlook}.

\section{Nucleon-nucleon sub-collisions in \boldmath\pA\ and \AA}
\label{sec:nucsel}

In high energy \pp\ collisions the real part of the amplitude $A$ is
small. If this can be neglected, we can define the real
quantity \begin{equation} T\equiv \operatorname{\mathbb{I}m}\{A\} =
  1-S.
\label{eq:T}
\end{equation}
If diffractive excitation also can be neglected, the elastic cross
section is just the shadow of the absorption, which in impact
parameter space is determined by the probability $1 - S^2$. 
The inelastic cross section is then simply the difference between 
the two. The elastic, total, and inelastic \pp\ cross sections are then given by
$d\sigNN\subel/d^2b = T(\mbf{b})^2$,
$d\sigNN\subtot/d^2b = 2T(\mbf{b})$, and
$d\sigNN\subabs/d^2b = 2T(\mbf{b})-T^2(\mbf{b})$ respectively.

The formulations of high energy nucleus collisions in terms of
individual nucleon--nucleon interactions was carried out by Glauber in
a pioneering paper in \citeref{Glauber:1955qq}. In this paper
several kinds of fluctuations were neglected. As pointed out by
Gribov, and discussed in the introduction, diffractive excitation of
individual nucleons is essential, both for cross sections and for
final state properties. The Glauber theory is formulated in impact
parameter space, where cross sections can be directly interpreted as
probabilities. It is then most convenient to include diffractive
excitation using the Good--Walker formalism \cite{Good:1960ba}, as the
result of fluctuations in the nucleon wave functions. In this section
we shortly discuss the Glauber and Good--Walker formalisms for
estimating scattering cross sections and distributions of wounded
nucleons and $NN$ sub-collisions. The discussion of effects on the
properties of exclusive final states will be presented in
\sectref{sec:woundet-to-fs}.

\subsection{Glauber formalism}
\label{sec:Glauber}

The Glauber formalism is based on the eikonal approximation in
transverse coordinate space. Here the projectile nucleon(s) travel
along straight lines, and undergo multiple sub-collisions with small
transverse momenta. Multiple interactions correspond to a convolution
of the individual $S$-matrices in transverse momentum space, which in
transverse coordinate space simplifies to a product.

We let $\mbf{b}_\mu$ and $\mbf{b}_\nu$ denote the set of positions in
impact parameter space for the nucleons in the projectile and target
nucleus respectively, and $\mbf{b}$ the separation between the centres
of the colliding nuclei. The $S$-matrix for scattering between nucleus
A and nucleus B is the given by
\begin{equation}
S^{AB}(b) = \prod_{\mu=1}^A \prod_{\nu=1}^B S^{(N_\mu, N_\nu)}(b_{\mu\nu}).
\label{eq:glauber}
\end{equation}
Here $\mbf{b}_{\mu\nu}=\mbf{b}_\mu+\mbf{b}-\mbf{b}_\nu$ is the
relative separation between the two colliding nucleons $N_\mu$ and
$N_\nu$. For \pA\ collisions the product over $\mu$ contains only the
projectile proton with $b_\mu=0$.

As mentioned above, fluctuations were neglected in Glauber's original
paper. In the \textit{optical limit}, with a smooth distribution of
nucleons in the nuclei, and where the size of the nuclei are large
compared to the range of the \NN\ interaction, the resulting
nucleus-nucleus cross sections can be calculated analytically.

\subsection{Fluctuations}

\subsubsection{Nucleus geometry}

The simplest way to include fluctuations in the nucleon positions,
$\mbf{b}_\mu$, within a nucleus, is to randomly distribute the $A$
nucleons in three-dimensional space according to a Woods--Saxon
distribution. More advanced models include correlations in form of a
hard repulsive core (\eg~\cite{Broniowski:2007nz,Rybczynski:2013yba}),
or a more sophisticated description of the two- (or three-) particle
correlations between the nucleons within the nucleus
\cite{Alvioli:2009ab,Alvioli:2011sk}. Fluctuations in the geometry is
taken into account, when new nucleus states are generated for each new
event.

\subsubsection{Fluctuations in the individual \boldmath\NN\ interactions, and the
  Good--Walker formalism} 
\label{sec:GW}

We here shortly describe the Good--Walker formalism for diffractive
excitation, assuming for simplicity first that a fluctuating
projectile collides with a non-fluctuating target. For a projectile
particle with an internal substructure, it is possible that the mass
eigenstates differ from the elastic scattering eigenstates. We denote
the mass eigenstates $\Psi_i$, with the projectile in the ground state
(\eg\ a proton) denoted $\Psi_0$, while $\Phi_l$ are the eigenstates
to the scattering amplitude $T$, with $T \Phi_l = t_l \Phi_l$. The
mass eigenstates are linear combinations of the scattering
eigenstates, $\Psi_i = \sum_l a_{il} \Phi_l$. The scattering can now
be regarded as a measurement, where the projectile "has to choose" one
of the eigenvalues $t_l$, with probability $|a_{0l}|^2$.

The elastic amplitude for the ground state projectile is then given by
$\langle \Psi_0|T|\Psi_0\rangle = \sum_l |a_{0l}|^2 t_l \equiv \langle
T \rangle$, where $\langle T \rangle$ is the expectation value for the
amplitude $T$ for the projectile. The elastic cross section is then
given by
\begin{equation}
d\sigma_{el}/d^2 b =\langle T(b) \rangle^2. 
\end{equation}
We here work in impact parameter space, and the amplitude depends on
$b$. The total diffractive scattering $\sigma\subdiff$ (including the
elastic) is the sum of transitions to all states $\Phi_l$:
\begin{equation}
d\sigma\subdiff/d^2 b = \sum_l \langle \Psi_0|T|\Phi_l\rangle \langle \Phi_l
|T|\Psi_0\rangle = \langle \Psi_0| T^2| \Psi_0 \rangle, 
\end{equation}
where we have used the fact that $\Phi_l$ form a complete set of
states. Subtracting the elastic cross section we then get the cross
section for diffractive excitation, which thus is given by the
fluctuations in the scattering amplitude:
\begin{equation}
d\sigma\subdtot/d^2 b = \langle T^2 \rangle - \langle T \rangle^2.
\end{equation}

In a nucleon-nucleon collision both the projectile and the target are
fluctuating, leading to single diffractive excitation of the
projectile or the target, as well as to double diffraction. The
different cross sections are then given by

\begin{eqnarray}
  \label{eq:nnxsecs}
  d\sigNN\subtot/d^2b &=& \llangle 2T(\mbf{b})\rrangle_{p,t}\nonumber\\
  d\sigNN\subabs/d^2b &=& \llangle 2T(\mbf{b})-
                          T^2(\mbf{b})\rrangle_{p,t}\nonumber\\
  d\sigNN\subel/d^2b &=& \llangle T(\mbf{b})\rrangle^2_{p,t}\nonumber\\
  d\sigNN\subsdt/d^2b &=& \llangle\llangle T(\mbf{b})\rrangle_p^2\rrangle_t -
                          \llangle T(\mbf{b})\rrangle^2_{p,t}\nonumber\\
  d\sigNN\subsdp/d^2b &=& \llangle\llangle T(\mbf{b})\rrangle_t^2\rrangle_p -
                          \llangle T(\mbf{b})\rrangle^2_{p,t}\nonumber\\
  d\sigNN\subdd/d^2b &=& \llangle T^2(\mbf{b})\rrangle_{p,t} -
                         \llangle\llangle T(\mbf{b})\rrangle_p^2\rrangle_t -
                         \llangle\llangle T(\mbf{b})\rrangle_t^2\rrangle_p +
                         \llangle T(\mbf{b})\rrangle^2_{p,t}.
\end{eqnarray}
Here $\llangle\cdots\rrangle_p$ and $\llangle\cdots\rrangle_t$ are
averages over projectile and target states respectively, and
subscripts $Dt$, $Dp$ and $DD$ stand for single diffractive excitation
of the target, the projectile, and double diffraction respectively.
We note here that while the total cross section depends only on the
average of $T(b)$, all other cross sections include also average of
$T^2$ over projectile and/or target states. However, if wounded target
nucleons include also diffractively excited nucleons, we see that the
corresponding cross section for a wounded target nucleon,
$\sigNN\subwt\equiv\sigNN\subabs+\sigNN\subsdt+\sigNN\subdd$, can be
written
\begin{equation}
  \label{eq:wtxsec}
  d\sigNN\subwt/d^2b=\llangle 2\llangle T(\mbf{b})\rrangle_t-
  \llangle T(\mbf{b})\rrangle^2_t\rrangle_p=1-\llangle 
  \llangle S(\mbf{b})\rrangle^2_t\rrangle_p.
\end{equation}

\subsubsection{Fluctuations in collisions with nuclei}
\label{sec:flucnuc}

The expression for the amplitude $T(\mbf{b}) = (1-S(\mbf{b}))$ in
\eqref{eq:nnxsecs} can be directly inserted into the amplitude for
collisions with nuclei in \eqref{eq:glauber} (as before we neglect the
real part of the amplitudes). The scattering probability can be
regarded as a measurement, after which a projectile nucleon is in one
of the eigenstates to the amplitude $T$, and thus also to the
probability for colour connection (the absorption probability)
$2T-T^2$. Thus all nucleons are frozen in the same state during the
scattering process. (We here neglect the modification when one or a
few partons have changed colour in the first encounter.) As a
consequence the average of the \AA\ amplitude in \eqref{eq:glauber}
will include also higher powers of $T$.
However, for \pA\ collisions the multiple sub-collisions imply that
the total and wounded nucleon cross sections contain higher moments
with respect to \textit{projectile} fluctuations, but still only the
average over the uncorrelated \textit{target nucleon} states. We also
note that these moments should be taken for fixed impact
parameters. Thus, to calculate the ratios of \eg\ the
\textit{integrated} elastic and total cross sections, it is also
necessary to know the $b$-distribution of the amplitude.

To visualise the effects of fluctuations and diffractive excitation we
can study a simple example with a proton colliding with two target
nucleons, with and without fluctuations. We assume in both cases that
the inelastic $NN$ cross section (including diffractive excitation) is
$d\sigma^{NN}_{\mathrm{inel}}/d^2b = 3/4$.

Case 1: No fluctuations. The \NN\ amplitude and $S$ matrix are
$T^{\mathrm{NN}}=1/2$ and $S^{\mathrm{NN}}=1-T^{\mathrm{NN}}=1/2$. The
inelastic cross section when hitting \emph{two} target nucleons is
then from \eqsref{eq:glauber}, \eq{eq:wtxsec} given by
$d\sigma_{\mathrm{inel}}/d^2b =15/16$, ($\sigma_{\mathrm{D}}=0$).

Case 2: With fluctuations. We neglect the fluctuations in the target,
and assume that the projectile state is given by
$\Psi_{0}= (1/\sqrt{2})(\Phi_1 + \Phi_2)$. The states $\Phi_1$ and
$\Phi_2$ are here diffractive eigenstates with eigenvalues $t_1=0$ and
$t_2=1$. From \eqref{eq:nnxsecs} we get for collision with \emph{one}
target nucleon $d\sigma_{abs}/d^2b =1/2$ and $d\sigma_{D}/d^2b
=1/4$. For \emph{two} target nucleons we get actually the same
result. If the projectile is in state $\Phi_1$ it misses both targets,
and if in state $\Phi_2$, it is absorbed already in the first
one. Thus the inelastic cross section is only 3/4 (1/2 for absorption
and 1/4 for diffractive excitation) compared to 15/16 in the
non-fluctuating case.\footnote{This case is actually essentially the
  ``fluctuating gray disk model'' discussed in section
  \ref{sec:fluctmod} and used in analyses of RHIC data by PHENIX.}

\subsection{From cross sections to probabilities}
\label{sec:probabilities}

The absorptive cross section in impact parameter space shown in
\eqref{eq:nnxsecs} is the average of the expression
$2T_{i,k}(\mbf{b})- T_{i,k}^2(\mbf{b}) \equiv 1 - S^2_{i,k}(\mbf{b})$,
where $T_{i,k}$ is the scattering amplitude (and $S_{i,k}$ the
$S$-matrix) for a projectile proton in state $i$ colliding with a
target in state $k$. This expression is always $\le 1$, and it can be
directly interpreted as the probability for an absorptive interaction
between the projectile and the target. (Such an interpretation is not
possible in transverse momentum space, where the cross section has the
dimension of momentum to the fourth power.)

We note, however, that neither the elastic cross section nor
diffractive excitation is the average of an expression depending on
only $i$ and $j$. (The elastic cross section can be written
$\sum_{i,j,k,l} T_{i,k}T_{j,l}$.) When the interaction is driven by
absorption, elastic scattering and diffractive excitation is the
result of interference between waves, which missed the absorbing
target. The cross section for this diffractive scattering is also
bounded by 1, and together with absorption it gives a total cross
section bounded by 2. A consequence of this feature is that to
properly generate events including diffractive excitation for \AA\
collisions in an event generator, it is necessary to, for every
projectile nucleon, $\mu$, in state $i$ calculate the average of the
amplitude $T_{i,k}(b_{\mu\nu})$ over all states of each target
nucleon, $\nu$, for all impact parameters $b_{\mu\nu}$ (and similarly
all averages over projectile states $i$ for every target state
$j$). This would give a very slow program, and in \sectref{sec:AAfluc}
we show how to obtain a good approximation.

In \pA\ collisions the picture is, however, much simplified. From
\eqref{eq:wtxsec} we note that although the wounded nucleon cross
section $d\sigNN\subwt/d^2b$ contains one piece from absorption and
one piece from diffraction, the sum is always bounded by 1. The
question whether a target nucleon $\nu$ will be a wounded target (with
this definition) in a sub-collision with a projectile in state $i$ can
only be answered by yes or no. Therefore the answer yes must have the
probability given by the cross section in \eqref{eq:wtxsec}. This is
used \eg\ in applications of the Glauber--Gribov model described in
\sectref{sec:fluctmod}.

\subsection{\boldmath\NN\ scattering models used in Glauber calculation Monte Carlos}


\subsubsection{Non-fluctuating models}

The simplest approximation for the \NN\ amplitude is the
\textbf{``black disk model''}, where the target acts as a black
absorber. This model has been frequently used in experimental analyses
(see \eg\ the review in \citeref{Miller:2007ri}). It is then assumed
that two colliding nucleons are interacting, if their separation in
impact-parameter is smaller than some radius $R$. The cross sections
are here given by
$\sigma_{inel}=\sigma_{el}=\sigma_{tot}/2 = \pi R^2$. As there are no
fluctuations, the cross section for diffractive excitation is zero.
It is then obvious that the model cannot reproduce the experimental
results, which satisfy
$\sigNN\subel \approx \sigNN\subdtot \approx \sigNN\subtot /4$. (Here
$\sigNN\subdtot$ denotes the sum of single and double diffractive
excitation.) In the literature it is common to set
$2\pi R^2=\sigma\subtot^{(exp)}$, which reproduces the experimental
total cross section, but neither the elastic nor the inelastic cross
section (when the latter includes diffractive excitation). In later
studies it has become more common to choose
$\pi R^2=\sigma\subinel^{(exp)}$, which reproduces the total inelastic
cross section, but gives
$\sigma\subtot^{(model)} = 2\, \sigma\subinel^{(model)} \approx 1.5\,
\sigma\subtot^{(exp)}$.

For most applications in \pA\ and \AA, the elastic cross section is
not very important, but we note that it could still be reproduced by
introducing a grayness or opacity of the collision, assuming that
within a radius $R$ the scattering amplitude is a constant $a$ between
0 and 1. $R$ and $a$ can then always be adjusted to reproduce both the
total and the elastic cross sections (and thus also the total
inelastic cross section). Diffractive excitation would, however, still
be absent.

\subsubsection{Models including fluctuations}
\label{sec:fluctmod}

In a variation of the opacity model above, the projectile is instead
fully absorbed with probability $a$. This obviously includes
fluctuations and thus also diffraction. With the value $a=1/2$ we get
the cross section ratios
$\sigNN\subel = \sigNN\subdtot = \sigNN\subtot /4$, in reasonable
agreement with experiments. As the model describes the combined
fluctuations of the projectile and the target, it is here not possible
to separate diffractive excitation of the projectile from that of the
target or from double diffraction.

In the introduction we mentioned the ``Glauber--Gribov'' model for
\pA\ collisions, developed by Strikman and
coworkers~\cite{Heiselberg:1991is, Blaettel:1993ah, Alvioli:2013vk,
  Alvioli:2014sba, Alvioli:2014eda}. It is there assumed that the
fluctuations in the \textit{projectile} can be described by a
distribution in the quantity
$\sigma \equiv \int d^2b\, \langle 2 T(\mbf{b}) \rangle_t$ of the form
\begin{eqnarray}
  \label{eq:strik}
  P\subtot(\sigma)&=&\rho\, \frac{\sigma}{\sigma + \sigma_0} 
\exp\left\{-\frac{(\sigma/\sigma_0 - 1)^2}{\Omega^2}\right\},\nonumber \\
 \sigNN\subtot&=&\int \sigma P\subtot(\sigma)d\sigma.
\end{eqnarray} 
(The second relation follows from \eqref{eq:nnxsecs}.) This formalism,
has been used in analyses of \pPb\ data from LHC, \eg\ by
\citeref{Aad:2015zza}, to estimate the number of wounded or
interacting nucleons, which in turn has been used to estimate the
centrality for the collision. The quantity $\sigma$ is then normally
rescaled so that the integral in \eqref{eq:strik} gives the inelastic
rather than the total cross section. We note that, as the fluctuating
quantity $\sigma$ includes the fluctuations over \textit{projectile}
states, but averages over \textit{target nucleon} states, we see from
\eqref{eq:wtxsec} that what is counted as wounded nucleons includes
diffractively excited nucleons.

As discussed in \sectref{sec:probabilities}, the cross section in
\eqref{eq:wtxsec} also determines the probability distribution for
wounded nucleons, but we want to emphasise that the differential cross
section $\langle T(\mbf{b}) \rangle_t$ is needed for all values of the
impact parameter $b$. In \citeref{Alvioli:2013vk} this is assumed
to be Gaussian $\propto \exp(-b^2/2B(\sigma))$, with a slope parameter
$B(\sigma)$ proportional to $\sigma$, in order to satisfy the
unitarity constraint $T(b)\leq 1$.

In \citeref{Bierlich:2016smv} we investigated the fluctuations in
the nucleon cross sections using Mueller's dipole approach to BFKL
evolution \cite{Mueller:1993rr,Mueller:1994jq} as implemented in the
\dipsy Monte Carlo program \cite{Avsar:2005iz, Avsar:2006jy,
  Flensburg:2011kk}. The model is formulated in impact parameter
space, and includes also a set of sub-leading corrections beyond the
leading-log BFKL approximation. Non-linear effects are introduced by
the "colour swing" mechanism, which suppresses large dipoles,
corresponding to $k_\perp$ below a saturation scale. BFKL evolution is
a stochastic process, and the result was here that the fluctuations
have a longer tail out to large cross sections compared to the
distribution in \eqref{eq:strik}. Rather than the Gaussian suppression
assumed in \cite{Heiselberg:1991is}, we found a distribution more
similar to a Log-normal for the $b$-integrated and target-averaged
$\sigma$:
\begin{equation}
  \label{eq:lognormal}
  P\subtot(\ln \sigma) = \frac{1}{\Omega\sqrt{2\pi}}
  \exp\left(-\,\frac{\ln^2(\sigma/\sigma_0)}{2\Omega^2} \right).
\end{equation}
To also describe the
$b$-dependence of $\langle T(b)\rangle_t$, we used a
semi-transparent disk approximation with the elastic amplitude
\begin{equation}
  \label{eq:greydisk}
  \llangle T(\mbf{b},\sigma)\rrangle_t=
  T_0\Theta\left(\sqrt{\frac{\sigma}{2\pi T_0}}-b\right).
\end{equation}
The parameters ($\Omega$ and $\sigma_0$) in $P\subtot(\sigma)$ and
$T_0$ in \eqref{eq:greydisk} could here be fitted to $\sigNN\subtot$,
$\sigNN\subel$ and $\sigNN\subwt$ taken from experimental data, to
obtain a Glauber-like calculation for \pA. Together with the
parton-level stacking also proposed in \cite{Bierlich:2016smv} we then
also obtained a fair description of \eg\ the observable used by ATLAS
in \cite{Aad:2015zza} for estimating centrality, as well as the
corresponding pseudo-rapidity distributions as a function of that
centrality.

We note that the stochastic nature of BFKL evolution has also been
studied by Iancu, Mueller and Munier in \citeref{Iancu:2004es}.
When the probability for a dipole splitting is small, the mean field
approximation in the Balitsky--Kovchegov equation does not properly
describe the probability for rare events with large cross section. In
\citeref{Iancu:2004es} they studied the fluctuations in the
saturation scale, $Q_s$, and showed that for asymptotic energies the
width of the distribution in $\ln(Q_s)$ is growing proportional to
$\sqrt{\bar{\alpha}\ln(s)}$, with a tail to large $Q_s$-values in
qualitative agreement with \eqref{eq:lognormal}.

\subsection{Nucleon fluctuations in \boldmath\AA\ collisions}
\label{sec:AAfluc}

As mentioned in section~\ref{sec:GW}, to study \pA\ collisions also
higher moments over projectile fluctuations are needed. When we now
want to generalise the formalism to \AA\ collisions, both projectile
and target nucleons are frozen under the collision (but still
uncorrelated). This implies that we must be able to calculate not only
$\langle \langle T(b)\rangle_t^n\rangle_p$, but any moment
$\langle \langle T(b)^{n_p}\rangle_p^{n_t}\rangle_t$. To cope with
this situation we need a formalism which can give the amplitude
$T_{ik}(b)$ for any combination of projectile state $i$ and target
state $k$.

We noted that the Log-normal distribution in \eqref{eq:lognormal} is
quite similar to a Gamma-function, and for technical reasons and the
fact that the sum of two Gamma distributed random variables is also
Gamma distributed, we will use that instead to model fluctuations in
the radius, $r$, of a nucleon:
\begin{equation}
  \label{eq:gammadist}
  P(r)=\frac{r^{k-1}e^{-r/r_0}}{\Gamma(k)r_0^k}.
\end{equation}
We then also use a slightly different elastic
amplitude
\begin{equation}
  \label{eq:newdisk}
  T(\mbf{b},r_p,r_t)=T_0(r_p+r_t)
  \Theta\left(\sqrt{\frac{(r_p+r_t)^2}{2 T_0}}-b\right). 
\end{equation}
where the opacity of the semi-transparent disk now depends 
on $r_p$ and $r_t$:
\begin{equation}
  \label{eq:varyT0}
  T_0(r_p+r_t)=\left(1-\exp\left(-\pi(r_p+r_t)^2/\sigma_t\right)\right)^\alpha.
\end{equation}
This introduces two more parameter, $\sigma_t$ and $\alpha$, (besides $k$ and $r_0$
in \eqref{eq:gammadist}) and this varying opacity makes it possible
to get a reasonable fit to all the cross sections in
\eqref{eq:nnxsecs}, as well as the elastic slope parameter
$B=-d\ln\,\sigNN\subel/dt|_{t=0}$, for a wide range of energies.
The result for $\sqrtsNN=5$~TeV is shown in table \ref{tab:ppxsecs}.

\subsubsection{Determining the interaction of nucleon sub-collisions}

We now want to take all pairs of colliding nucleons in an \AA\
collision, and for each of these select which kinds of interactions
are possible. At high energies all nucleons are frozen in their
(random) states during the passage through the opposite nucleus. The
probability for an absorptive interaction between nucleon $\mu$ (in a
state $i$ with radius $r_{i\mu}$) in the projectile and nucleon $\nu$
(in state $k$ with radius $r_{k\nu}$) in the target, is then directly
given by \eqref{eq:nnxsecs} as $P_{abs}=2T_{ik}-T_{ik}^2$, with
$T_{ik}=T(\mbf{b},r_{i\mu},r_{k\nu})$ given by \eqref{eq:newdisk}. To
estimate the probability for a diffractive excitation of a given
nucleon is more difficult, as diffractive excitation is part of the
shadow scattering caused by absorption, to which all encountered
nucleons contribute.

We showed in \citeref{Bierlich:2016smv} that for a given state of a
projectile nucleon the probability that a given target nucleon is
absorptively or diffractively wounded in the interaction is given by
the average over the possible states of the target (\cf\
\eqref{eq:wtxsec}) and that this probability factorises for all
nucleons in the target nucleus. However, in \AA\ the symmetry between
projectile and target complicates things further, as we need both a
specific state and the average over all states for all nucleons.

In \angantyr this is handled by generating two states (one primary,
$r$, and one auxiliary,~$r'$) for each nucleon in the nuclei. The
primary one is used to calculate the probability of an absorptive \NN\
interaction, while the secondary is used to statistically sample the
average state of each nucleon. The algorithm ensures that on average
(over the four possible combinations of states in an \NN\ interaction)
we get the correct probability of the projectile and target nucleon
being absorptively and diffractively wounded.

The technical details of this algorithm is presented in
\appref{sec:gener-absorpt-diffr}, while here we will only show that it
works as expected. In \tabref{tab:ppxsecs} we give an example where we
have fitted the parameterisation of the fluctuations according to
\eqsref{eq:gammadist} -- \eq{eq:varyT0} to the default
parameterisation of the semi-inclusive cross sections in \pytppp.
This default parametrisation \cite{Schuler:1993wr,Donnachie:1992ny}
does not necessarily agree well with cross section measurements from
LHC \cite{Rasmussen:2018dgo}, and it is possible for a user to easily
supply their own cross sections as input to the fit. The last line
denoted ``generated'' shows the results from generating \NN\
collisions in \angantyr for $\sqrts=5$~TeV. We see that the absorptive
cross section comes out close to the input one, and also the wounded
cross sections, $\sigma\subwp$ and $\sigma\subwt$ are reasonably well
reproduced. However, we see that the individual diffractive excitation
cross sections are not reproduced, nor is the elastic ones. However,
for the final states in \AA\ collisions, we are mainly interested in
getting the absorptive and wounded cross section right, so even if our
procedure probably can be improved, we are quite satisfied with this
result.

\TABLE[t]{
  \centering
  \begin{tabular}{|l|c|c|c|c|c|c|c|c|}
    \hline
    & $\sigma\subabs$ & $\sigma\subwp$ & $\sigma\subwt$
    & $\sigma\subsdp$ & $\sigma\subsdt$ & $\sigma\subdd$
    & $\sigma\subel$ & $B$\\
    & (mb) & (mb) & (mb) & (mb) & (mb) & (mb) & (mb) & (GeV$^{-2}$)\\
    \hline
    input     & 47.7 & 61.5 & 61.5 & 6.1 & 6.1 & 7.7 & 18.4 & 20.8 \\
    model     & 47.8 & 61.4 & 61.5 & 5.7 & 5.8 & 7.9 & 18.7 & 24.1 \\
    generated & 47.8 & 61.3 & 61.3 & 11.4 & 11.4 & 2.2 & - & - \\
    \hline
  \end{tabular}
  \caption{Fitting the values of input cross sections for \pp\
    collisions at $\sqrts=5$~TeV and using the resulting fluctuations
    in a generation and different collision types. $B$ is the elastic
    slope $-d\log\,\sigma\subel/dt|_{t=0}$. The cross sections used as
    ``input'' were taken from the default parameterisation in
    \pytppp. The line ``model'' shows the results of a fit to the
    model in \eqsref{eq:gammadist} -- \eq{eq:varyT0}. The line
    ``generated'' finally shows the result of the approximation
    discussed in this subsection and in the appendix. The fitting
    procedure assumed a 5-10\% uncertainty on the input values, and
    the statistical uncertainty on the presented output values are
    around and below 0.5\%. The resulting parameters values in
    \eqsref{eq:gammadist} and \eq{eq:varyT0} were $k=1.80$,
    $r_0=0.407$~fm, $\sigma_t=13.88$~fm$^2$, and
    $\alpha=0.22$.\label{tab:ppxsecs}}}

\section{From wounded nucleons to exclusive final states}
\label{sec:woundet-to-fs}

In the wounded nucleon model, as formulated by Bia\l as and Czyz
\cite{Bialas:2004su}, each wounded nucleon contributes to the final
state multiplicity distribution, according to a single nucleus
emission function $F(\eta)$, giving a total multiplicity of:
\begin{equation}
	\label{eq:woundedMult}
	\frac{dN_{ch}}{d\eta} = w_pF(\eta) + w_tF(-\eta).
\end{equation}
Here $w_{p|t}$ denotes the number of wounded nucleons from left and
right respectively, calculated for a given centrality class, defined
by impact parameter. In the wounded nucleon model, $F(\eta)$ must be
extracted from data, and depends on centrality class
\cite{Barej:2017kcw}, but a crucial feature of the model is that
\eqref{eq:woundedMult} reduces to the \pp\ multiplicity distribution for
$w_p = w_t = 1$.

The \angantyr prescription for generating exclusive final states has
conceptual similarities with the wounded nucleon model. But instead of
extracting an emission function from data, MPI events from \pytppp are
used. We will in this section briefly review the \pytppp MPI model,
and motivate the addition of additional MPIs from multiple wounded
nucleons to the model.

\subsection{Multiparton interactions in \pp\ collisions}
In the \pytppp MPI model \cite{Sjostrand:1987su}, all partonic
sub-collisions are to a first approximation treated as separate QCD
$2\rightarrow 2$ scatterings\footnote{The MPIs are not fully
  uncorrelated, as momentum conservation needs to be obeyed, and the
  parton density corresponding to the extracted parton, is rescaled by
  a factor $(1-x)$.}. Since the cross section diverges at low
$p_\perp$, it is regularised using a parameter $p_{\perp 0}$ which
depends on the collision energy, giving:
\begin{equation}
\label{eq:mpi-model}
  \frac{d\sigma_{2\rightarrow 2}}{dp_\perp^2} \propto
  \frac{\alpha_s^2(p_\perp^2)}{p_\perp^4} \rightarrow
  \frac{\alpha_s^2(p_\perp^2 + p_{\perp 0}^2)}{(p_\perp^2 + p_{\perp 0}^2)^2}.
\end{equation}

\input{figures/MPIladders.tex}

\FIGURE[ht]{
  \begin{picture}(450,220)(10,30)
    \put(0,0){\usebox{\ppMPIladders}}
  \end{picture}
  \caption{\label{fig:mpi-pp} Schematic pictures of multi-parton
    interactions in a \pp\ collision. The $y$-axis should be
    interpreted as rapidity. All initial- and final-state radiation
    has been removed to avoid cluttering. Each gluon should be
    interpreted as having two colour lines associated with it, which
    in the subsequent string hadronisation will contribute to the soft
    multiplicity. In (a) the colour lines for both sub scatterings
    stretches all the way out to the proton remnants, while in (b) and
    (c) the secondary scattering is colour-connected to the primary
    one.}}

This cross section is then folded with parton densities to get a
relative probability for each additional sub-scattering. The densities
are rescaled according to an overlap function using some assumption
about the matter distribution in the colliding protons and an assumed
impact parameter.

In \figref{fig:mpi-pp}a there is an illustration of an event with two
sub-scatterings (in red and black) which we have assumed are both of
the type $gg\to gg$. Note that in the \pythia MPI model all incoming
and outgoing partons would be dressed up with initial- and final-state
radiation, but these have been left out of the figure to avoid
cluttering. With completely uncorrelated sub scattering, one would
assume the colours of the incoming gluons would also be uncorrelated,
and since each gluon carries both colour and anti-colour one would
naively think that in the subsequent hadronisation phase, there would
be four strings stretched between the proton remnants and giving rise
to particle production over the whole available rapidity range. Again
to avoid cluttering of the figures, we ask the reader to simply
imagine two colour lines (strings) stretched along each gluon and that
the vertical axis can be loosely interpreted as rapidity.

Already in the original paper \cite{Sjostrand:1987su} it was realised
that it was basically impossible to reproduce data if each
sub-scattering was allowed to add particles in the whole available
rapidity range. Especially sensitive to this was the multiplicity
dependence of the average particle transverse momenta, and to rectify
this the MPI model in \pythia was modified so that additional
sub-scatterings almost always was colour connected to outgoing partons
in previous sub-scatterings. This is illustrated in
\figref{fig:mpi-pp}b and c, where the colour correlation between the
two sub-scatterings gives rise to a colour flow \emph{as if} they were
(perturbatively) connected. In this way the multiple scatterings can
give rise to increased average transverse momentum from the partons
coming from extra sub-scattering, without increasing the multiplicity
of soft particles due to the strings stretched all the way out to the
proton remnants.

\subsection{Multi-parton interactions in a \pA\ collision}

We now turn to the case of a \pA\ collision and imagine the projectile
proton interacting absorptively with two nucleons in the nuclei. To be
true to the \pythia MPI model we should simply redefine the overlap
function using the matter distribution of the two target nucleons. In
principle this can surely be done, however, technically we found it
almost forbiddingly difficult.

\FIGURE[ht]{
  \begin{picture}(450,220)(0,30)
    \put(0,0){\usebox{\pAMPIladders}}
  \end{picture}
  \caption{\label{fig:mpi-pA} A schematic picture (\cf\
    \figref{fig:mpi-pp}) of multiple scattering between one projectile
    and two target nucleons (\eg\ in a \pd\ collisions). In (a) the
    second interaction is directly colour connected to the first one,
    while in (b) the second nucleon is only diffractively excited by a
    \pomeron exchange. Both cases give rise to final string
    configurations that will contribute in the same way to the final
    state hadron distribution.}}

Instead we note that the handling of colour correlations in the \pp\
model would typically result in string topologies corresponding to the
sketch in \figref{fig:mpi-pA}a. The primary scattering looks like
normal scattering between the projectile and one of the target
nucleons, while the secondary scattering is now between the projectile
and the other target nucleon. Since both target nucleons have been
found to be absorptively wounded, the secondary scattering must be
colour connected to the second target nucleon, while in the direction
of the projectile it looks like a normal secondary scattering.

We also note that we would get the same colour topology, and hence the
same distribution of particles, if the second sub-scattering was a
separate single (high-mass) diffractive excitation event, which in
\pytppp is handled as a \pomeron-proton collision. This is illustrated
in \figref{fig:mpi-pA}b, where the \pomeron is shown as a green zigzag
line. A secondary absorptive wounded nucleon thus
contributes to the final state \emph{as if} the final state particles
were produced in a single diffractive excitation. This similarity
is what we, in the following, will exploit to build up a final state
from primary absorptive interactions and secondary absorptive interactions,
the latter being modelled as single diffractive excitation.

The procedure will therefore be to decide which of the two
absorptive interactions is to be considered the primary one, and treat
this as a completely normal non-diffractive multiple scattering event
in \pythia. The secondary scattering will be generated as a single
diffractive excitation event in \pythia. Also here there may be
additional multiple parton scatterings, but they will be treated as
multiple scatterings in the \pomeron--proton system, which is standard
in the high-mass diffraction machinery in \pythia.

Referring back to \eqref{eq:woundedMult}, this means that we are
modelling the single nucleus emission function $F(\eta)$ using
high-mass diffractive excitation events. We do not expect them to
necessarily look like ordinary diffractive event, but we nevertheless
use the diffractive machinery in \pytppp. In \sectref{sec:sasd} we
will describe how we modify this machinery in order to try to fulfil
the requirement that $F(\eta) + F(-\eta)$ (\ie\ $w_p = w_t = 1$ in
\eqref{eq:woundedMult}) would reproduce the distribution in a normal
non-diffractive \pp\ event in \pytppp.

The two different sub-events are then merged together so that the
elastically scattered proton in the diffractive event is discarded,
and the momentum of the \pomeron is instead taken from remnants of the
projectile proton.

The assumption in \cite{Bierlich:2016smv} was that the momentum
fraction of the \pomeron in such diffractive events can be taken to be
distributed approximately as $dx_{\pom}/x_{\pom}$, which means that the
mass of the diffractive system is given by $dM_X^2/M_x^2$. This is
approximately what one has found for normal high-mass diffractive
events and it is the same assumption as in the old \fritiof model. We
do not have a solid explanation why this should be the case. In
\cite{Bierlich:2016smv} we gave some handwaving arguments based on AGK
cutting rules and the similarity between triple-\pomeron diagrams in
diffractive \NN\ scatterings and (doubly) non-diffractive
proton--deuterium scattering, but in the end the best argument for
this choice is that it seems to work very well.

\subsection{Multi-parton interactions in an \boldmath\AA\ collision}

\FIGURE[ht]{
  \begin{picture}(450,260)(0,30)
    \put(0,0){\usebox{\AAMPIladders}}
  \end{picture}
  \caption{\label{fig:mpi-AA} A schematic picture (\cf\
    \figref{fig:mpi-pA}) of multiple scatterings between two projectile
    and two target nucleons in an \AA\ collision. In (a) there are two
    separate \NN\ collisions, while in (b) and (c) there is one
    primary sub-collision and two secondary ones.}
}

Going one step further in complexity we now consider \AA\
collisions. In \figref{fig:mpi-AA} we illustrate the situation when
two nucleons in one nucleus collides with two nucleons in the other,
and all four possible \NN\ interactions are absorptive. We find that
there is three ways of doing this which are consistent with our \pA\
model. Either, as in \figref{fig:mpi-AA}a, we can model it as two
primary absorptive interactions, or as one primary and two secondary
interactions, where the second of these can either be coupled to the
primary interaction (b) or to the first secondary one (c).

All three cases will give us four absorptively wounded nucleons, and
in \fritiof and the original wounded nucleon model there would be no
distinction between the cases. In the \angantyr model we do, however,
want to differentiate between these, and in the following section we
will describe a procedure to classify all \NN\ interactions in a \AA\
collisions.

\section{Generating and combining parton-level \boldmath\NN\ events}
\label{sec:stacking}

In general, each nucleon in the projectile nucleus may interact with
several nucleons in the target nucleus an vice versa. When building
up the final state by stacking parton level nucleon--nucleon events we
need to concentrate on the most important ones first. This is in line
with the general philosophy in \pytppp, that harder processes always
are considered before softer ones. After having gone through all pairs
of projectile--target nucleons and determined their interactions as
outlined in the previous section, we therefore order all these
interactions in increasing nucleon--nucleon impact parameter,
$b_{\mu\nu}$.

We will then go through this list several times, treating one kind of
interaction at the time, starting with the absorptive interactions, as
they will give the back bones around which we will build up the full
event. As soon as an \NN\ interaction has been selected a
corresponding sub-event will be generated with the standard \pytppp
minimum bias model, and the corresponding nucleons are marked as
already interacted. If an \NN\ interaction is found in the list where
one of the nucleons has already interacted, this will be labelled
\emph{secondary} and the generated sub-event will be added to the
sub-event to which the already interacted nucleon belongs, as
described in \cite{Bierlich:2016smv} and detailed below.

\subsection{Selecting primary absorptive collisions}
\label{sec:gener-prim-absorpt}

The first pass over the potential \NN\ interactions, we will only look
at absorptive interactions. This will give us a set of $N^{'}\subabs$
primary absorptive collisions and a set of $N^{''}\subabs$ secondary
ones (where one of the nucleons already has already been absorptively
wounded in another interaction).

For each of the primary ones we now generate an inelastic
non-diffractive minimum bias event in \pytppp, each of which will give
a separate sub-events. However, since the procedure takes
sub-collisions with small \bNN\ first, the primary absorptive events
should typically be a bit harder and have higher multiplicity than the
secondary ones. In \cite{Bierlich:2016smv} this was handled by telling
\pytppp to generate $N^{'}\subabs+N^{''}\subabs$ events, but only
keeping the $N^{'}\subabs$ ones with smallest impact parameter (as
reported by \pytppp). For the method described here we have instead
implemented directly in \pytppp a way to specify by hand which impact
parameter you want a given minimum bias event to have, which makes
thing a bit more efficient, and also gives a noticeable improvement on
the description of some observables, as discussed below in
\sectref{sec:results}.

Just as in standard \pytppp\ it is easy to specify signal processes
rather than only consider minimum bias events. This may be used to
simulate triggers on hard jets more efficiently, or to \eg\ produce 
$Z$-tagged jets in central \AA\ collisions \cite{Sirunyan:2017jic} or
top events in \pA\ collisions \cite{Sirunyan:2017xku} or \AA. The
way this is done is simply to substitute the hardest absorptive
primary event with a corresponding signal event, and reweighting the
event with a factor
\begin{equation}
  \label{eq:1}
  w\sub{signal}=\frac{(N^{'}\subabs+N^{''}\subabs)\sigNN\sub{signal}}%
  {\sigNN\subabs}
\end{equation}
to get the correct cross section. For signal processes with a large
cross section the possibility to have additional signal processes in
the same event is also taken into account, however for technical
reasons at most $N^{'}\subabs$ signal sub-events can be included in
each event\footnote{For most use cases this should be adequate, as 
$\sigNN\sub{signal} \ll \sigNN\subabs$ for most processes of interest}.

In the current implementation we assume that minimum bias processes
are basically iso-spin invariant, and all such sub-events are
generated as \pp\ events in \pytppp, flipping by hand the iso-spin of
a remnant quark or di-quark afterwards in case the corresponding
nucleon was actually a neutron, to conserve total charge. Signal
processes are, however, not necessarily isospin invariant. To account
for this, we generate \pp, \pn, \np, and \nn\ collisions separately
for all signal processes. To decide what type of collision should be
generated, all nucleons in the colliding nuclei are marked as either
protons or neutrons, under the assumption that neutrons and protons
are distributed evenly in the nucleus.

One should note that measurements of proton and neutron distributions
in \eg\ lead at low energies \cite{Tarbert:2013jze} have indicated
that the neutron distribution reaches further out than the proton
distribution, giving rise to a "neutron skin" effect. It has been
pointed out \cite{Helenius:2016dsk} that this could give rise to
effects at the 10\% level in selected observables in peripheral \PbPb\
collisions. It has also been pointed out \cite{Paukkunen:2015bwa} that
one could in principle use this effect to design different centrality
measures, especially in the case of asymmetrical collision systems.
Currently we know of only one very recent Glauber calculation
including such effects \cite{Loizides:2017ack}, and in the present
version we have left them out entirely\footnote{An interested user
  can, however, plug in their own Glauber MC including neutron skin
  effects.}.

\subsection{Adding secondary absorptive interactions.}
\label{sec:adding-second-absorp}

Once the back-bone sub-events have been generated we go through the
list again, this time only looking at the secondary absorptive
interactions, in which one of the participating nucleons has already
been included in a generated primary absorptive sub-event. As
described in \cite{Bierlich:2016smv} we will generate these secondary
absorptive sub-collisions as if they were single diffractive
excitation events. We here use the standard \pytppp diffraction
machinery, but with important modifications detailed in
\sectref{sec:sasd} below.

The final state generated for a given secondary absorptive interaction
is then added to a primary absorptive sub-event. The elastically
scattered proton is removed and the energy and momentum it had given
to the excited nucleon is instead taken from the remnants of the
nucleon in the primary sub-events.

It may very well happen that there is not enough energy left in the
remnants in the primary sub-event to allow for the addition of a
diffractively excited state. In that case it is possible to try again
and maybe generate a diffractive event with lower $M_X$. There is a
parameter in the program the limits the number of tries allowed, and
if the maximum is passed, the corresponding secondary absorptive
interaction is simply discarded (although the corresponding nucleon
still has the chance to become wounded in another secondary
interaction).

The way secondary nucleon interactions are selected according to the
\NN\ cross sections, does not take into account possible effects of
energy-momentum conservation, therefore it makes sense to try to take
such effects into account in this \textit{a posteriori} way. The parameter we
introduced should not be taken as the final word in the matter, but at
least it allows us to investigate the effects of energy-momentum
conservation.

\subsection{Adding diffractive interactions}
\label{sec:adding-other-scatt}

Having taken care of all absorptive interactions we continue with diffractive
interactions in much the same way. For each type we again 
go through the impact-parameter-ordered list of \NN\ interactions
twice. In the first round, we only consider \textit{primary}
interactions, \ie\ where neither of the nucleons have previously
been included in a sub-event, and generate a sub-event which could be a single
or a double diffractive excitation. These are treated as (soft) diffractive
events in \pytppp, as discussed in \sectref{sec:sasd}. 

In the second round we also consider \textit{secondary}
interactions, where one of the nucleons has already been treated, and
an appropriate contribution from the other nucleon (which we will here
call a \textit{half} event) is generated and added to the
corresponding previous sub-event.

As an example consider an already wounded nucleon in the projectile nucleus,
which interacts with a previously unwounded nucleon in the target. The wounded
nucleon is already connected to another target nucleon, and cannot be further
excited. There are three possibilities for the diffractive interaction: 
\begin{enumerate}
\item The new interaction is a single diffractive excitation of the
  \textit{target} nucleon. The interaction is then treated as a normal single
  diffractive excitation of the target nucleon. 
\item The new interaction is a single diffractive excitation of the
  \textit{projectile} (already wounded) nucleon. In this case the target nucleon
  is elastically scattered. 
\item The new interaction is a double diffractive excitation. In this case the
  already wounded projectile nucleon is not modified, and the
  interaction is again treated as a single diffractive excitation of the
  \textit{target} nucleon. 
\end{enumerate}

In a final iteration\footnote{Note that central diffraction is not handled
properly in the current version of the program.} also purely elastic
interactions are considered, and here again the half events are single
elastically scattered nucleons. In each case energy and momentum
conservation is handled in the same way as for secondary absorptive
interaction.

Modulo the effects of secondary interactions being discarded due to
energy-momentum conservation, this procedure will correctly handle
the probability that a given nucleon is wounded in some way.
Note however that, as discussed in \sectref{sec:AAfluc},
although some nucleons
in the program are classified as elastically scattered, elastic scattering is
not included properly. As elastic scattering 
is a coherent effect of shadowing due to absorption, the Good--Walker
formalism can be used to calculate the cross 
section for elastic scattering of the incoming \emph{nuclei}, but not for
individual nucleons in a nucleus. \footnote{Naturally electromagnetic
interaction, not included here, is responsible for most of the coherent
elastic nucleus scattering.} 
Diffractive excitation of individual nucleons can, however, be calculated via
the trick described in \sectref{sec:AAfluc}.  

In the end we have generated a set of parton-level sub-events, which
we now can join together in a single parton-level \AA\ event. This event
is then handed back to \pytppp for hadronisation and decay of unstable hadrons.
Finally the non-interacting projectile and target nucleons are bunched together in
two remnant nuclei.\footnote{The nucleus remnants are in the event record 
given the name \emph{NucRem} and PDG-id codes on the form $100ZZZAAA9$, 
which in the PDG standard corresponds to a highly excited nucleus.}

\section{Modifications of single diffractive to secondary absorptive}
\label{sec:sasd}

In \sectref{sec:woundet-to-fs} and in \citeref{Bierlich:2016smv} we
argued that secondary absorptive interactions will contribute to
particle production in the same way as a single diffractive (SD)
excitation event (\cf\ \figref{fig:mpi-pA}). Assuming that such SD events
produce a simple flat string with mass distributed as
$dM_X^2/M_X^2$, this would naively give a triangular shape of the
$F(\eta)$ wounded nucleon emission function in \eqref{eq:woundedMult}.

\FIGURE[t]{
  \begin{minipage}{1.0\linewidth}
    \begin{center}
      \includegraphics[width=0.50\textwidth]{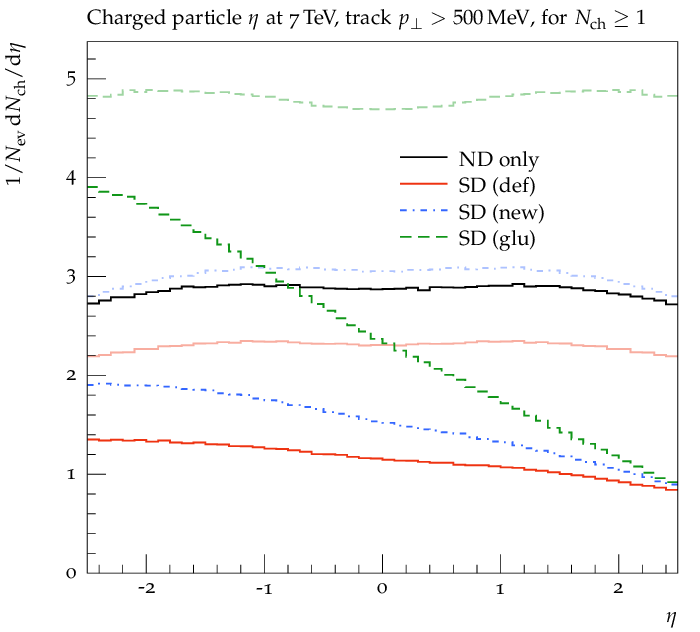}
    \end{center}
  \end{minipage}
  \caption{\label{fig:FofEta}Illustration of the shape of the
    multiplicity function in \eqref{eq:woundedMult} using the $\eta$
    distribution as measured by ATLAS in \cite{Aad:2010ac}. The black
    and red lines are the shapes of standard non-diffractive and single
    diffractive events from \pytppp respectively. The green dashed and
    blue dash-dotted lines are single diffractive events generated by
    \pytppp using the modifications presented in
    \cite{Bierlich:2016smv} and the modifications presented in this
    article respectively. For each single diffractive line there is
    also a pale line corresponding to adding the mirror image to
    emulate a non-diffractive distribution \`a la \fritiof.}
}

We will use the SD excitation machinery in \pytppp, where at high
energies the diffractive systems are much more complicated than a
single string. As described in more detail below, it models the
diffractive excitation as a non-diffractive (ND) interaction between
the target nucleon and a \pomeron emitted from the projectile (in the
spirit of Ingelman and Schlein \cite{Ingelman:1984ns}), and this is
then treated with the full MPI machinery as if the \pomeron was a
hadronic object with parton densities. In \figref{fig:FofEta} we show
the average multiplicity as a function of pseudo-rapidity for ND
events, and compare it to SD events from \pytppp using the default
settings. Clearly we get a somewhat triangular shape for the SD events
(SD(def) in the figure), and adding the multiplicity from target and
projectile excitation, we get a shape similar to the ND shape, fully
in accordance with \eqref{eq:woundedMult} in the case of
$w_p = w_t = 1$.

In \citeref{Bierlich:2016smv} we noticed that using the default
\pytppp SD machinery for secondary absorptive collisions resulted in
too low activity in \pA\ and tried different modifications to increase
the multiplicity. One of these modifications included increasing the
gluon density in the \pomeron, which is also shown in
\figref{fig:FofEta} (SD(glu)).

Here we will try to be more systematic in our approach to modify the
default \pytppp SD machinery. Looking at \figref{fig:mpi-pA}b, it is
clear that the rapidity region close to the one close to the direction
of the two target nucleons will be our main focus. Here we note that
we could equally well have chosen the second nucleon to be in the
primary interaction and the first nucleon to be in the secondary, and
would then want to have the same distribution of particles. This means
that we want the single diffractive event to look as much as possible
as a non-diffractive event close to the direction of the two
nucleons. We have therefore investigated several different
modifications of the SD model and for different diffractive masses we
have studied particle distributions in different pseudo-rapidity
intervals and compared these with the corresponding particle
distributions in the same intervals for ND events.

In the end we settled for a new modification (labelled SD(new) in
\figref{fig:FofEta}), which is the default way of generating secondary
absorptive interactions as of version 8.235 of \pytppp
\footnote{Normal diffractive interactions between projectile and
  target nucleons are treated by the usual \pytppp diffraction
  set-up.}. To motivate this, we first need to take a closer look at
the SD machinery in \pythia.

\subsection{High-mass diffractive excitation and secondary absorptive}
\label{sec:high-mass-diff}

There are more than one way of generating diffractive events in
\pytppp, but here we will only concern ourselves with the
\textit{soft} diffraction used for minimum bias events. Also
here there are two treatments depending on the mass, $M_X$. For low
masses, $\lesssim10$~GeV, the excited system is modelled as a simple
longitudinally stretched string. In an \AA\ collision, such small
excitations will typically be mixed up with the nucleus remnants in
the very forward and backward regions and we will here mainly
concentrate on high-mass diffraction, which contributes also in the
central rapidity region as seen in \figref{fig:FofEta}.

For high-mass diffraction, \pythia treats a proton-\pomeron
collision as a normal non-diffractive (ND) hadron-hadron collision and uses
the whole MPI machinery with initial- and final-state parton showers.
This means that there will be multiple $2\to2$ semi-hard partonic
scatterings given by
\begin{equation}
  \label{eq:dsigpartonicpom}
	d\sigma_{ij}^{\ppom}(p_\perp^2)=\frac{dx_{\pom}}{x_{\pom}}\frac{dx_1}{x_1}\frac{d\beta}{\beta}
  F(x_{\pom})x_1f_i^\prot(x_1, p_\perp^2)\beta f_j^{\pom}(\beta, p_\perp^2)
  d\hat{\sigma}_{ij}(p_\perp^2).
\end{equation}
Here $x_{\pom}$ denotes the fraction of the target proton momentum
taken by the \pomeron; $\beta$ is the fraction of the \pomeron momentum
taken by the parton $j$; and $x_1$ is the fraction of the projectile
proton momentum taken by parton $i$. Furthermore we have the parton
densities in the proton, $f_i^\prot$, and the corresponding densities in
the \pomeron, $f_j^{\pom}$. Finally we have the flux factor
$F(x_{\pom})$ controlling the diffractive mass given by
$M_X^2=x_{\pom}s$. In the following we will assume a flat distribution
in $\log{(M_X^2)}$, in which case $F(x_\pom)$ is just a constant.

The partonic cross section $d\hat{\sigma}_{ij}(p_\perp^2)$ diverges
for small $p_\perp^2$, and although it is regularised as in
\eqref{eq:mpi-model} the integrated partonic cross section may still
exceed the total non-diffractive \ppom\ cross section for a given
$M_X$. In the \pythia MPI model this is then interpreted as the
possibility of having several sub-scatterings in each collision, with
the average number of sub-scatterings given by
\begin{equation}
  \label{eq:avNscatpom}
  \langle N_{sc}^{\ppom}(M_X)\rangle = \frac{1}{\sigma\sub{ND}^{\ppom}(M_X)}\int\frac{dx_1}{x_1}
  \int\frac{d\beta}{\beta}\int dp_\perp^2\sum_{ij}
  x_1f_i^\prot(x_1, p_\perp^2)\beta f_j^{\pom}(\beta, p_\perp^2)\frac{d\hat{\sigma}_{ij}}{dp_\perp^2}.
\end{equation}
Here the default value of the of the non-diffractive \ppom\ cross
section, $\sigma\sub{ND}^{\ppom}(M_X)$, is just set to a constant
10~mb. The $p_\perp^2$ integral is over the full available phase
space, all the way down to zero, but with the $\hat{\sigma}_{ij}$
regulated as in \eqref{eq:mpi-model}. The parameter $p_{\perp0}$
here varies as a small power of $M_X^2$, in the same way as the
$p_{\perp0}$ in normal \pp\ scatterings varies with $s$.

\FIGURE[ht]{
  \includegraphics[width=0.33\textwidth]{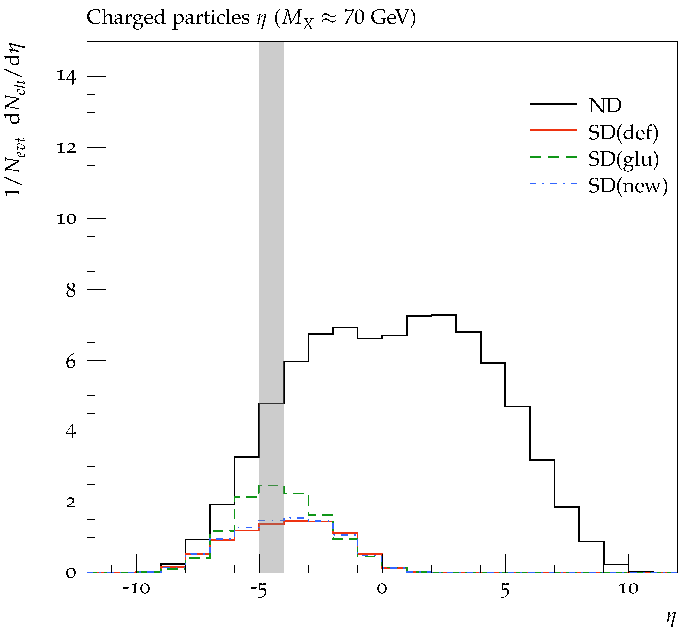}%
  \includegraphics[width=0.33\textwidth]{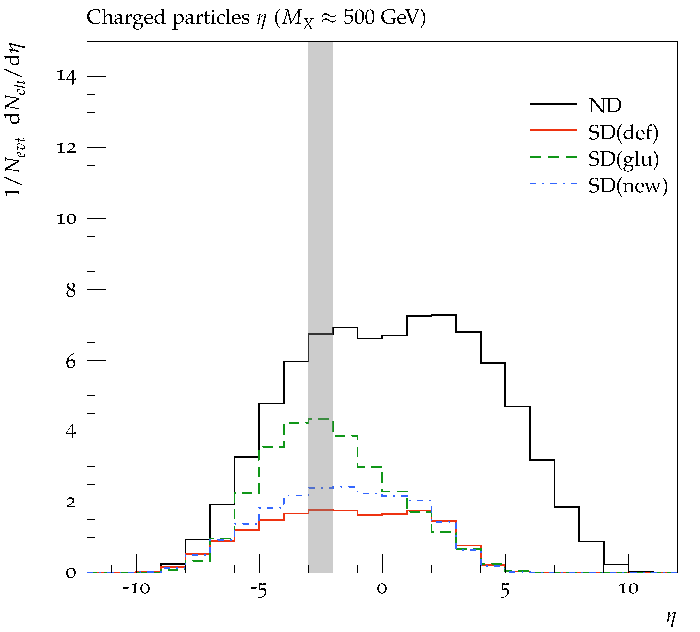}%
  \includegraphics[width=0.33\textwidth]{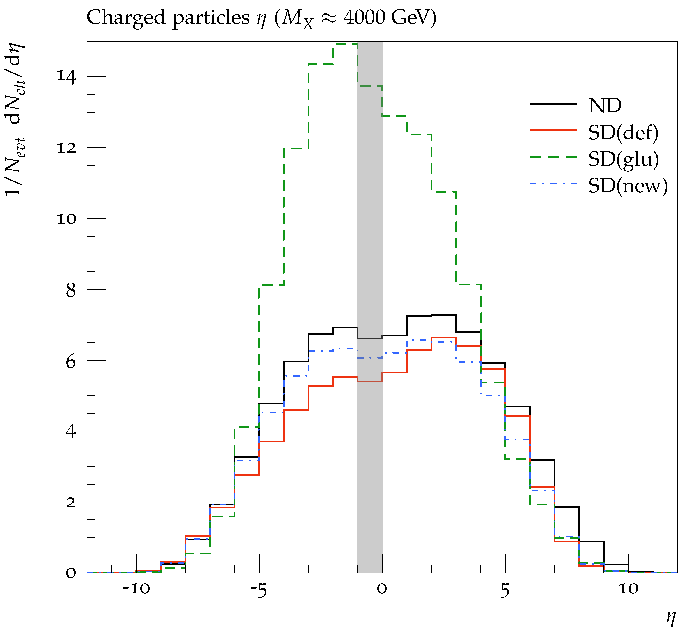}
  \caption{\label{fig:HF2}Pseudo-rapidity distribution of charged
    particles for different diffractive masses for the default single
    diffraction in \pytppp (red solid lines), the modifications made
    in \cite{Bierlich:2016smv} (green dashed lines) and the new
    modifications presented here (blue dash-dotted lines). Left,
    centre and right histograms correspond to $M_X$ values of
    $\approx70$, $500$, and $4000$~GeV respectively. For comparison
    the results from non-diffractive \pythia events at $\sqrts=5$~TeV
    is shown as the solid black line. The shaded areas in the figure
    indicate the pseudo rapidity intervals where the comparisons
    between SD and ND particle distributions in
    \sectref{sec:valid-second-absorpt} were studied. }
}

In \figref{fig:HF2} we show the resulting pseudo-rapidity distribution
of charged particles for different values of $M_X$ for diffractive
events from \pytppp with $\sqrts=5~TeV$.\footnote{The kinematics is
  given by the LHC \pPb\ run, giving a slightly tilted distribution in
  $\eta$.} Here we see the expected behaviour with a large rapidity gap
for smaller $M_X$, typical for diffraction. When we want to use the
diffractive excitation in \pythia to model the secondary absorptive
interactions, we want to make the event in the target proton direction
to look as much as a normal non-diffractive \pp\ event as possible,
and in particular we want the whole event to look approximately the
same in the limit $M_X^2\to s$. From the figure we see that this is
not quite the case for the default diffraction parameters in
\pytppp. We also see that the modifications we presented in
\cite{Bierlich:2016smv} seems to be a bit too forceful.

Looking at \eqsref{eq:dsigpartonicpom} and \eq{eq:avNscatpom} it is
easy to see that we can increase the multiplicity by either increasing
the general activity by modifying the \pomeron parton densities (as is
done in SD(glu) in \figrefs{fig:FofEta} and \ref{fig:HF2}), or we can
try to increase the number of sub-scatterings by \eg\ adjusting the
free parameter $\sigma\sub{ND}^{\ppom}(M_X)$. We will here look at
both these options by studying \eqref{eq:avNscatpom} more closely.
Studying the average number of sub-scatterings for a fixed rapidity,
$y=\log(x_1/\beta x_{\pom})/2$, we get
\begin{eqnarray}
  \label{eq:avNyscatpom}
  \frac{d\langle N_{sc}^{\ppom}\rangle}{dy} &=& \frac{1}{\sigma\sub{ND}^{\ppom}(M_X^2)}\int\frac{dx_1}{x_1}
  \int\frac{d\beta}{\beta}\int dp_\perp^2\sum_{ij}
  x_1f_i^\prot(x_1, p_\perp^2)\beta f_j^{\pom}(\beta, p_\perp^2)\nonumber\\
  & & \qquad\qquad\qquad\qquad\qquad\qquad\qquad\qquad\times\frac{d\hat{\sigma}_{ij}}{dp_\perp^2}
  \delta\left(y - \log{\frac{x_1}{\beta x_{\pom}}}\right).
\end{eqnarray}
If we now compare this to the same for standard non-diffractive \pp\ events,
\begin{eqnarray}
  \label{eq:avNyscatpp}
  \frac{d\langle N_{sc}^{\pp}\rangle}{dy} &=& \frac{1}{\sigma\sub{ND}^{\pp}(s)}\int\frac{dx_1}{x_1}
  \int\frac{dx_2}{x_2}\int dp_\perp^2\sum_{ij}
  x_1f_i^\prot(x_1, p_\perp^2)x_2f_j^{\prot}(x_2, p_\perp^2)\nonumber\\
  & & \qquad\qquad\qquad\qquad\qquad\qquad\qquad\qquad\times\frac{d\hat{\sigma}_{ij}}{dp_\perp^2}
  \delta\left(y - \log{\frac{x_1}{x_2}}\right),
\end{eqnarray}
we see immediately that if we modify the \pomeron parton density and
make it $x_\pom$-dependent,
$\beta f_j^\pom(\beta,p_\perp^2)\to x_\pom\beta f_j^\prot(x_\pom\beta
,p_\perp^2)$, and at the same time make the total non-diffractive
\ppom\ cross section as well as the soft regulator, $p_{\perp0}$,
independent of $M_X$, \ie,
$\sigma\sub{ND}^{\ppom}(M_X^2)\to\sigma\sub{ND}^{\pp}(s)$ and
$p_{\perp0}(M_X^2)\to p_{\perp0}(s)$, we will get very similar
expressions. They will not be exactly the same, since the kinematical
limits $p_{\perp}$ will differ, especially for small $M_X$. Also, for
technical reasons, \pytppp will adjust the selected $p_{\perp0}$ for
each $M_X$ value to ensure that the average number of scatterings is
always larger than one, effectively making low $M_X$ events softer.

The resulting modification is shown in \figref{fig:HF2} as the lines
labelled SD(new), and we see that the multiplicity in the proton
direction is not much improved at small $M_X$, but at large $M_X$ it
traces the non-diffractive quite well.

In the next section we will look in more detail on the particle distributions in
the rapidity regions where we want the secondary absorptive sub-events
to resemble normal non-diffractive events in \pythia.

\subsection{Comparing primary and secondary absorptive sub-events}
\label{sec:valid-second-absorpt}

From \figref{fig:HF2} we see that SD final state particles only
populate the rapidity region corresponding to the colour exchange
between the \pomeron and the proton (\cf\ \figref{fig:mpi-pA}b). We
will here investigate further to what extent the SD events generated
by \pythia (with or without modifications) look the same as the ND
events in this region. To do this we will study the distribution of
particles in different pseudo-rapidity slices for different values of
the diffractive mass, $M_X$. In these slices we have looked at
standard minimum bias observables based on charged particles, such as
average multiplicity (shown in \figref{fig:HF2}), the distribution in
multiplicity ($N_{ch}$), the transverse momentum distribution
($p_\perp$), the distribution in summed ($\sum p_\perp$) and average
($\langle p_\perp\rangle$) transverse momentum for particles within
one unit of $\eta$, and average transverse momentum as a function of
multiplicity ($\langle p_\perp(N_{ch})\rangle$).

Naturally, we do not expect these observables to look the same for a
diffractively excited system and a full non-diffractive event. Close
to the rapidity gap, we are in the fragmentation region of the \pomeron
remnant, and here the transverse momentum of final state particles are
severely restricted by the kinematics. Also close to the proton
fragmentation region, the transverse momenta are limited by
kinematics, but here we expect the SD and ND events to look very
similar, and indeed we find that they do.

\FIGURE[ht]{
  \includegraphics[width=0.33\textwidth]{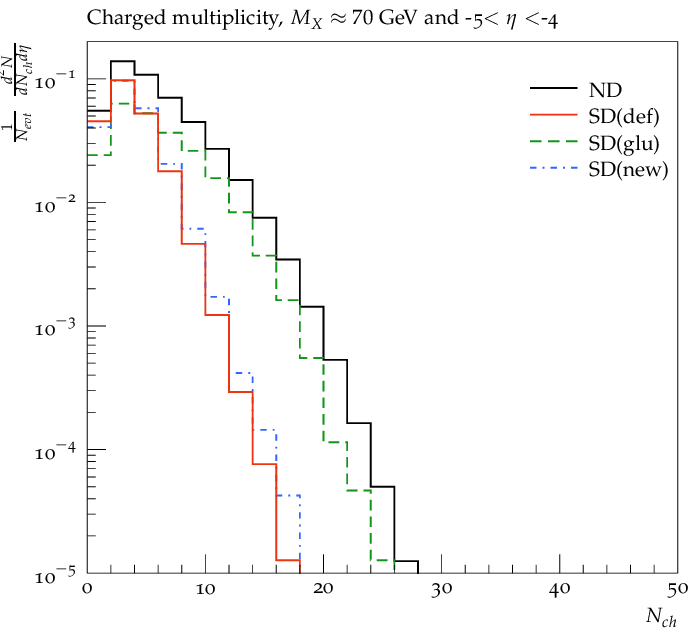}%
  \includegraphics[width=0.33\textwidth]{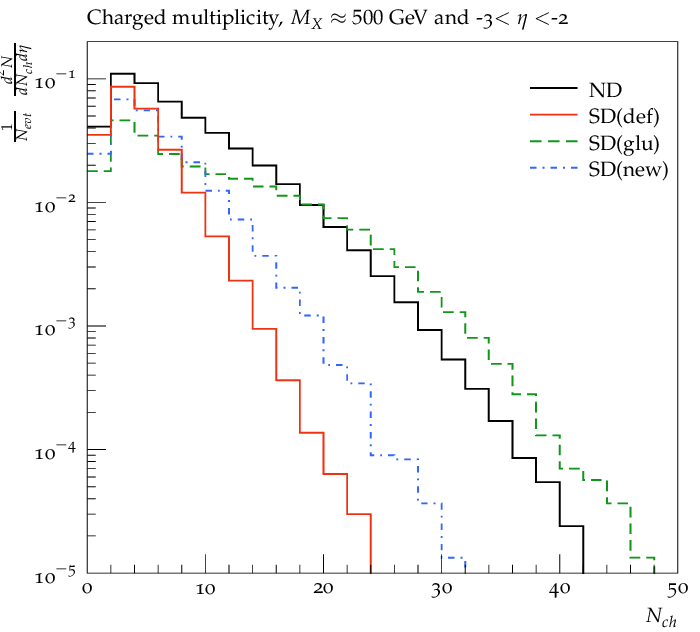}%
  \includegraphics[width=0.33\textwidth]{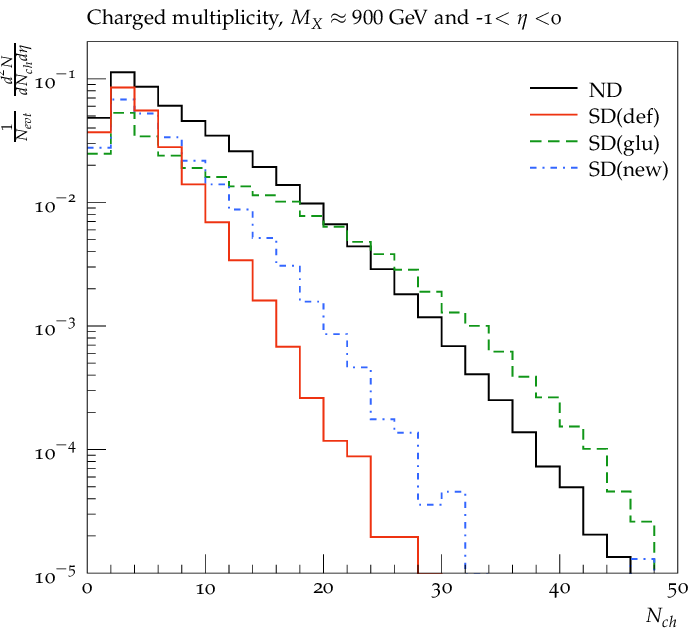}\\
  \includegraphics[width=0.33\textwidth]{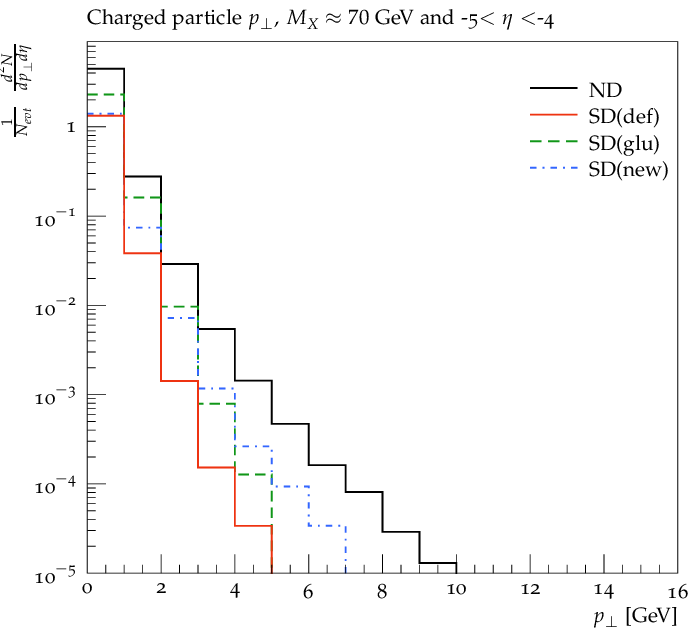}%
  \includegraphics[width=0.33\textwidth]{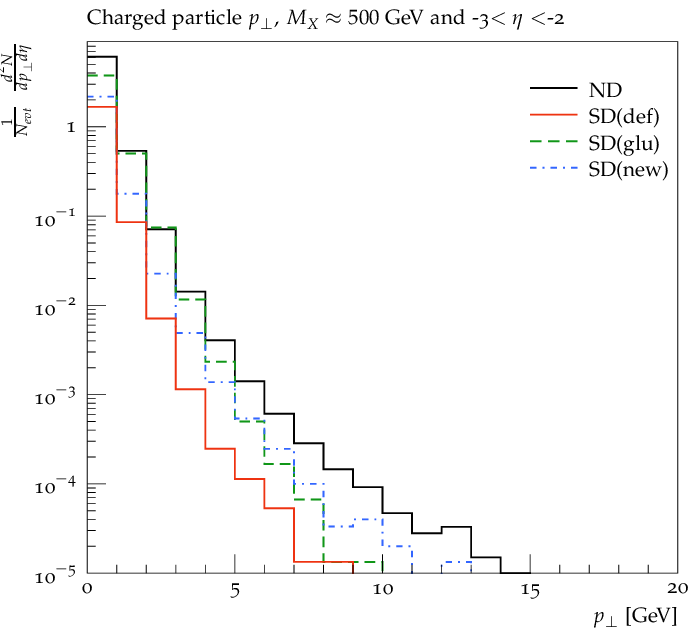}%
  \includegraphics[width=0.33\textwidth]{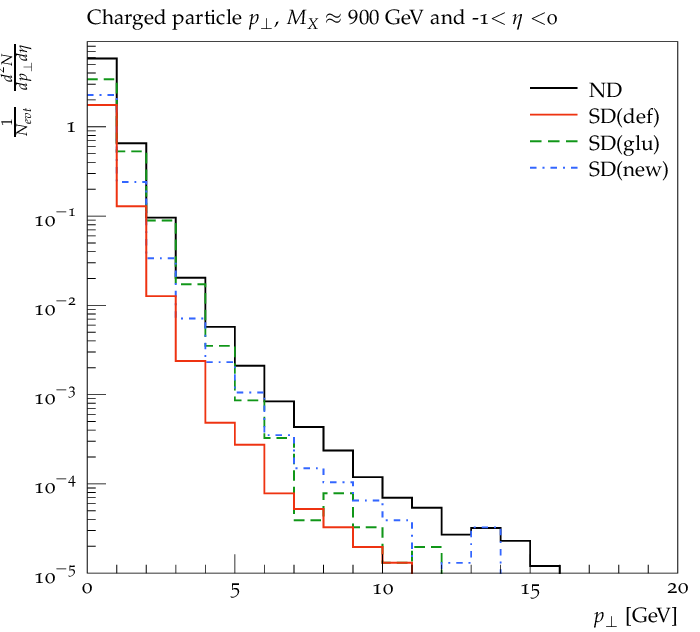}
  \caption{\label{fig:HF3}         
    Charged particle distributions in non-diffractive events (black
    lines marked ND) compared to different options (SD(def): red
    lines, SD(glu): green dashed lines, and SD(new): blue dash-dotted
    lines) for single diffractive excitation events in different
    rapidity slices and different excitation masses, $M_X$. The top
    panel shows the multiplicity of charged particles, and the bottom
    panel their transverse momentum distribution.}
}

\FIGURE[ht]{
  \includegraphics[width=0.33\textwidth]{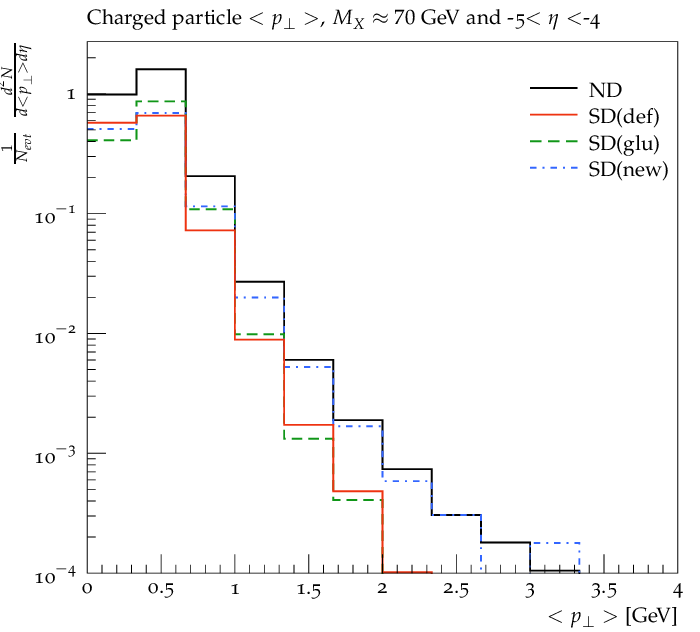}%
  \includegraphics[width=0.33\textwidth]{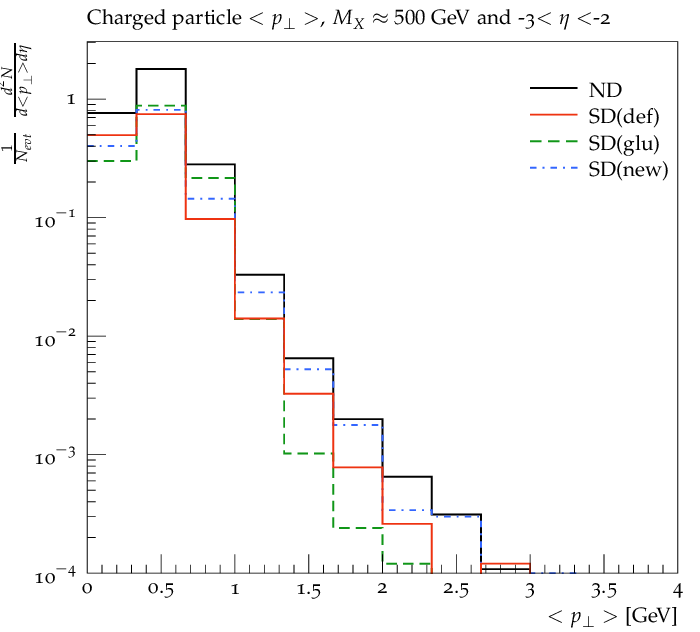}%
  \includegraphics[width=0.33\textwidth]{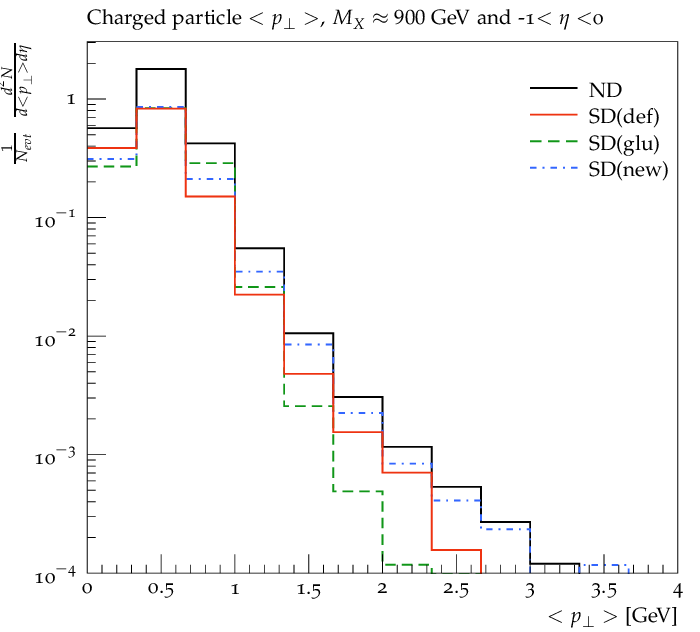}\\
  \includegraphics[width=0.33\textwidth]{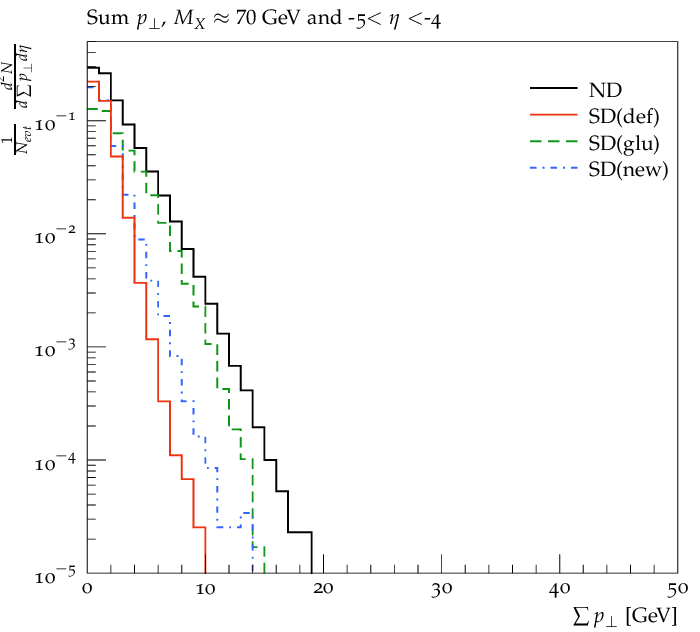}%
  \includegraphics[width=0.33\textwidth]{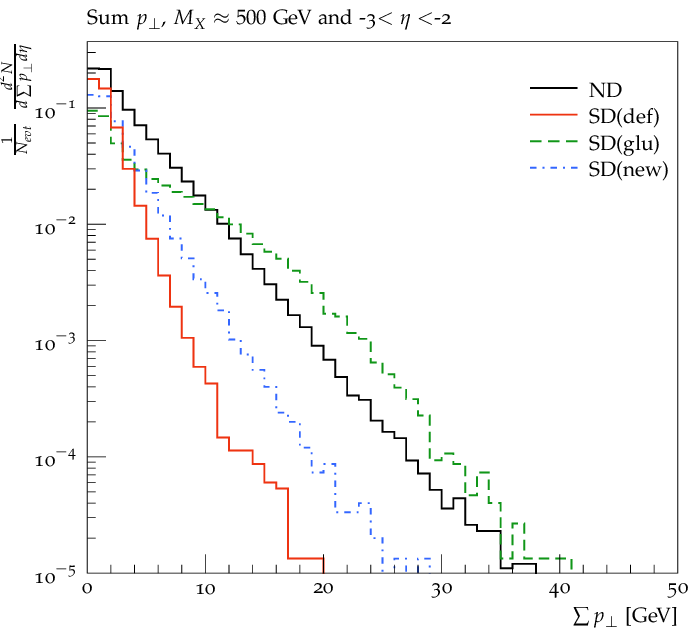}%
  \includegraphics[width=0.33\textwidth]{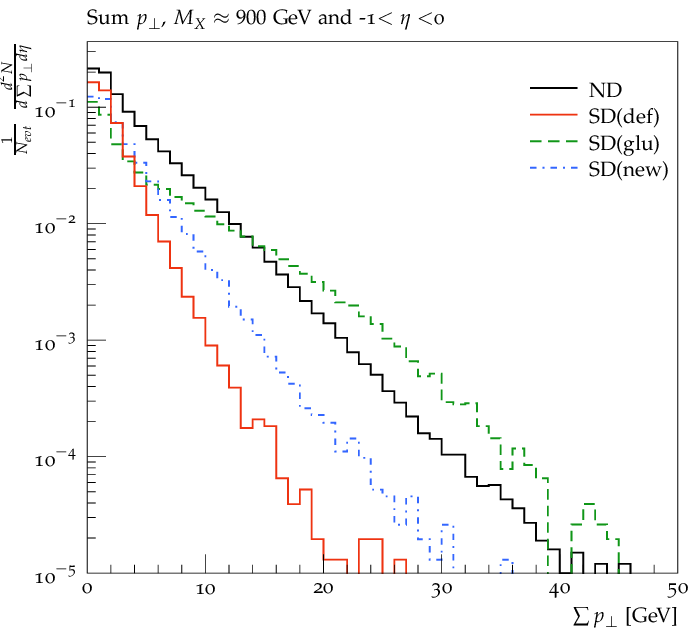}\\
  \includegraphics[width=0.33\textwidth]{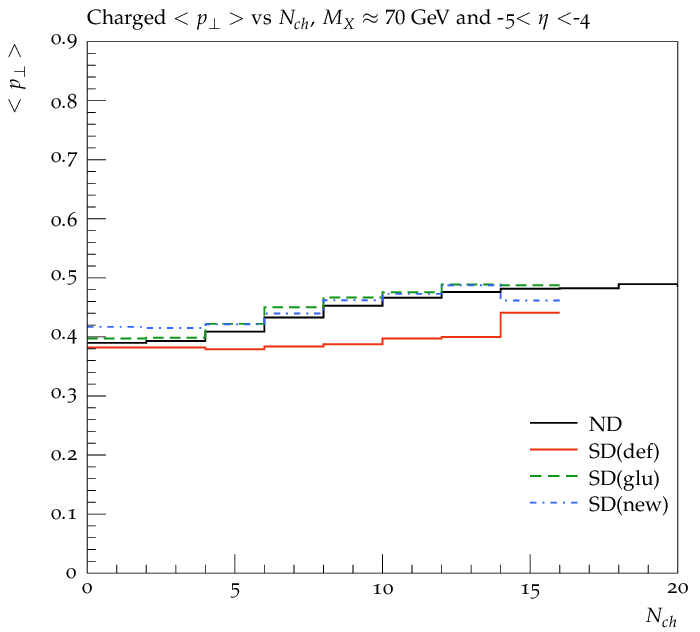}%
  \includegraphics[width=0.33\textwidth]{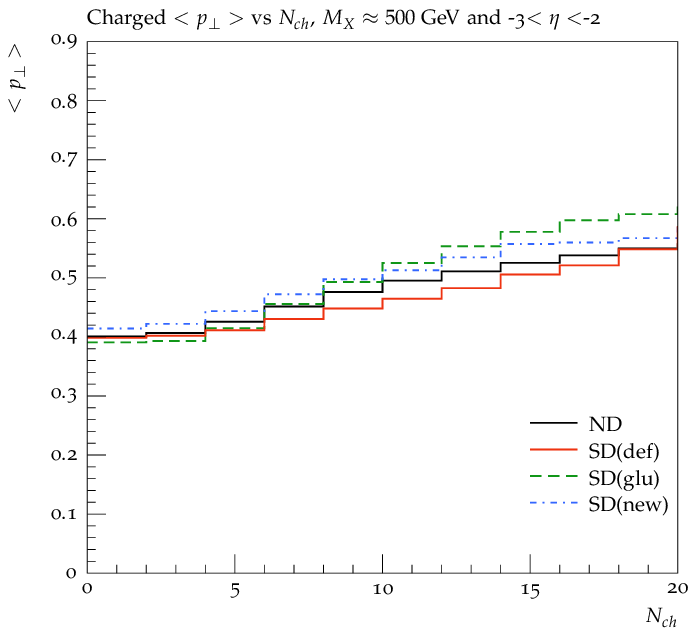}%
  \includegraphics[width=0.33\textwidth]{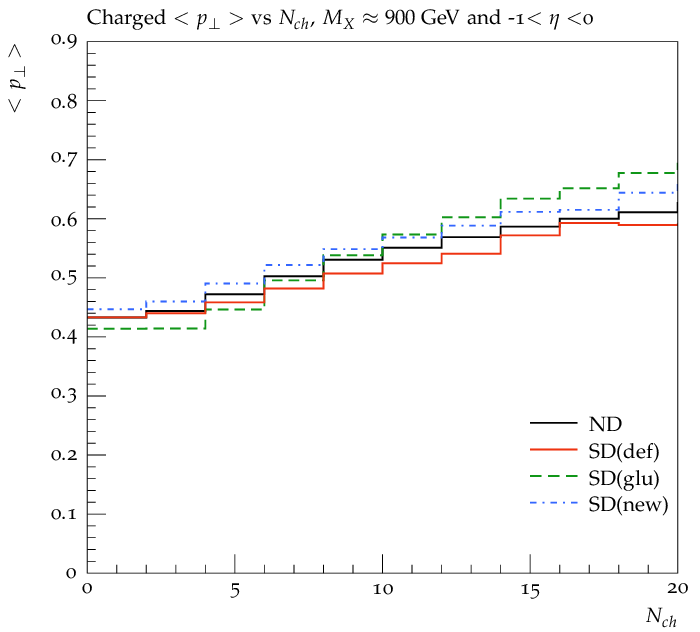}
  \caption{\label{fig:HF4}
    As \figref{fig:HF3}, but different observables. The top panel
    shows the average transverse momentum, while the middle one
    shows the summed transverse momentum and the bottom panel the
    average transverse momentum as a function of the multiplicity.}
}

Here we will concentrate on the rapidity regions around the plateau of
each $M_X$, and in \figrefs{fig:HF3} and \ref{fig:HF4} we show some
distributions in the slices $\eta\in[-5,-4]$, $[-3,-2]$ and $[-1,-0]$
(the shaded regions in \figref{fig:HF2}) for mass bins with
$M_X\approx70$, $500$ and $900$~GeV respectively. As for the overall
multiplicity we find that the default SD machinery, (SD(def)), is
quite far from the ND observables in the same rapidity slice. The
SD(glu) modification is much closer, but overshoots quite
significantly at large $M_X$ in the multiplicity distribution
(\figref{fig:HF3}) and $\sum p_\perp$ (\figref{fig:HF4}). 
The SD(new) curve gives a slightly better description of $p_\perp$
in \figref{fig:HF3} and the average $p_\perp$ observables in \figref{fig:HF4},
but no improvement -- or even a slightly worse performance -- in the
two remaining observables. The choice of which option to use 
can therefore only be based on an assessment of what types of observables 
are deemed most important to reproduce correctly. 
In particular the dependence of
the average transverse momentum on the multiplicity is known to be
very sensitive to the handling of the multi-parton interactions
\cite{Sjostrand:1987su}, and here we see that SD(new) is quite close
to the ND curves here, as may be expected from comparing
\eqsref{eq:avNyscatpp} and \eq{eq:avNyscatpom}.

The fact that the $\sum p_\perp$ distributions in SD(glu) in
\figref{fig:HF4} is much harder than in standard ND events would be a
problem for the description of the centrality observables used in \pA\
and \AA, which are often based on the total transverse activity in the
forward/backward region (see \sectref{sec:pa-results}).

It is, however, clear that we could have put more emphasis on
charged multiplicity and $\sum p_\perp$ in the regions where the SD(glu)
option outperforms SD(new), and thereby made another
choice of recommended option. In \sectref{sec:uncertainties} we will compare the
three different choices against each other for \pA\ results.

In \sectref{sec:gener-prim-absorpt} we explained how the impact
parameter obtained for each \NN\ sub-collision is used as input to
\pytppp. Here small impact parameters will lead to more multiple
scatterings for primary absorptive sub-events. The same impact
parameter dependence is also used for secondary absorptive
sub-events. It is therefore interesting to compare the SD events with
ND events for a specific impact parameter. In \figref{fig:HF56} we
show typical examples of such comparisons for impact parameters
slightly smaller and larger than average. Comparing with the
corresponding distributions in \figrefs{fig:HF3} and \ref{fig:HF4}, we
see that the difference between the SD and ND curves tend to diminish
with increasing impact parameter, which is good, since by construction
the secondary absorptive interactions are at larger impact parameter
than the primary ones.

\FIGURE[ht]{
  \includegraphics[width=0.33\textwidth]{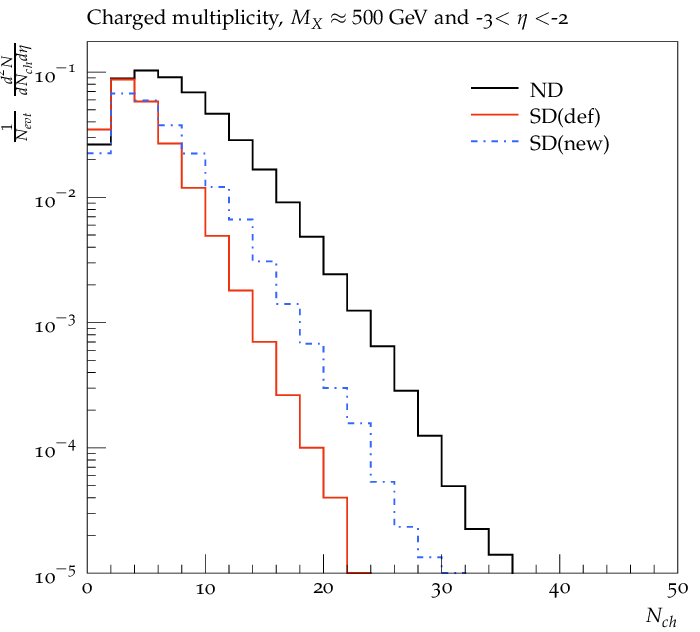}%
  \includegraphics[width=0.33\textwidth]{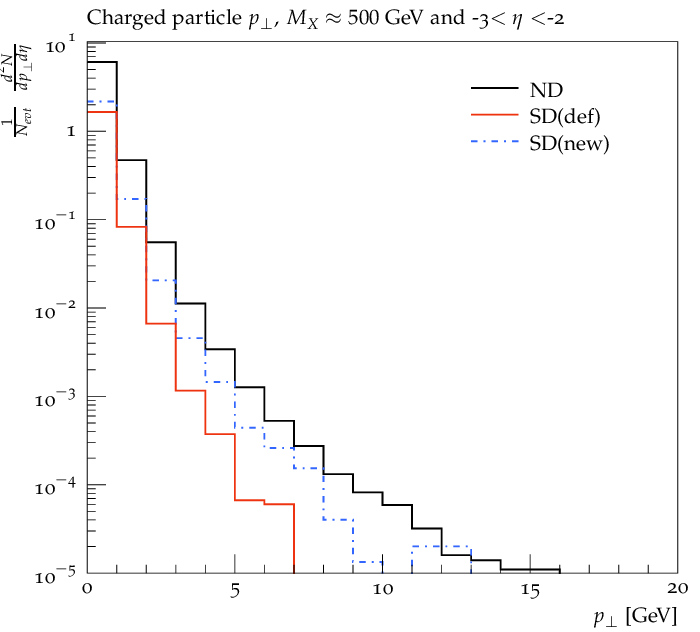}%
  \includegraphics[width=0.33\textwidth]{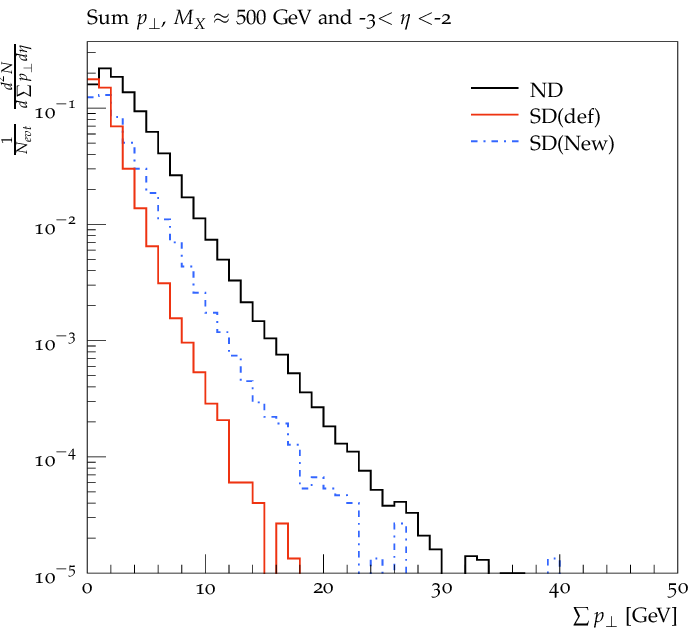}\\
  \includegraphics[width=0.33\textwidth]{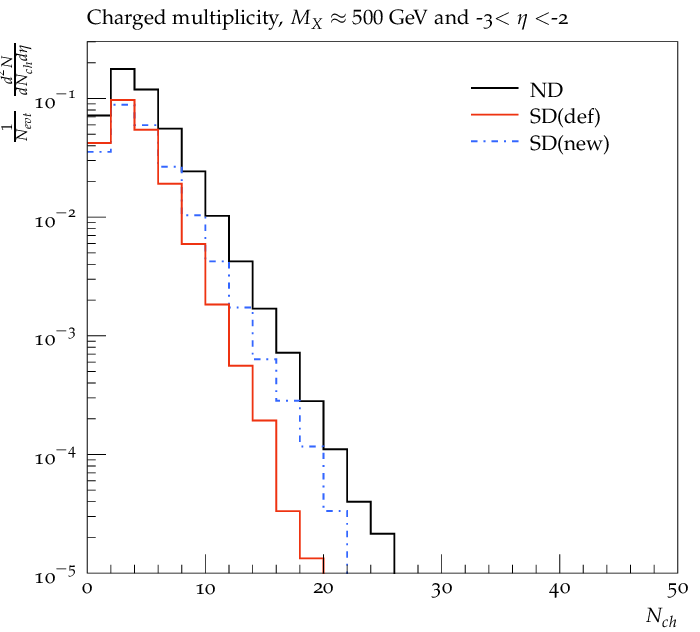}%
  \includegraphics[width=0.33\textwidth]{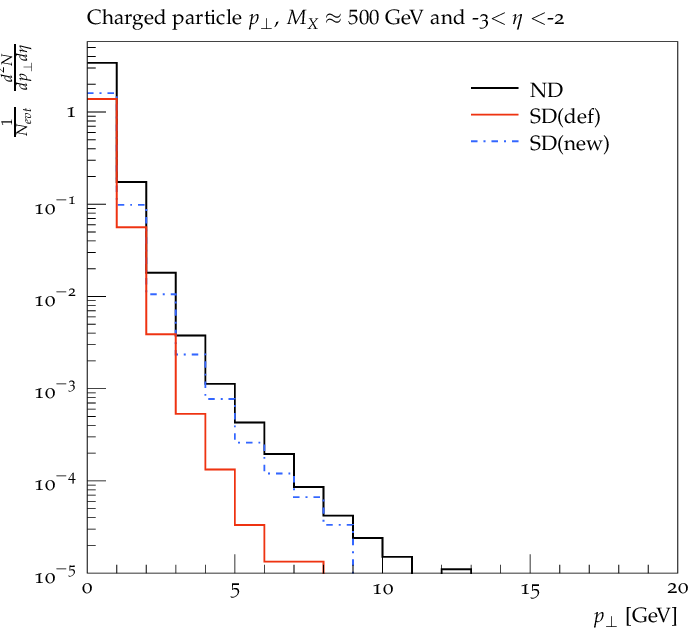}%
  \includegraphics[width=0.33\textwidth]{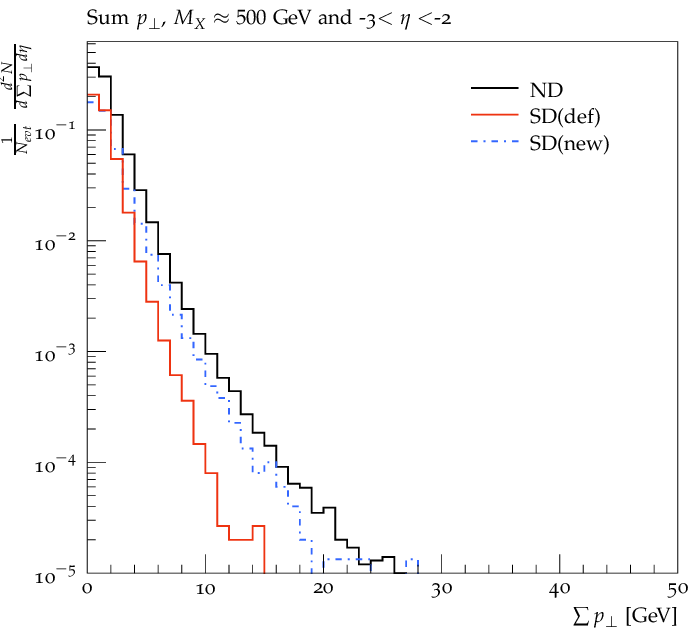}
  \caption{\label{fig:HF56}Multiplicity, $p_{\perp}$ and
    $\sum p_{\perp}$ of charged particles for different modifications
    of SD events with $M_X\approx500$~GeV compared to ND events in the
    pseudo-rapidity interval $-3<\eta<-2$. All events were generated
    at fixed impact parameter, $b/\langle b\rangle=0.9$ (top panel)
    and $1.3$ (bottom panel). The lines are as in \figref{fig:HF3}.}
}

In \figref{fig:HF56} we did not show curves for SD(glu), and in the
following we will disregard this option completely. The modifications
there are too severe and somewhat ad-hoc, resulting in far too large
effects especially on particle production at high $M_X$
(\figref{fig:HF2}). We will also disregard the SD(def) option, as it
produces too few particles in the nucleus' fragmentation region in
\pA\ collisions \cite{Bierlich:2016smv}. SD(new) does not give a
perfect reproduction of the ND distributions, and we do not expect any
SD model to do that, due to phase space constraints. 

The conclusion from the analyses in this section is that
SD(new) provides an overall fair description, as well as being more
theoretically appealing than the other variations. The SD(new) is
therefore, since version 8.235 of \pythia, the default model for
secondary absorptive sub-events in \angantyr.

\section{Sample results}
\label{sec:results}

All results presented here are generated with \pytppp version 8.235
using default settings\footnote{Since \angantyr is the default
  heavy-ion model in \pythia, it suffices to specify suitable nuclei
  as beam particles to reproduce the results presented here. }. This
means in particular that:
\begin{itemize}
\item the nucleon distributions in the nuclei are generated according
  to the formulae in \cite{Rybczynski:2013yba} using the hard-core
  option, where parameters are tuned to low-energy \eA\ data;
\item the impact parameter is sampled using a Gaussian distribution
  with a width large enough to have fairly uniform weights;
\item the fluctuations in the nucleons were modelled according
  \eqsref{eq:gammadist} -- \eq{eq:varyT0}, fitted the default
  parameterisation of semi-inclusive cross sections in \pytppp;
\item the different \NN\ interactions were classified using the
  procedure described in \sectrefs{sec:AAfluc} and
  \ref{sec:stacking};
\item the sub-events were generated with the default \pytppp
  minimum-bias machinery, except for the secondary absorptive ones,
  where the modifications in \sectref{sec:sasd} was used.
\end{itemize}

As with most things in \pytppp, there are many options beyond the
default behaviour in \angantyr, and there are also so-called user
hooks where the user can implement alternative models for \eg\ the
nucleon distribution, impact-parameter sampling and modelling of
fluctuations. There are also a number of parameters in \angantyr that
influences the generation of collisions involving nuclei, but most of
these can be fitted to \pp\ data. In fact, there are only two
parameters that clearly influences the results presented here, which
cannot be tuned to \pp\ data. One is the distribution of diffractive
masses used in the generation of secondary absorptive sub-events. Here
we have assumed a distribution $\propto dM_X^2/M_X^{2(1+\Delta)}$
where we have simply chosen $\Delta=0$ as in the original wounded
nucleon model as implemented in \fritiof. The other was mentioned in
\sectref{sec:adding-second-absorp} and is related to energy-momentum
conservation when adding secondary sub-events. The default is to
simply veto a secondary \NN\ interaction if there is not enough energy
left in the corresponding remnant nucleon in the primary sub-event. An
alternative is to instead generate a new secondary sub-event
(regenerating $M_X$) to see if that one can be included.\footnote{The
number of attempts allowed for this is governed by the parameter
\texttt{Angantyr:SDTries.}} Below in \sectref{sec:uncertainties} 
we will study the effects of these choices.

\subsection{\pp\ results}
\label{sec:pp-results}

\FIGURE[ht]{
  \includegraphics[width=0.50\textwidth]{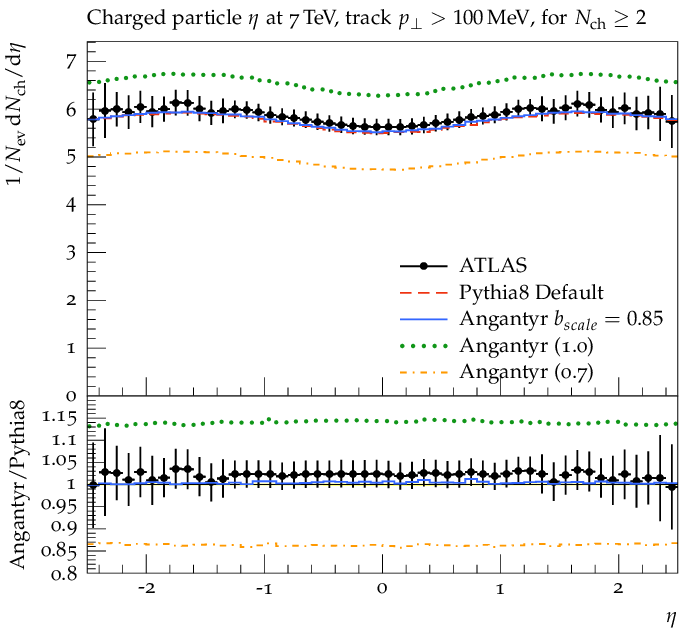}%
  \includegraphics[width=0.50\textwidth]{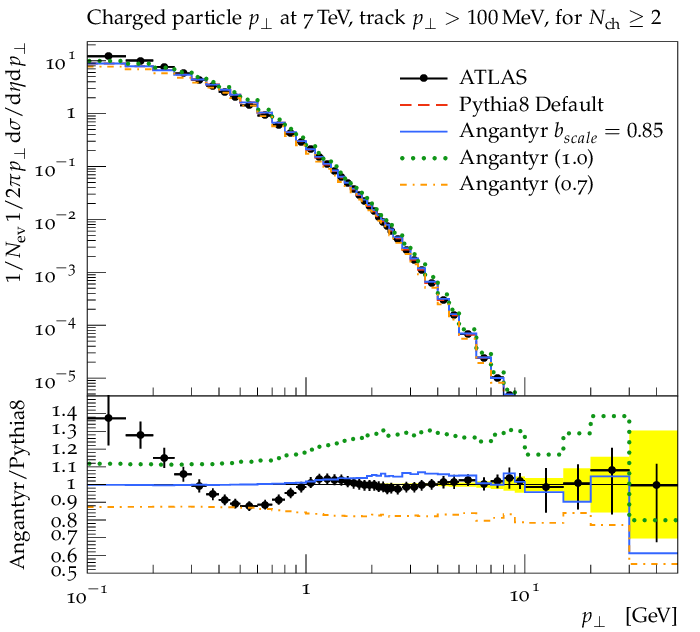}\\
  \includegraphics[width=0.50\textwidth]{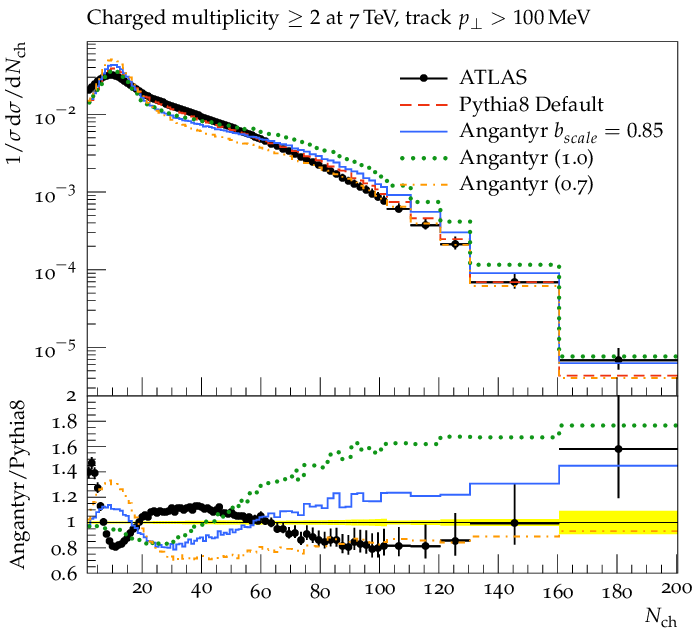}%
  \includegraphics[width=0.50\textwidth]{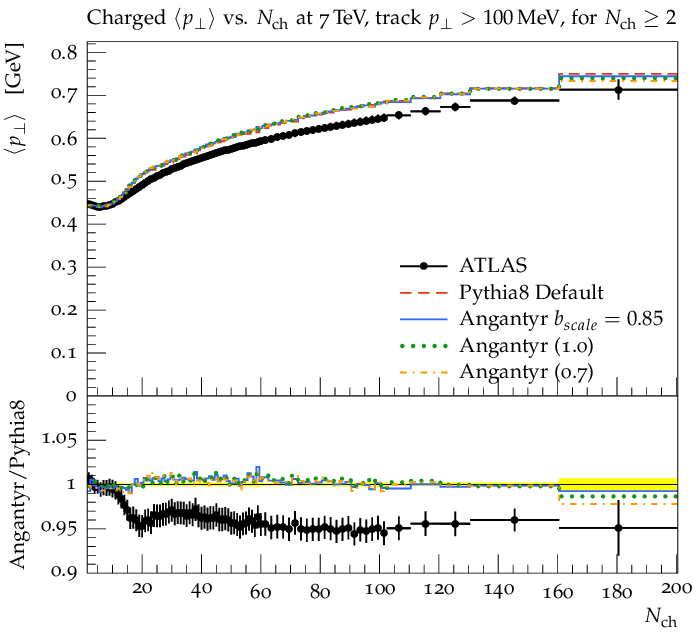}
  \caption{\label{fig:ppmb}The default \pytppp description of some
    typical minimum-bias observables in \pp, compared to the
    description using the \angantyr machinery. The latter is given for
    a range of values of $b\sub{scale}$ (as quoted in parenthesis in 
    the figure legend). For comparison we show data from ATLAS 
    \cite{Aad:2010ac} as implemented in Rivet \cite{Buckley:2010ar}. }}

We begin by using the \angantyr generation for the simplest of nuclei,
\ie\ for \pp\ collisions. Since we actually use the \pytppp minimum
bias machinery, we need to make sure that typical minimum-bias
observables are reproduced as well when using \angantyr. We expect
some differences since all semi-inclusive cross sections are not
exactly reproduced in the generation, as explained in
\sectref{sec:AAfluc}. Furthermore
the distribution in impact parameter is not the same, and since this
directly affects the amount of MPI it is important to make sure that
the translation between the two works, at least on average.

In \pytppp, the impact parameter is by default chosen according to an
exponentially falling overlap function, while in \angantyr it is
determined by the fluctuations and opacity functions in
\eqsref{eq:gammadist} -- \eq{eq:varyT0}, and it is not straight
forward to translate directly between the two. In principle one could
try to implement the \angantyr distribution as an option in the
\pytppp MPI machinery, which then would require a full retuning to
\pp\ data. Here we have decided to instead implement a simple scaling
factor, $b\sub{scale}$, so that for absorptive (non-diffractive)
events,
\begin{equation}
  \label{eq:bfudge}
  b\sub{Pyt}=
  \frac{\langle b\sub{Pyt}\rangle}{\langle b\sub{Ang}\rangle}
  \frac{b\sub{Ang}}{b\sub{scale}},
\end{equation}
which is set to a value ensuring that \angantyr gives
approximately the same results as \pytppp for typical \pp\
minimum-bias observables. In \figref{fig:ppmb} we see that our tuned
value of $b\sub{scale}=0.85$ fairly well reproduces the \pytppp
results and gives approximately the same level of agreement with data.
For comparison, the figure also shows the effect of varying this scale
to $b\sub{scale} = 1.0$ and $0.7$, as indicated in the parenthesis in the
figure legend.
      
\subsection{\boldmath\pA\ results}
\label{sec:pa-results}

Comparing to \pA\ data means that we need to consider the concept of
\emph{centrality}, which is used in almost all published experimental
heavy ion results. Centrality is based on a final-state observable
that is assumed to be correlated with the overall impact parameter of
a collision. Typically, this observable involves the activity
(multiplicity, transverse energy) close to the direction of the
nuclei, and other observables are then conventionally presented in bins of
percentiles of this centrality observable.

\FIGURE[ht]{
  \begin{minipage}{0.5\textwidth}
    \includegraphics[width=1.0\textwidth]{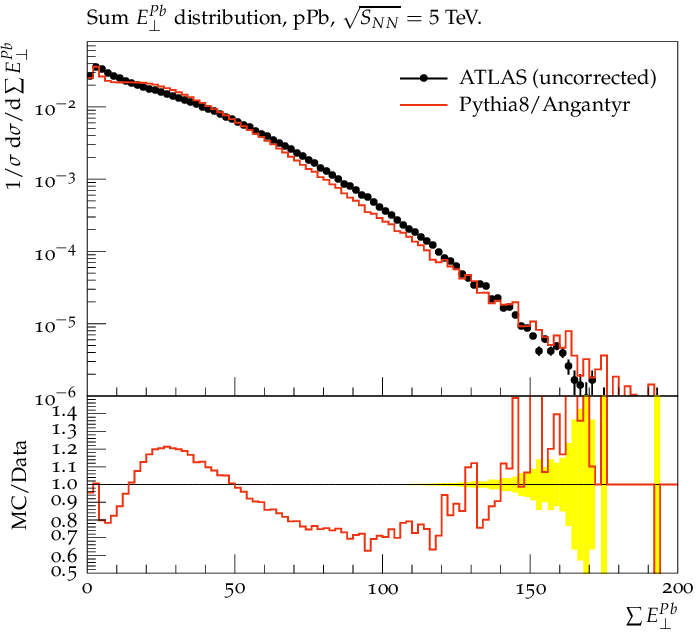}
  \end{minipage}\begin{minipage}{0.5\textwidth}
    \begin{center}
      \begin{tabular}{|l|l|l|}
        \hline
        percentile & ATLAS & \angantyr\\
        (\%) & (GeV) & (GeV) \\
        \hline
        0 --  1 & 90 & 87 \\
        1 --  5 & 66 & 62 \\
        5 -- 10 & 53 & 51 \\
        10 -- 20 & 41 & 39 \\
        20 -- 30 & 32 & 32 \\
        30 -- 40 & 24 & 26 \\
        40 -- 60 & 13 & 16 \\
        60 -- 90 &  6 &  3 \\
        \hline
      \end{tabular}
    \end{center}
  \end{minipage}
  \caption{\label{fig:ATLASpPbCent} The summed transverse energy in
    the lead direction ($-4.9<\eta<-3.2$) for \pPb\ collisions at
    \sqrtsNN=5~TeV. Data from ATLAS \cite{Aad:2015zza} is compared to
    results from \angantyr. The table shows the resulting bin edges
    when dividing up in percentiles for the experimental and generated
    data respectively.} }
      
We will here use the centrality observable defined by ATLAS in
\cite{Aad:2015zza}, which is based on the summed transverse energy in
the pseudo-rapidity interval $[-4.9,-3.2]$. As seen in
\figref{fig:ATLASpPbCent}, \angantyr is able to reproduce fairly well
the measured distribution. However, it should be noted that the
experimental distribution has not been corrected for detector effects,
so it is difficult to draw firm conclusions about the performance of
the model.

When we want to use this centrality measure we now have the option to
divide it into percentile bins using the measured distribution or the
generated distribution, and since they do not exactly agree we will
get somewhat different bins, as is shown in table in
\figref{fig:ATLASpPbCent}.

\FIGURE[ht]{
  \includegraphics[width=0.50\textwidth]{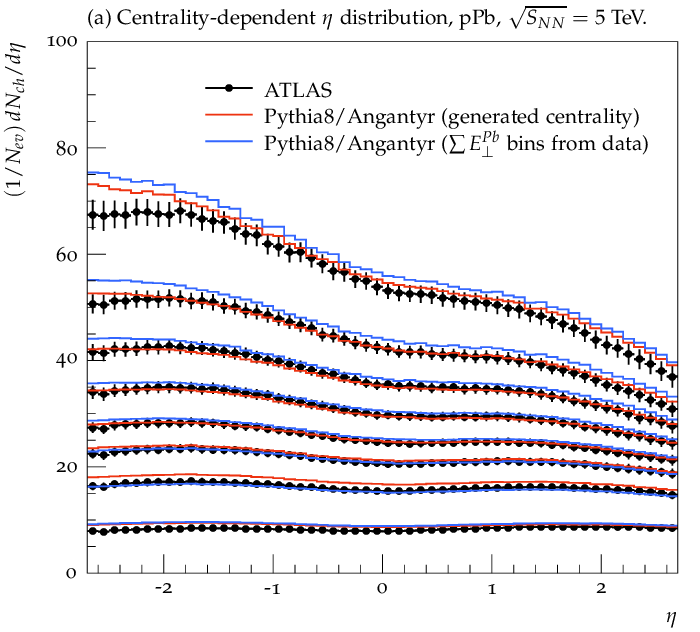}%
  \includegraphics[width=0.50\textwidth]{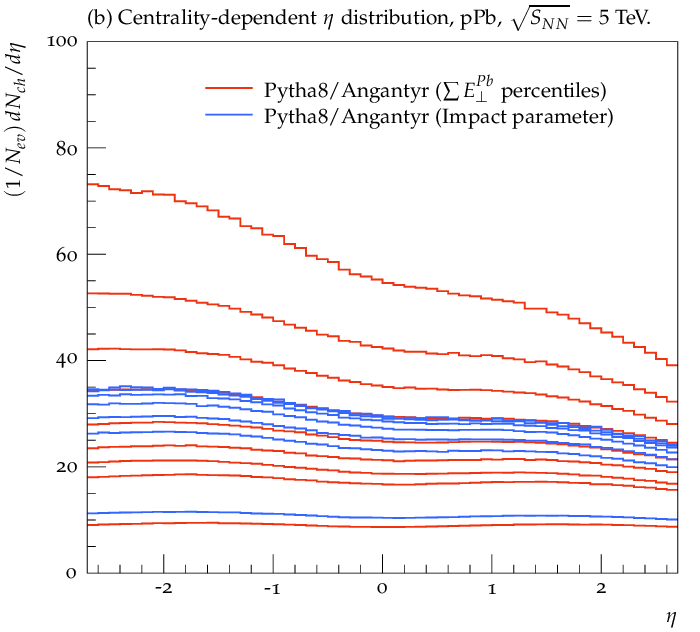}
  \caption{\label{fig:ATLASpPbEta}Comparison between the average
    charged multiplicity as a function of pseudo rapidity in
    percentile bins of centrality for \pPb\ collisions at
    $\sqrtsNN=5$~TeV. In (a) data from ATLAS \cite{Aad:2015zza}
    is compared to results from \angantyr. The lines correspond to
    the percentile bins in \figref{fig:ATLASpPbCent} (from top to
    bottom: 0--1\%, 1--5\%, \ldots, 60--90\%). The red line is
    binned using percentiles of the generated $\sum E_\perp^{Pb}$, and
    the blue line according to the experimental distribution (\cf\ the
    table in \figref{fig:ATLASpPbCent}). In (b) the red line is the
    same as in (a), but here the blue line uses percentile bins based
    on the generated impact parameter in \angantyr. }}
      
In \figref{fig:ATLASpPbEta}(a) we show the average charged particle
multiplicity as a function of pseudo-rapidity measured in the
centrality bins defined in \figref{fig:ATLASpPbCent}. It is important
to remember that even if this is presented as the centrality
dependence of the pseudo-rapidity distribution, what is in fact
measured is the correlation between the transverse energy flow in the
direction of the nuclei and the central multiplicity. In the figure we
therefore show two sets of lines generated with \angantyr with the two
different binnings presented in \figref{fig:ATLASpPbCent}. Clearly the
difference between the two is not significant, which is an indication that
\angantyr fairly well reproduces the centrality measure. And the fact
that neither curve is far from the experimental data\footnote{The
  $\eta$-distributions in \figref{fig:ATLASpPbEta}(a) has been
  corrected for detector effects.} gives a strong indication that the
\angantyr is a reasonable way of extrapolating \pp\ final states to
\pA.

Comparing to the results we presented in \cite{Bierlich:2016smv},
the description of data has been much improved. The main reason for
this is the more careful treatment of secondary absorptive sub-events,
but the new handling of the impact-parameter dependence in the
primary absorptive events has also somewhat improved the description of
data.

\FIGURE[ht]{
  \begin{minipage}{1.0\linewidth}
    \begin{center}
      \includegraphics[width=0.50\textwidth]{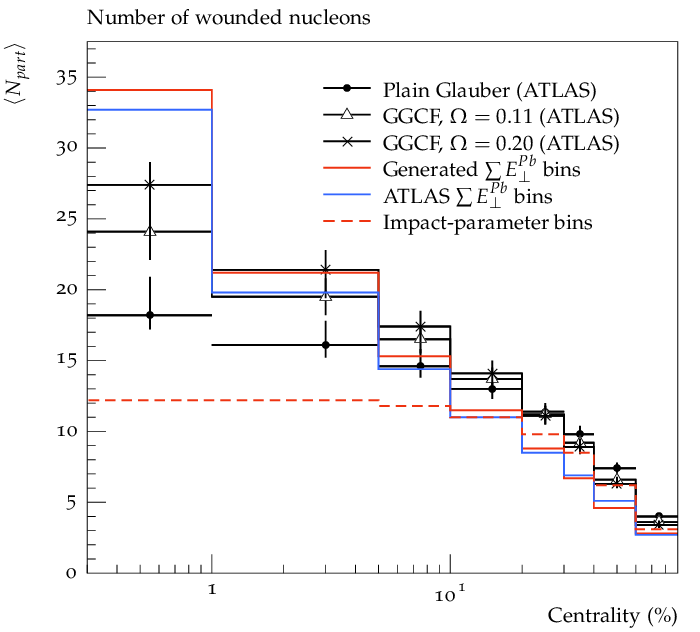}
    \end{center}
  \end{minipage}
  \caption{\label{fig:ATLASpPbNpart} Average number of wounded
    nucleons as a function centrality for \pPb\ collisions at
    $\sqrtsNN=5$~TeV. The point are taken from
    \cite{Aad:2015zza} where the numbers were calculated using three
    different Glauber calculations: filled circles used a standard
    calculation without fluctuations, while triangles and crosses used
    the model in \eqref{eq:strik} with fluctuations controlled by
    $\Omega=0.11$ and $\Omega=0.20$ respectively. The solid lines are
    generated with \angantyr binned in generated (red) and
    experimental (blue) $\sum E_\perp^{Pb}$ percentiles. The dashed
    line is also from \angantyr, but binned in impact-parameter
    percentiles.}}

Within our model it is possible to look at the actual centrality of an
event in terms of the generated impact parameter, and in
\figref{fig:ATLASpPbEta}(b) we show a comparison between the
pseudo-rapidity distribution when binned in percentiles of the
generated impact parameter and when binned in the generated
$\sum E_\perp^{Pb}$ distribution. Clearly, in the \angantyr model, the
binning in $\sum E_\perp^{Pb}$ is not very strongly correlated with the actual
centrality in impact parameter. This is especially the case for the most 
central collisions. The reason for this is the fluctuations modelled
in \angantyr, both in the number of wounded nucleons and in the
correlation between the number of wounded nucleons and the activity in
the direction of the nucleus.

To study the fluctuations further we show in
\figref{fig:ATLASpPbNpart} the average number of wounded nucleons as a
function of $\sum E_\perp^{Pb}$-centrality, both for \angantyr and for
three Glauber-model fits performed by ATLAS in \cite{Aad:2015zza}: one
using standard calculation without fluctuations, and two using the
fluctuating cross sections in \eqref{eq:strik} with different
$\Omega$-parameters. Clearly we see that \angantyr has larger
fluctuations than these standard calculations. In
\figref{fig:ATLASpPbNpart} we also show the number of wounded nucleons
in percentile bins of generated impact parameter. As expected the
dependence is very weak for the most central bins ($0 - 30$\%),
confirming here that the ATLAS centrality measure mainly picks up the
fluctuations in the number of wounded nucleons in this region, and
does not correlate very well with the actual impact parameter. The
number of participant nucleons is a thus highly model dependent
quantity, especially considering \pA\ collisions.

\FIGURE[ht]{
  \begin{minipage}{\linewidth}
    \begin{center}
      \includegraphics[width=0.5\textwidth]{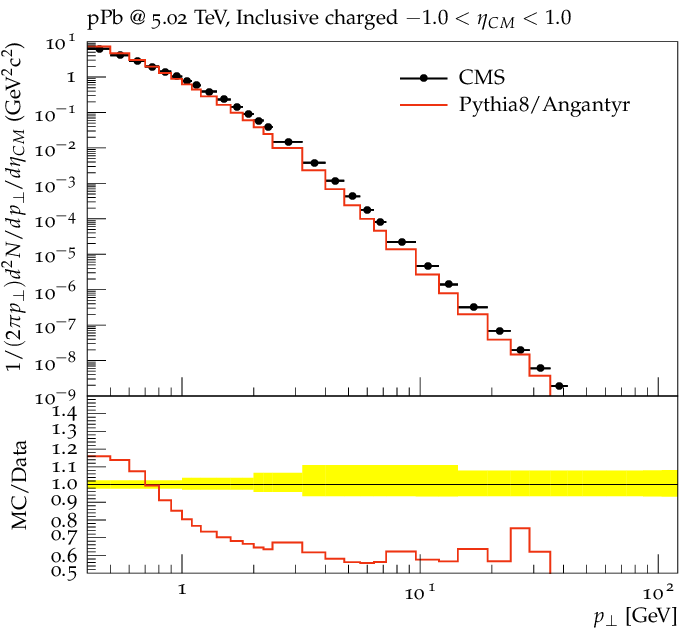}%
    \end{center}
  \end{minipage}
  \caption{\label{fig:CMSpPb}The transverse momentum distribution of
    charged particles in the central pseudo-rapidity region in
    inclusive \pPb\ events.} }

Another way of studying possible nuclear effects in \pA\ is to study
particle production as a function of $p_\perp$. In \figref{fig:CMSpPb}
we show a comparison to CMS data. The model is clearly not perfect,
but nevertheless gives a fair description of the shape over the ten
orders of magnitudes shown. Comparing to the results in
ref.~\cite{Bierlich:2016smv} we again see an increased agreement due to the
more careful treatment of secondary absorptive sub-events.

\subsection{\boldmath\AA\ results}
\label{sec:aa-results}

When we now turn to \AA\ collisions, we expect the fluctuations to
have less influence on the centrality measure, since at small impact
parameters there are so many \NN\ sub-collisions that most
fluctuations will average out. It is therefore reasonable to assume
that basically any centrality observable based on multiplicity or
energy flow in the nuclei directions will be well correlated with the
number of wounded nucleons and the actual impact parameter. Since we
will now compare simulation to results from the ALICE experiment, we
must in principle use the ALICE experimental definition of
centrality, rather than the one from ATLAS used in the previous
chapter. In ALICE centrality is defined as percentiles of the
amplitude distribution obtained in the two V0 detectors, placed at
$-3.7 < \eta < -1.7$ and $2.8 < \eta < 5.1$. Since this amplitude is
not unfolded to particle level, and cannot be reproduced by \angantyr
without realistic detector simulation, we instead construct a
reasonable particle level substitute for this measure. We assume that
the V0 amplitude is proportional to the total $\sum E_\perp$ from
charged particles with $p_\perp > 100$~MeV in that region.

\FIGURE[ht]{
  \includegraphics[width=0.7\textwidth]{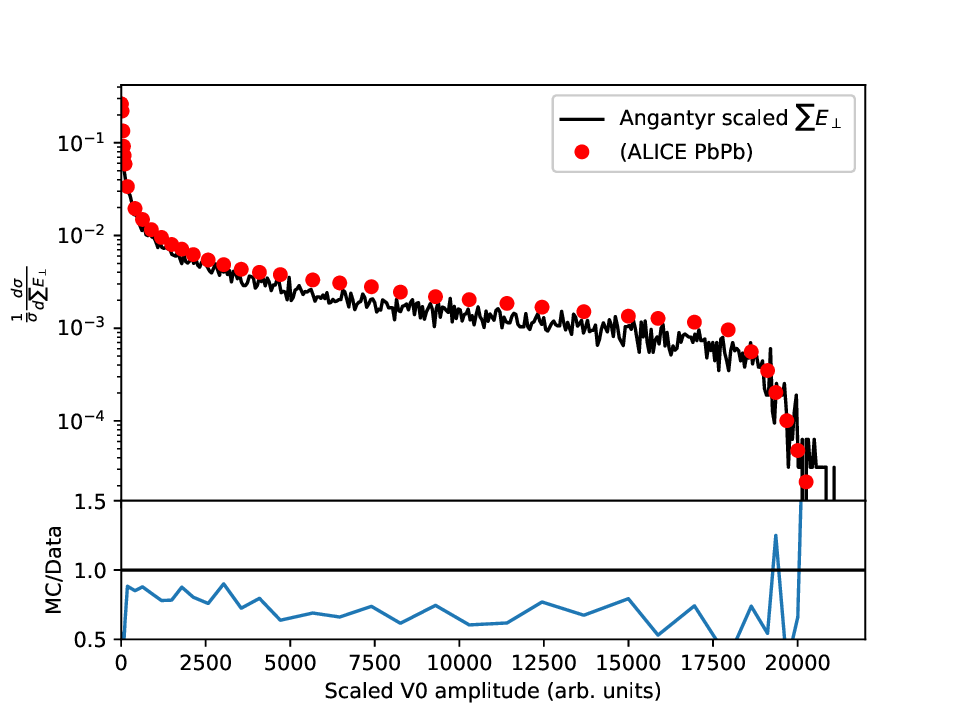}
  \caption{\label{fig:alice-cent}Scaled $\sum E_\perp$ of charged
    particles at $-3.7 < \eta < -1.7$ and $2.8 < \eta < 5.1$ from
    \angantyr, compared with the ALICE V0 amplitude, data taken from
    \citeref{Aamodt:2010cz}.}
}

In \figref{fig:alice-cent} we compare the measured V0 amplitude 
\cite{Aamodt:2010cz} with the substitute observable,
scaled to match the bin just before the distribution drops sharply at
high amplitudes. The shape of the distribution is described quite
well, while the normalisation is a bit off. This is likely due to
difficulties extracting the data for very low amplitudes. We will
throughout this section use this as a centrality observable, combined
with the trigger setup described in \citeref{Aamodt:2010cz}.
Furthermore, all experiments have some definition of what a primary
particle is. In \figref{fig:ATLASpPbEta} we used the ATLAS definition
where all particles with $c\tau>10$~mm are considered as
primary\footnote{This means \eg, that a pair of $\pi^+\pi^-$ which
  comes from the decay of a $K_S^0$, will not be included in the
  charged multiplicity}. The ALICE definition is at its heart very
similar, but has been described in more detail in
\citeref{ALICE-PUBLIC-2017-005}. This definition has been
conveniently implemented in Rivet \cite{ALICE-PUBLIC-2017-005}, and we
use this definition instead of a cut on $c\tau$.

\FIGURE[ht]{
  \includegraphics[width=0.50\textwidth]{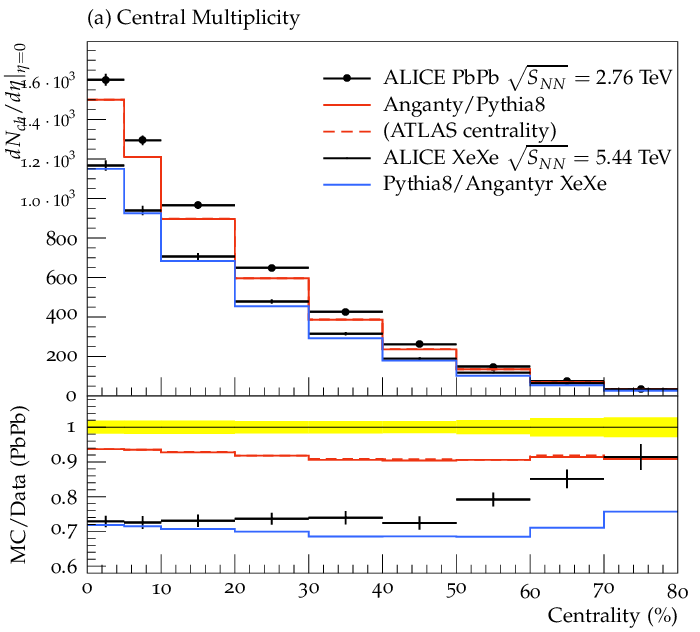}%
  \includegraphics[width=0.50\textwidth]{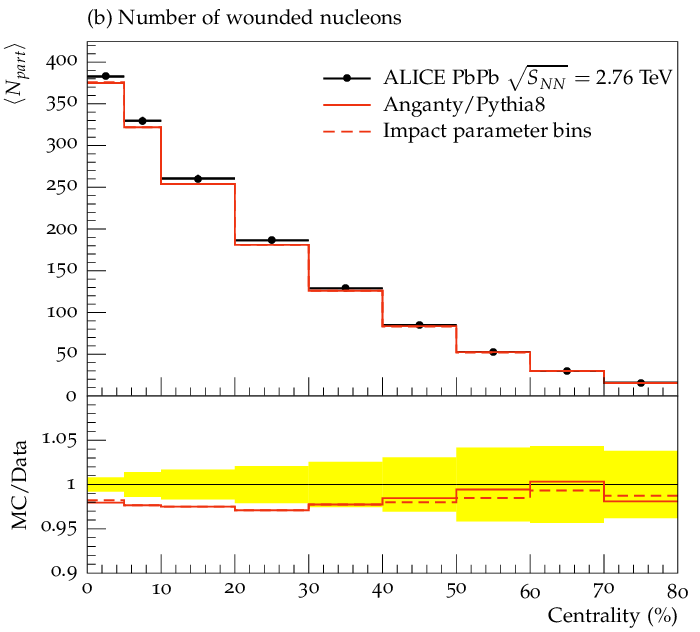}
  \caption{\label{fig:ALICEcmult}(a) The centrality dependence of the
    average charged multiplicity in the central pseudo-rapidity bin for
    \PbPb\ collisions at $\sqrtsNN=2.76$~TeV and \XeXe\ collisions
    at $\sqrtsNN=5.44$~TeV. Data points (for \PbPb) are from
    ALICE \cite{Aamodt:2010cz}, while red (\PbPb) and blue (\XeXe)
    lines are from \angantyr. (b) Shows the averaged number of
    wounded nucleons as a function of centrality. The points are from
    a Glauber-model calculations from ALICE \cite{Aamodt:2010cz}, while
    the red line is the result from \angantyr. For comparison the
    dashed line shows the number of wounded nucleons as a function of
    percentiles in generated impact parameter in \angantyr.}
}

In order to finish the discussion on the centrality measure, we show
in \figref{fig:ALICEcmult}(a) the ALICE results on the centrality
dependence of the average charged multiplicity in the central
pseudo-rapidity bin for \PbPb\ collisions at
$\sqrtsNN=2.76$~TeV~\cite{Aamodt:2010cz} using the measured
centrality, and in \figref{fig:ALICEcmult}(b) with impact parameter
bins. The agreement between these two results are clearly much better
in \PbPb\ than for \pPb, confirming the initial statement in this
section.

In \figref{fig:ALICEcmult}(a) we also show our
predictions\footnote{Although we present this after the data was
  published we still consider it a prediction, as the program was
  released before the data was analysed.} for Xenon--Xenon collisions
at $\sqrtsNN=5.44$~TeV compared to the \alice data that were published
in \cite{Acharya:2018hhy}.

In \figref{fig:ALICEeta} we show the charged multiplicity compared to ALICE
data \cite{Abbas:2013bpa,Adam:2015kda,Adam:2016ddh}
over a much wider $\eta$ range, for both $\sqrtsNN=2.76$~TeV and
$\sqrtsNN=5.02$~TeV. 
\FIGURE[ht]{
  \includegraphics[width=0.50\textwidth]{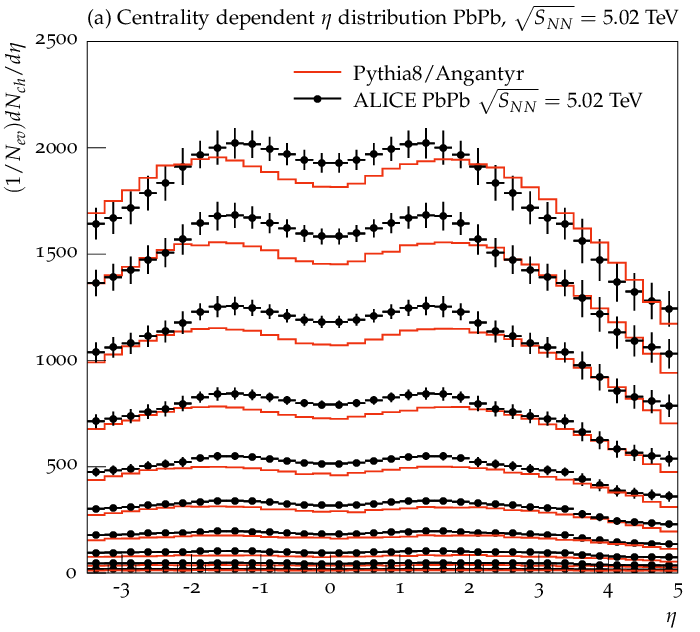}%
  \includegraphics[width=0.50\textwidth]{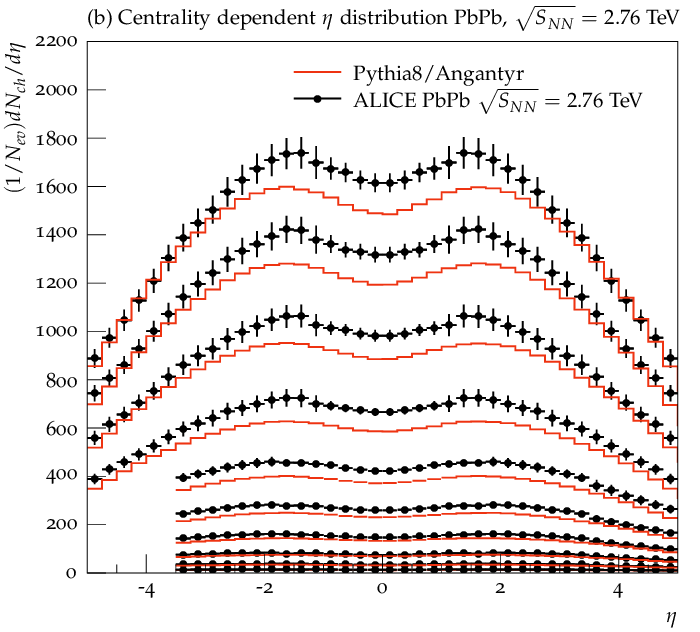}
  \caption{\label{fig:ALICEeta}The centrality dependence charged
  multiplicity over a wide $\eta$ range in \PbPb\ collisions at 
  $\sqrtsNN=5.02$~TeV (a) and $\sqrtsNN=2.76$~TeV (b). Both for
  centralities 0-5\%, 5-10\%, 10-20\%, 20-30\%...80-90\%.
  Data from ALICE \cite{Abbas:2013bpa,Adam:2015kda,Adam:2016ddh}.}
} The trend, also visible in \figref{fig:ALICEcmult}, is that
\angantyr produces somewhat too few particles at central $\eta$;
the multiplicity is systematically 5-10\% too low. We regard this as
surprisingly good, considered that no tuning of any kind to \AA\ data
has been done.

\FIGURE[ht]{
  \includegraphics[width=0.50\textwidth]{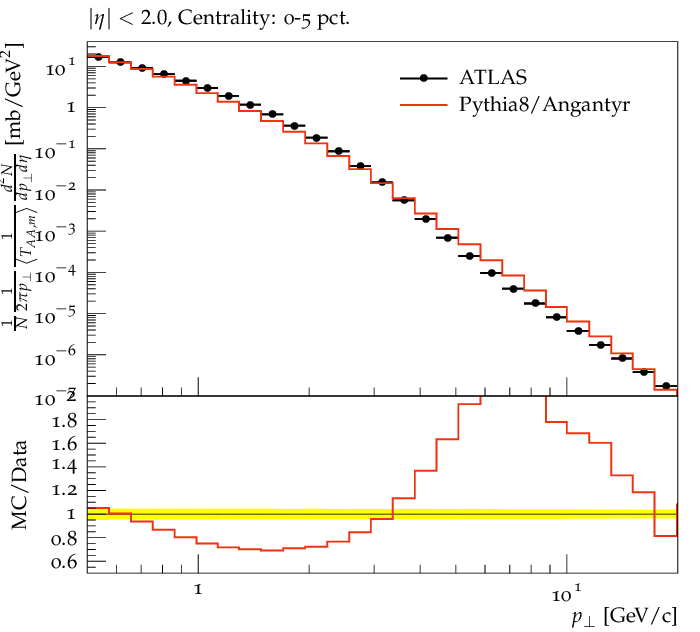}%
  \includegraphics[width=0.50\textwidth]{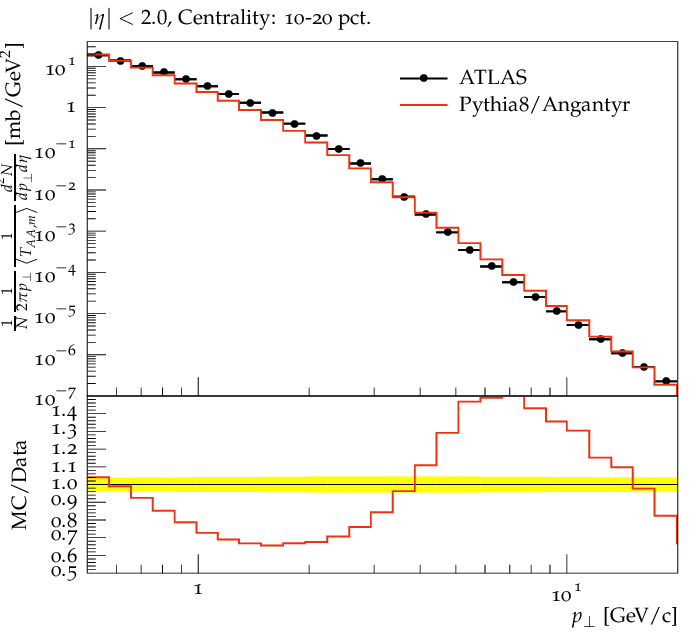}
  \includegraphics[width=0.50\textwidth]{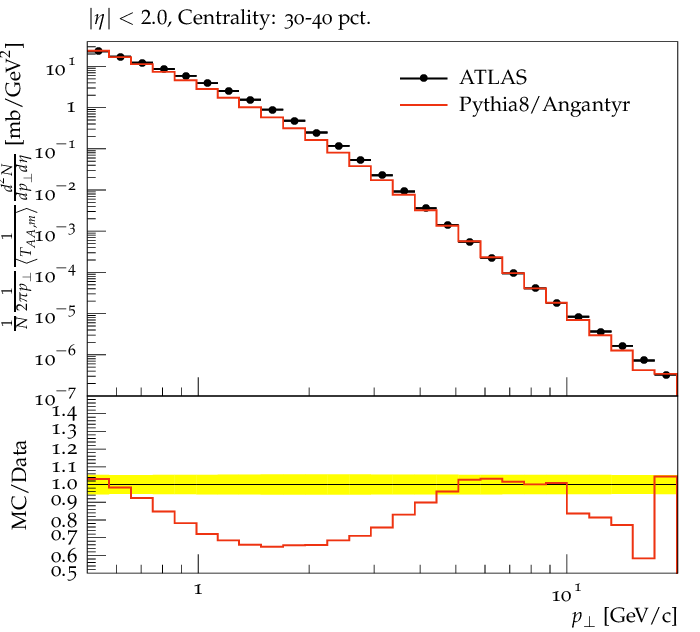}%
  \includegraphics[width=0.50\textwidth]{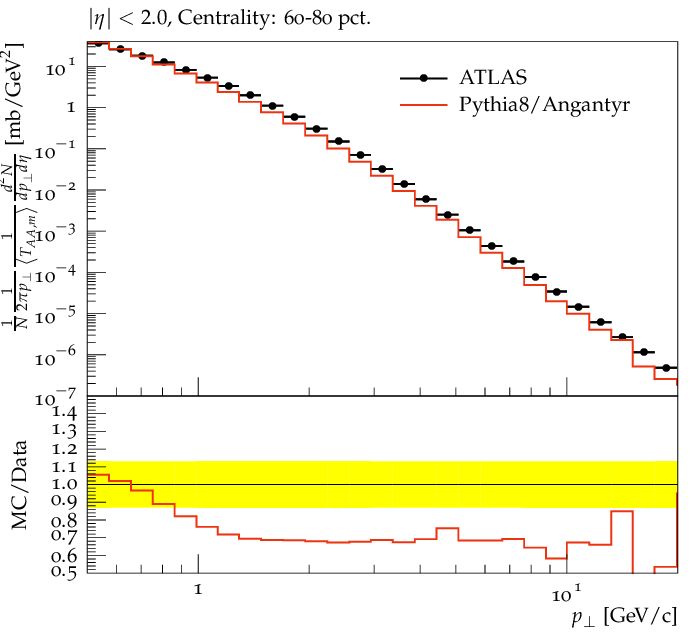}
  \caption{\label{fig:ATLASpt} Transverse momentum distributions of charged particles
  in \PbPb\ collisions at $\sqrtsNN = 2.76$~TeV in four centrality bins,
  compared to \angantyr. Data from ATLAS \cite{Aad:2015wga}.}
}

We now turn to transverse momentum spectra in \AA\ collisions. In
\figref{fig:ATLASpt} we show results from ATLAS \cite{Aad:2015wga}
compared to our model. The published $p_\perp$ spectra was scaled with
the average number of wounded nucleons, calculated using a black disk
Glauber model. We have not used the number of wounded nucleons as
input to \angantyr, just scaled our result with the same number (as
published in the article) to obtain comparable spectra. Hence, the
results are not scaled to match, as both are simply scaled with the
same number.

Finally we want to add a comment about the low multiplicity in the
central region, shown in figs.~\ref{fig:ALICEcmult}(a) and
\ref{fig:ALICEeta}. One of the main features of \angantyr is that
tuning of MPI model, shower and hadronisation should only be carried
out using $e^+ e^-$, \ep\ and \pp\ data. However, looking at the
comparison to \pp\ in \figref{fig:ppmb}, we see that even the \pp\
model undershoots the multiplicity at very low $p_\perp$ (below
$~500$~MeV). Since ALICE measures charged particle multiplicity all
the way down to zero transverse momentum\footnote{The multiplicity
  below $~50$~MeV is extrapolated, but this does not contribute to the
  total multiplicity by more than a few percent.}, it is not clear if
the default \pytppp behaviour should even be applicable here. The
transverse momentum of such low-$p_\perp$ particles does not origin in
the (perturbative) parton shower, but rather in the dynamics of string
breakings. As seen from the comparison to \pp\, this is not yet fully
understood. The validity of this point is underlined by comparing to
the ATLAS data shown in \figref{fig:ATLASpt}, where multiplicity is
measured with low-$p_\perp$ cut--off of 500~MeV. In
\figref{fig:ATLASmult} we show the multiplicity distribution obtained
by integrating the distributions measured by ATLAS, and see that the
description improves.

\FIGURE[ht]{
  \includegraphics[width=0.65\textwidth]{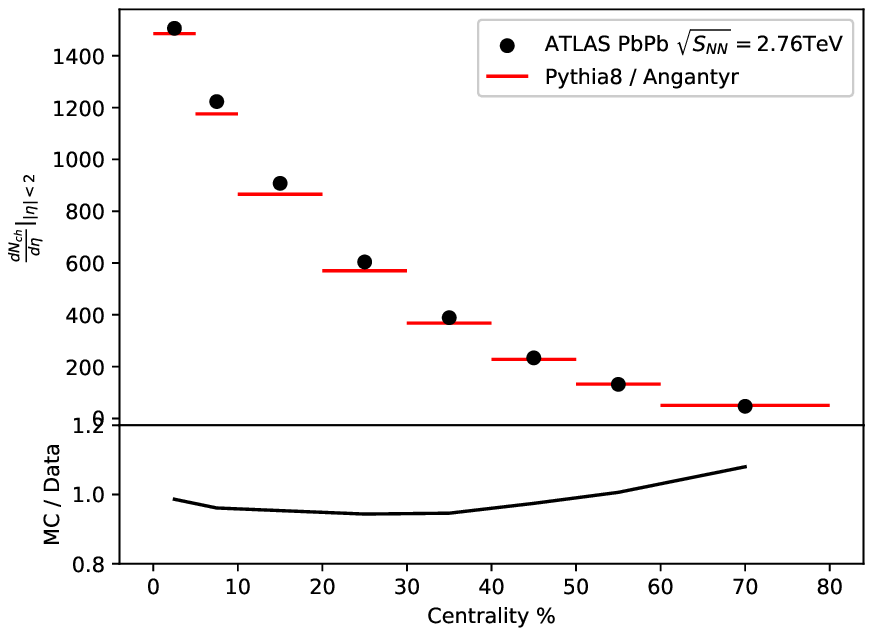}%
  \caption{\label{fig:ATLASmult} Comparison to total multiplicity at
    mid--rapidity in \PbPb\ collisions at $\sqrtsNN = 2.76$~TeV, with
    a minimum $p_\perp$ cut of 500~MeV, obtained by integrating the
    $p_\perp$ distributions measured by ATLAS \cite{Aad:2015wga}.}
}

We want to make clear that (part of) this discrepancy could of course
be due to a faulty comparison to data, where triggers, centrality
measure \etc~ is not implemented in exactly the way as it is done by
experiments. But if it is not, it points to an interesting point for
improvement of the underlying model for soft particle production, also
in \pp. We will return to this subject in a future paper, but
meanwhile we note that it would be interesting if experiments like
ALICE, who can measure very near zero $p_\perp$, will extend their
publications to also include data with a minimum $p_\perp$ cut--off,
which could serve as an important aide in further understanding.

\subsection{Collectivity and non-flow estimation}

One of the primary goals of the heavy ion programs at RHIC and LHC, is
to investigate the collective behaviour of final state particles
produced in collisions of nuclei accelerated to relativistic
energies. The anisotropic flow measures the momentum anisotropy of the
final state particles. As such, it is sensitive to both the initial
geometry of the nuclear overlap region, as well as the transport
properties of the final state before hadronisation.

The anisotropic flow is quantified in flow coefficients $v_n$ and
corresponding symmetry planes $\Psi_n$, defined by a Fourier series
decomposition of the azimuthal distribution of final state particles:
\begin{equation}
  \frac{dN}{d\phi} \propto 1 + 2 \sum_{n=1}^{\infty} v_n \cos\left[n(\phi - \Psi_n)\right].
\end{equation}
In practise, the flow coefficients are calculated using cumulants
\cite{Zhou:2015iba,Bilandzic:2010jr,Bilandzic:2013kga}, which we also
employ here. When flow coefficients are calculated using two-particle
cumulants, the calculated coefficient also picks up azimuthal
correlations not related to collectivity, but from \eg\ resonance
decays and intra-jet correlations. Such "non-flow" effects can be
suppressed by requiring a gap in $\eta$ between particle pairs.

\FIGURE[ht] {
  \includegraphics[width=0.8\textwidth]{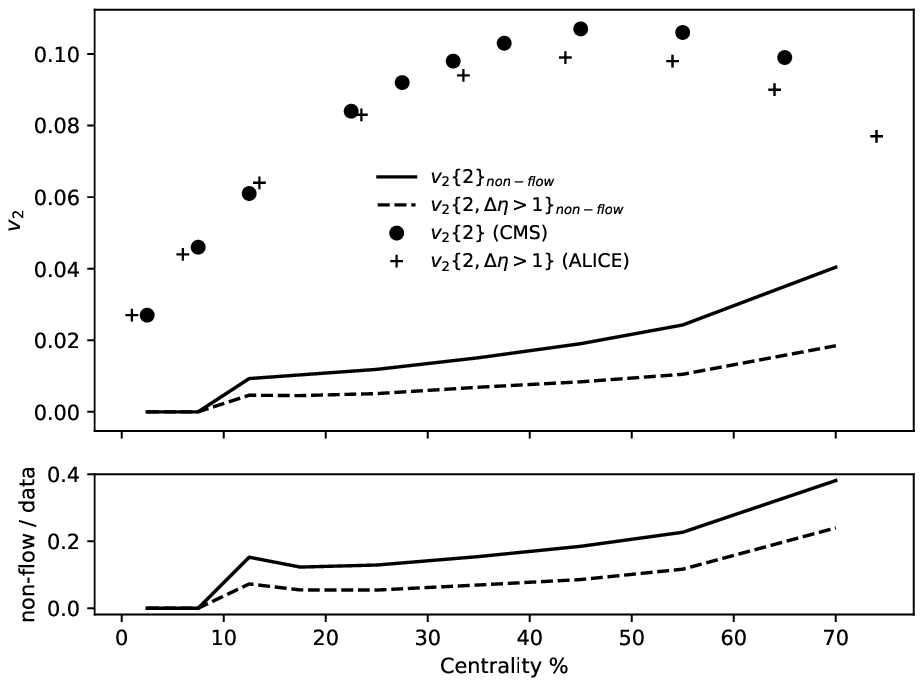}
  \caption{\label{fig:flow-nonflow}The elliptic flow coefficient
    $v_2\{2\}$ at $\sqrtsNN=2.76$~TeV, as measured by CMS
    \cite{Chatrchyan:2012ta} (without $\Delta \eta$-gap) and ALICE
    \cite{ALICE:2011ab,Adam:2016izf} (with $\Delta \eta = 1$),
    compared to the non-flow contribution calculated by \angantyr. In
    the ratio plot it is seen that the non-flow contribution without
    $\Delta \eta$-gap is nearly 40\%. This is reduced to 20\% when
    applying a gap.}
}

In \figref{fig:flow-nonflow} we show $v_2\{2\}$ as function
of centrality\footnote{Using the aforementioned adapted version of
  ALICE centrality.} measured with and without a $\Delta \eta$ gap of
1.0, by ALICE \cite{ALICE:2011ab,Adam:2016izf} and CMS
\cite{Chatrchyan:2012ta} respectively. Since \angantyr produces a full
final state, it allows for the construction of the same observable,
even in the absence of collective effects, giving an estimate of the
non-flow present. We see that the non-flow contribution in the most
central collisions is negligible (as one would expect), but rise to
about 40\% of the measured result for $v_2\{2\}$ without gap for
peripheral collisions. This number falls to 20\% when a gap is
included, indicating that the method of applying a gap can remove some
non-flow effects, but not all.

We want to emphasise that at this point, \angantyr does not make any
attempt at modelling collective effects, and can therefore be used to
estimate the contribution of non-flow. It is our plan to introduce a
microscopic model for collectivity, based on string--string
interactions to \angantyr, which has shown promising results in
\pp. The increased energy density from overlapping strings would here
give a transverse pressure, leading to strings "shoving" each other
before hadronisation \cite{Bierlich:2016vgw,Bierlich:2017vhg}.

\section{Model uncertainties}
\label{sec:uncertainties}
The main idea behind \angantyr~is to extrapolate \pp\ dynamics, as described by
the model for MPIs/underlying event in the \pytppp MC,
to heavy ion collisions, retaining as
much as possible from \pp. This principle was outlined already in the
introduction, especially \figref{fig:flowchart}, but as the model
has now been presented, as well as results from \pA~and \AA~collisions,
we will here also discuss the model uncertainties related to this extrapolation
procedure.

Primary interactions correspond directly to inelastic non-diffractive
pp collisions. Here \pytppp, is known to reproduce most features of 
both soft and hard pp collisions at LHC fairly well, 
and the extrapolation to primary interactions in a heavy ion collision
is therefore mainly a source of model uncertainty up to \pytppp's shortcomings 
in describing such collisions in pp. 
We already discussed some of those shortcomings in the previous section, but as they are not
uncertainties directly related to the \angantyr~model (but rather the underlying
\pytppp~model) we will not discuss them further here.

The largest uncertainty comes instead from our treatment of secondary absorbed
nucleons. The main reason is that secondary absorption has no pp equivalent. 
In \sectref{sec:sasd} we outlined the procedure of modifying single 
diffractive collisions to describe secondary absorbed nucleons, and we
will investigate uncertainties related to this treatment in \sectref{sec:secondary-uncertainties}.

\FIGURE[ht]{
  \begin{minipage}{\linewidth}
    \begin{center}
      \includegraphics[width=0.5\textwidth]{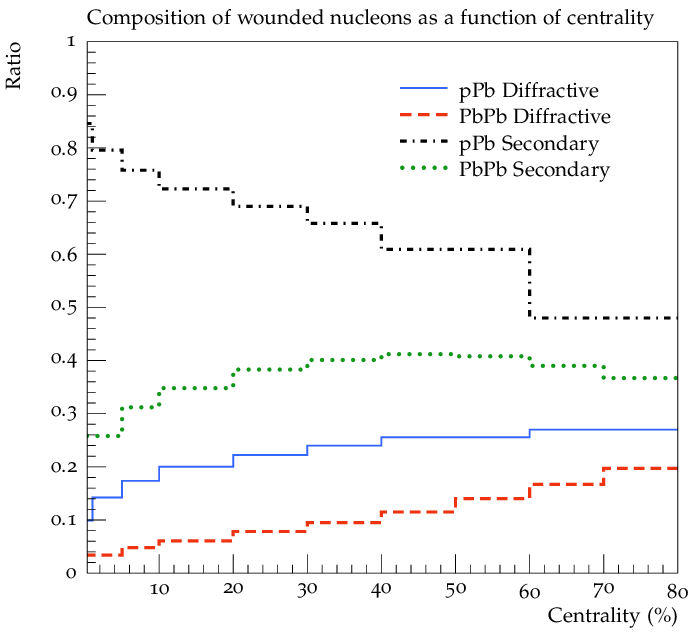}%
    \end{center}
  \end{minipage}
  \caption{\label{fig:Diffrat} The fraction of the wounded nucleons in
    the \angantyr model that are diffractively excited as a function
    of centrality for \pPb\ at $\sqrtsNN=5$~TeV (blue line) and \PbPb\
    at $\sqrtsNN=2.76$~TeV (red dashed line). Also shown is the
    fraction of wounded nucleons that come from secondary absorptive
    interactions in \pPb\ (black dash-dotted line) and \PbPb\ (green
    dotted line).} }

Diffractively excited nucleons give a comparatively small contribution
in collisions with nuclei, especially in central \AA\ collisions, as
illustrated in \figref{fig:Diffrat}. Diffractive excitation of nucleons can in
principle be determined from \pp\ collisions, but as we will discuss in
\sectref{sec:diffractive-uncertainties}, this is not straight forward.

\subsection{Uncertainties in treating secondary wounded nucleons} 
\label{sec:secondary-uncertainties}

A core feature of the \angantyr\ model, is that the contribution from a
secondary absorbed nucleon is similar to the contribution from an excited
nucleon in a single diffraction event. This corresponds to the black pieces in
figures \ref{fig:mpi-pA}a and 
\ref{fig:mpi-pA}b respectively. This assumption has two components:
\begin{enumerate}[ {(}i{)}]
	\item The distributions in the rapidity range covered, $\Delta y$, and 
	the corresponding mass, \mbox{$M\approx \exp(\Delta y/2) \times (1 \mathrm{GeV})$}, 
	are similar.
	\item The distribution of partons from the projectile nucleon,
        involved in the interaction with the secondary absorbed nucleon in
        \figref{fig:mpi-pA}a,  
	is similar to the partons in the Pomeron in \figref{fig:mpi-pA}b.
\end{enumerate}
Naturally none of these assumed similarities can be exact. 
Extracting the relevant properties in diffractive excitation in pp collisions
from data at LHC has also large uncertainties, as we will discuss further in 
\sectref{sec:diffractive-uncertainties}.
We also note that:
\begin{enumerate}[ {(}i{)}]
	\setcounter{enumi}{2}
	\item Energy--momentum conservation has generally important effects in
        high energy reactions, and has to be satisfied when nucleons suffer
        multiple \NN\ sub-collisions. 
\end{enumerate}
Also this point is associated with some model uncertainty, as discussed
in \sectref{sec:adding-second-absorp} and in \sectref{sec:EM-cons} below.

In the following we will discuss the uncertainties associated with all
three choices in the treatment of secondary wounded nucleons, and
their impact on model predictions. We will focus on \pA\
collisions, where there can at most be a single
primary interaction, and the treatment of secondary interactions consequently
has a relatively larger effect. This is illustrated in
\figref{fig:Diffrat}, where we see that secondary absorbed nucleons correspond
to about 80\% of all wounded nucleons in central \pPb\ collisions, but only
about 25\% in central \PbPb\ collisions.

\FIGURE[t]{  \centering
\scalebox{0.8}{\mbox{ 
  \begin{picture}(300,220)(0,50)
    \GOval(50,220)(7,7)(0){1}
    \GOval(10,80)(7,7)(0){1}
    \GOval(90,80)(7,7)(0){1}
    \def\axowidth{1.5 }    
    \Line(10,220)(43,220)
    \Line(57,220)(90,220)
    \Line(-20,80)(3,80)
    \Line(17,80)(45,70)
    \Line(55,70)(83,80)
    \Line(97,80)(120,80)
    \def\axowidth{0.5 }
    \ZigZag(50,213)(50,150){6}{6}
    \ZigZag(50,150)(10,87){6}{6}
    \ZigZag(50,150)(90,87){6}{6}
    \SetColor{Red}
    \DashLine(50,250)(50,50){4}
    \SetColor{Black}

    \LongArrow(150,220)(150,153)
    \Text(160,185)[c]{$y_1$}
    \LongArrow(150,150)(150,220)
    \LongArrow(150,150)(150,80)
    \Text(160,115)[c]{$y_2$}
    \LongArrow(150,80)(150,147)
    \Text(70,185)[c]{$M_D$}
    \Text(290,115)[c]{$M_A$}

    \GOval(250,220)(7,7)(0){1}
    \GOval(210,88)(7,7)(0){1}
    \GOval(290,72)(7,7)(0){1}
    \def\axowidth{1.5 }    
    \Line(210,220)(243,220)
    \Line(257,220)(290,220)
    \Line(180,88)(203,88)
    \Line(217,88)(273,88)
    \Line(291,88)(320,88)
    \Line(180,72)(195,72)
    \Line(180,72)(283,72)
    \Line(297,72)(320,72)
    \def\axowidth{0.5 }
    \ZigZag(250,213)(250,150){6}{6}
    \ZigZag(250,150)(212,95){6}{6}
    \ZigZag(250,150)(287,79){6}{6}
    \SetColor{Red}
    \DashLine(250,250)(250,150){4}
    \DashLine(250,150)(202,76){4}
    \DashLine(198,70)(191,60){4}
    \DashLine(250,150)(300,50){4}
    \SetColor{Black}
    \Text(50,40)[]{(a)}
    \Text(250,40)[]{(b)}
  \end{picture}}}

\caption{Pomeron diagrams with cuts indicated for (a) single
  diffractive excitation in proton--proton and (b) doubly absorptive
  proton--deuteron scattering.}
\label{fig:cutpom}
}

\subsubsection{Mass distribution}
We begin by discussing point (i), the mass distribution of secondary wounded nucleons.
The picture in \figref{fig:mpi-pA}a has the structure of a triple-Pomeron diagram. 
This similarity is somewhat symbolic, as each chain in this figure includes 
the multiple parton scatterings in \figref{fig:mpi-pp}, which correspond to 
Pomeron loops in a Reggeon field theoretical approach (see \textit{e.g.}
\citerefs{Gotsman:2014pwa, Khoze:2014aca, Ostapchenko:2010vb}). The
triple-Pomeron diagrams shown in \figref{fig:cutpom} would have a weight
proportional to: 
\begin{eqnarray}
dy_1 dy_2 \delta(y_1+y_2-Y) \exp(\Delta (y_1+2y_2))&=& \nonumber \\
	= \frac{ds}{s^{(1-2\Delta)}}\, \frac{d M_D^2}{(M_D^2)^{(1+\Delta)}}
       && \mbox{for diffractive excitation}, \nonumber \\
       = \frac{ds}{s^{(1-\Delta)}}\, \frac{d M_A^2}{(M_A^2)^{(1-\Delta)}}
       && \mbox{for secondary absorption}.
\label{eq:triplepomeron2}
\end{eqnarray}
Here $y_1$ and $y_2$ are the rapidities indicated in the figure,
and $Y=y_1+y_2 \propto\ln(s)$ is the total allowed rapidity range. 
The quantity $M_D \propto \exp(y_1/2)$ is the diffractively excited mass to the
left, and $M_A \propto \exp(y_2/2)$ is the mass of the secondary
absorbed nucleon to the right. Finally the expression $1+\Delta = \alpha_\mathbb{P}(0)$ 
is the intercept of the Pomeron trajectory.

As discussed above, in the default version of \angantyr we assume a mass 
distribution $\propto dM^2/M^2$ for both diffractively excited and secondary 
absorbed nucleons, corresponding to a critical Pomeron with $\Delta=0$.
With a hard BFKL-like Pomeron one could imagine a positive $\Delta$ in the 
range $0 <\Delta < 0.2$. In \figref{fig:pPbMXvar} we show the result of
generating the secondary absorptive sub-events with $\Delta=0$, $0.1$,
and $0.2$. From the $\sum E_\perp^{Pb}$ distribution in \pPb\ at $\sqrtsNN=5.02$~TeV
(used by ATLAS as centrality measure) shown in \figref{fig:pPbMXvar}a, we see a
noticeable effect already below 50~GeV. The effect follows the expectation that
a larger $\Delta$ will give larger $M_A$ values and thus more
activity. However, we also see that above 50~GeV the distributions for 
larger $\Delta$ seem to run out of steam, which we attribute to the fact
that higher $M_A$ values mean that the energy available from the
projectile proton is used up faster. This means that fewer secondary
absorptive interactions are accepted. In \figref{fig:pPbMXvar}b we also show the 
resulting pseudo-rapidity distribution of charged particles for two
centrality bins (using the experimentally determined bin edges in
$\sum E_\perp^{Pb}$). The larger values of $M_A$ are 
also reflected in the $\eta$-distributions, where the effect is that 
the distribution becomes too flat to describe data, especially for central events.

\FIGURE[ht]{
  \begin{minipage}{\linewidth}
    \begin{center}
      \includegraphics[width=0.5\textwidth]{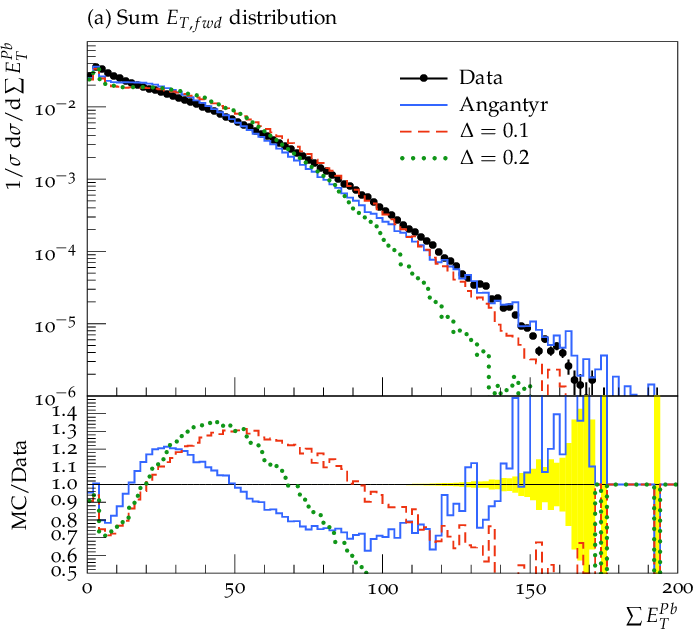}%
      \includegraphics[width=0.5\textwidth]{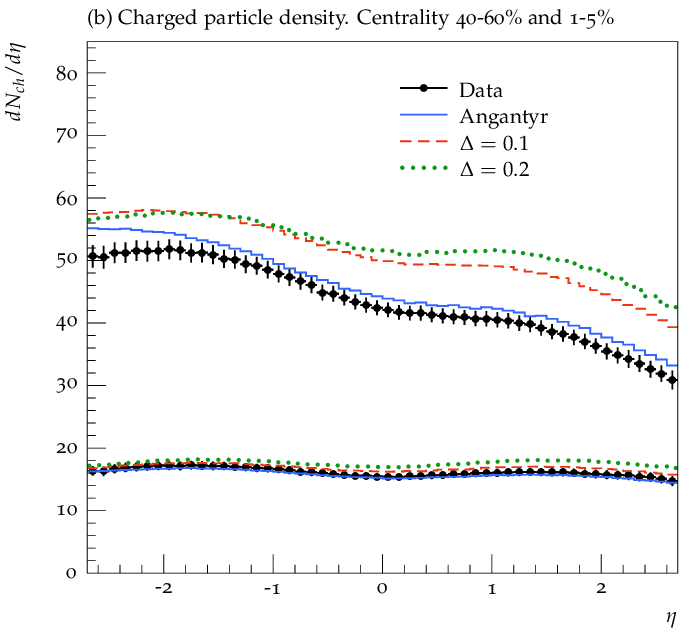}%
    \end{center}
  \end{minipage}
  \caption{\label{fig:pPbMXvar} Comparison between different choices
    of $\Delta$ for generation of secondary absorptive
    sub-events. Variations shown for (a) the summed transverse energy
    in the Pb direction ($-4.9<\eta<-3.2$) and (b) the average charged
    multiplicity as a function of pseudo--rapidity for \pPb\
    collisions at $\sqrtsNN=5.02$~TeV. Data points are from ATLAS
    \cite{Aad:2015zza}. Full blue line is the default choice of
    $\Delta=0$, while the red dashed and green dotted lines
    corresponds to $\Delta=0.1$ and $0.2$ respectively. In (b) the
    lines on the bottom and top corresponds to the 40-60\% and 1-5\%
    centrality bins respectively, using the experimentally defined bin
    limits in $\sum E_\perp^{Pb}$.} }

\subsubsection{Parton distribution in the projectile}

As discussed in \sectref{sec:sasd}, the secondary absorptive interaction in
\figref{fig:mpi-pA}a may involve several partons coming from the projectile
nucleon, in a way similar to how diffractive excitation is described by a
Pomeron PDF in the Ingelman--Schlein model. Point (ii)
concerns the distribution of these partons. In \sectref{sec:sasd} we studied three
different distributions, SD(new) (which is the default for secondary absorption),
SD(def) (which is the \pytppp~default for diffractive excitation), and
SD(glu) (which is the modified PDF for increased gluon activity introduced in 
ref.~\cite{Bierlich:2016smv}).
In \figref{fig:pPbSDvar}a we show the effect on the $\sum E_\perp^{Pb}$
distribution. Below 50~GeV, where the bulk of the events are found, all 
three options are reasonably close to each other, but the tail of the
distributions diverges considerably, in a way consistent with the differences
found in \sectref{sec:sasd}. The resulting pseudo--rapidity distributions
shown in \figref{fig:pPbSDvar}b do not show so dramatic differences. It
is, however, clear that our default choice gives the best description of
data. As discussed in \sectref{sec:sasd} our default choice is the one
that makes most sense on theoretical grounds, and it is satisfying to
see that it also makes sense in comparison to data.

\FIGURE[ht]{
  \begin{minipage}{\linewidth}
    \begin{center}
      \includegraphics[width=0.5\textwidth]{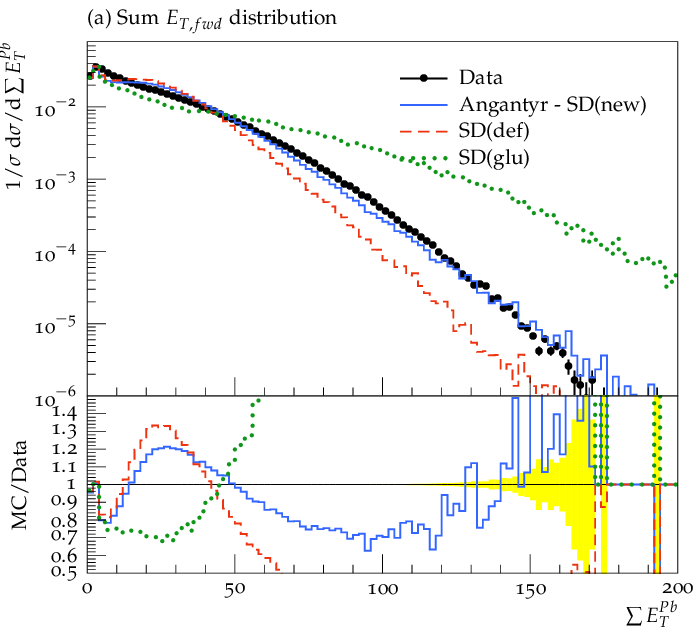}%
      \includegraphics[width=0.5\textwidth]{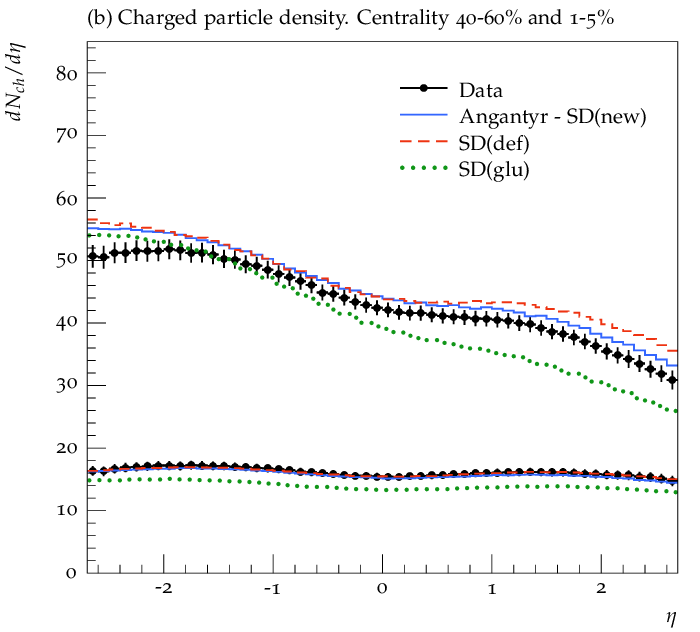}%
    \end{center}
  \end{minipage}
  \caption{\label{fig:pPbSDvar} Same as \figref{fig:pPbMXvar},
    but comparing different choices in the treatment of secondary 
    absorptive interactions. The lines corresponds to the models in
    \figref{fig:HF2}. } }

\subsubsection{Energy-momentum conservation}
\label{sec:EM-cons}

Energy-momentum conservation is frequently seen to have a very large impact
in high energy reactions. Here its effect could be seen in
\figref{fig:pPbMXvar}a. It is not clear from first principles
if energy--momentum conservation should prohibit a sub--collision, if a single 
sampling of the $M_A$ distribution turns out to require more than what is
available, or if it is possible to simply try again.
To further study the effects of this ambiguity, we show in
\figref{fig:pPbEMvar}, what happens if we allow \angantyr to retry adding
secondary 
sub-events, which fail due to energy-momentum conservation (as discussed
in \sectref{sec:adding-second-absorp}). We see that it does have an
impact on the most central collisions in the $\sum E_\perp^{Pb}$
centrality measure, while the effect on the resulting
$\eta$-distribution is barely visible. It is interesting to note that
the effect of allowing more attempts seem to saturate quickly, and
going from 2 to 4 attempts makes a much smaller change than
allowing two attempts instead of one (which is the default).

\FIGURE[ht]{
  \begin{minipage}{\linewidth}
    \begin{center}
      \includegraphics[width=0.5\textwidth]{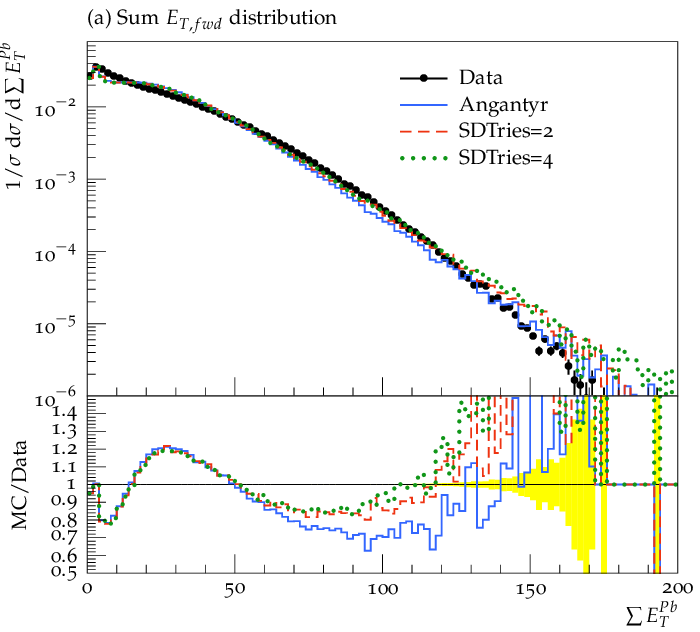}%
      \includegraphics[width=0.5\textwidth]{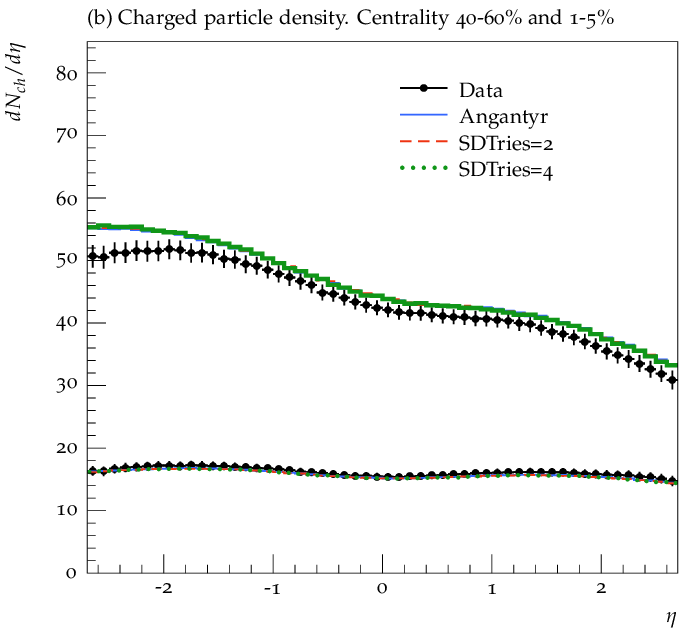}%
    \end{center}
  \end{minipage}
  \caption{\label{fig:pPbEMvar} Same as \figref{fig:pPbMXvar}, but now
    varying the number of attempts (parameter
    \texttt{Angantyr:SDTries}) allowed to generate secondary
    sub-events that can be added without violating energy--momentum
    conservation before giving up and vetoing the secondary
    interaction. The default version allows only a single attempt and
    is shown as the blue lines, while allowing two or four attempts is
    shown as dashed red and dotted green lines respectively.} }

\subsection{Diffractively excited nucleons}
\label{sec:diffractive-uncertainties}

In contrast to the secondary absorbed nucleons, a positive $\Delta$ in
\eqref{eq:triplepomeron2} means lower masses for diffractively excited
nucleons. In principle the $M_D$-distribution could be measured in pp
collisions at LHC, but it is quite challenging to isolate single diffraction
from the experimental distribution in the size of a gap in rapidity (see
refs. \cite{Aad:2012pw, Khachatryan:2015gka}).

In collisions with nuclei, multiple \NN\ interactions imply that, the
probability for absorption is enhanced and, as a consequence, the
probability for diffractive excitation is reduced. From \figref{fig:Diffrat}
we see that in \pA\ collisions about 20\% of the wounded nucleons are
diffractively excited, dropping to 10\% in central \pPb\ collisions.
In \AA\ collisions this fraction is further reduced to an average about 10\%,
and below 4\% for central \PbPb\ collisions. This implies that a reasonable
variation of the diffractive component will have comparatively small effect. 
For this reason we have here chosen to keep the
default setting in the \pytppp MC, with a distribution $\propto dM_D^2/M_D^2$. 

One could here also imagine including a Reggeon contribution $\propto
dM_D^2/(M_D^2)^{1.5}$. This contribution is concentrated to low
masses, and would not affect the results in most of the rapidity range,
including the forward detectors used to measure the centrality. It could,
however, give a contribution in the very forward region, and thus it might be
of importance \textit{e.g.} for interactions with cosmic rays.

\subsection{Uncertainties in \AA\ collisions}

Above we have discussed model uncertainties in \pA\ collisions. We have also pointed out that the corresponding uncertainties are significantly smaller in \AA\ collisions, in particular in central \AA\ collisions. In \figref{fig:Diffrat} we showed that the fraction of wounded nucleons which are secondary absorbed is about 70\% in \pPb\ but about 35\% in \PbPb\ collisions. For central collisions these ratios are about 80\% in \pPb\ and only about 25\% in \PbPb. We have checked
that a corresponding reduction of the uncertainties is obtained in the MC results for \AA\ collisions.

\section{Relation to other models}
\label{sec:other-models}

As \angantyr is a new model, it is instructive to compare it to
existing models, and we here discuss the most commonly used ones, also
mentioned in the introduction, HIJING \cite{Wang:1991hta}, AMPT
\cite{Lin:2004en}, and EPOS-LHC \cite{Pierog:2013ria}. Here HIJING is
most similar to \angantyr. Like \angantyr it is constructed as an
extrapolation of \pp\ dynamics, with the explicit motivation that
differences between the model and experimental results may indicate
effects of collective behaviour. In contrast AMPT and EPOS are both
assuming collective expansion of a thermalised medium.

The HIJING generator is built with a similar starting point as
\angantyr, thus it is inspired by the \fritiof model, using \pythia
for generating multiple hard partonic sub-collisions and the Lund
string model in \pythia for the hadronisation. Similarly to
\angantyr, HIJING relies on a Glauber calculation to determine the
number of inelastic sub-collisions, which are of two types: soft
nucleon-nucleon collisions treated as in \fritiof, and hard
parton-parton collisions treated as in \pythia. A new version written
in C++ was recently presented \cite{Barnafoldi:2017jiz}.

In contrast to \fritiof the interacting nucleons are in HIJING excited
to higher masses, covering most of the available rapidity range, but
just as in the later \fritiof version \cite{Andersson:1992de,
  Pi:1992ug}, gluon radiation is added using the soft radiation scheme
\cite{Andersson:1988gp} implemented in Ariadne
\cite{Lonnblad:1992tz}. The hard partonic scatterings are determined
via nucleus PDFs, where the parton density is suppressed by a
shadowing factor $R_{a/A}$, compared to $A$ independent nucleon
PDFs. To avoid double counting, emitted gluons in the soft component
is allowed only for $p_\perp$ below a scale $p_0$ (chosen to be
$\approx 2$~GeV), while the hard partonic collisions have a lower cut
at $p_\perp=p_0$.

Another difference between \angantyr and HIJING is that in HIJING
fluctuations are neglected both in the initial states of the
individual nucleons and in the position of nucleons within the
nuclei. The soft $NN$ amplitude is then chosen to reproduce the
inelastic cross section including diffraction. The probability for
multiple scattering is determined by the nuclear overlap function in
impact parameter space. In \angantyr we find that fluctuations plus
the distinction between primary and secondary absorptively wounded
nucleons, have a quite significant effect for the final state
multiplicity. In HIJING, the same effect may partly be due to the
introduction of the shadowing factor $R_{a/A}$. The shadowing factor
is a geometry dependent "k-factor", which accounts for nucleons
shadowing for each other during the collision, thus reducing to
nucleon--nucleon cross section from the result obtained from \pp\
collisions, to a lower, effective cross section. This suppresses the
hard partonic cross section with up to 50\% in AuAu collisions at
$\sqrtsNN=200$~GeV \cite{Wang:1991hta}. In the end all partons are in
HIJING connected by strings, and hadronised with \pythia. As an option
it is possible to include a model for jet quenching, and also a jet
trigger, enhancing the rate for events with high-$p_\perp$ jets.

As mentioned above, AMPT presumes that a hot dense medium is
formed. It uses the parton state obtained in HIJING as initial
conditions. The partons then evolve in a partonic cascade up to
freeze-out. After freeze-out the partons are connected in strings,
which hadronise according to the Lund model in \pythia. Finally the
obtained hadrons form a secondary cascade until the density is low
enough, when they continue as free particles. As an option the
hadronisation can also be calculated via quark-antiquark coalescence.

Finally the EPOS model works on different principles than the other
two, as no explicit Glauber calculation is performed. Instead partonic
sub-collisions are calculated using parton-based Gribov--Regge theory
\cite{Drescher:2000ha}. An elementary scattering is here represented
by a cut \pomeron or ``parton ladder''. This ladder is interpreted as a
flux tube, or a string, where the intermediate gluons provide a
transverse motion. The strings then break up into segments by
quark-antiquark pair production. In the central region with high
density, the ``core'', the segments within a bin in $\eta$ form a
cluster, which expands longitudinally and radially until
freeze-out. In regions of low density, called the ``corona'', the
strings fragment instead directly to hadrons. This is mainly the case
in the fragmentation regions. In a recent version, called EPOS LHC
\cite{Pierog:2013ria}, a new flow parametrisation is introduced, which
does not take advantage of the complete hydrodynamical calculation
followed by the hadronic cascade as in EPOS2 \cite{Werner:2010aa} or
EPOS3 \cite{Werner:2013tya}. One consequence is here that the time for
one \PbPb\ event is reduced from one hour to a few tenths of a
second. According to the authors, this also implies that this version
should not be used for a precise study of $p_\perp$ distributions or
particle correlations in HI collisions.

In \figref{fig:gencomp} we compare the multiplicity spectra at $\sqrtsNN
= 2.76$~TeV from \figref{fig:ALICEeta}(b) with \angantyr and the three
generators discussed above.
(For HIJING jet quenching is disabled in \figref{fig:gencomp}, but this
should not have a major impact on the result.) 

\FIGURE[ht] {
  \includegraphics[width=0.8\textwidth]{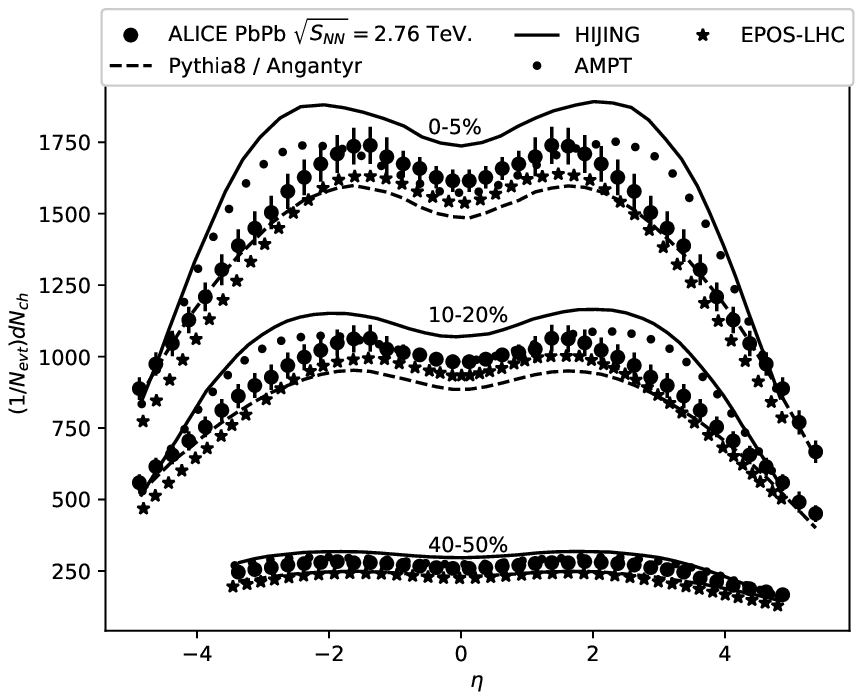}
  \caption{\label{fig:gencomp}Comparison of \angantyr to the
    generators EPOS-LHC, AMPT and HIJING. The figure shows charged
    particle production as function of pseudo-rapidity in \PbPb\ at
    $\sqrtsNN=2.76$~TeV as measured by ALICE \cite{Adam:2015kda}.  }
}

We note that with all differences mentioned above, all four generators
produce quite similar results for the centrality dependence of the
charged particle distribution.

Comparing first HIJING to \angantyr, we see that while \angantyr
undershoots at mid-$\eta$, HIJING overshoots on the full interval, and
produces a too wide shape for the distribution. The likely source of
this difference is the different way of handling secondary absorptive
events, as described in \sectref{sec:woundet-to-fs}. HIJING treats
all absorptive events on a similar footing, but the nuclear shadowing
included in HIJING implies that it produces an overall lower amount of
hard sub-collisions.

AMPT uses HIJING for initial conditions, but compared to HIJING the
overall multiplicity is reduced by the partonic and the hadronic
cascades. However, although the central density agrees with data, the
distribution is too wide. We also note that AMPT reproduces
multiplicity at mid-$\eta$ better than \angantyr, and refer to our
discussion about possible retuning of \angantyr to low-$p_\perp$ \pp\
data in \sectref{sec:aa-results}. Finally EPOS-LHC also does a better
job than \angantyr at mid-$\eta$, but worse away from the central
region. We note that AMPT and EPOS-LHC, which both include the
hydrodynamic expansion of a hot medium, do not describe data better
than \angantyr over the full $\eta$-range.

It is, however, clear that if one wants to pin down the physics of a
possible plasma phase, more exclusive observables than particle
production must be used. This is indeed also the case in contemporary
studies at the LHC and RHIC. Considering the precision obtained by the
current tools, we see that there is a need for improved tools for
comparing theory to data in heavy ion physics. To account for the
final $~10\%$ discrepancy shown by all four generators, analysis
specific effects like choice of centrality measure, trigger selection,
primary particle definition \textit{etc.} all play a major role. In
the present paper all comparisons of \angantyr to data are carried out
using the Rivet tool \cite{Buckley:2010ar}, which has proved highly
successful for this task in \pp. This has, however, been done using
our own implementation of the experimental analyses. It is crucial for
the further development of Monte Carlo event generators for heavy ion
physics, that present and future heavy ion data is released using
Rivet (or a similar tool), and we are pleased to note that experiments
are now starting to commit to this task, also in heavy ion physics.

\section{Conclusion and Outlook}
\label{sec:conclusion-outlook}

We have introduced a new model called \angantyr for generating
exclusive final states in proton-nucleus and nucleus-nucleus
collisions. It extrapolates \pp\ dynamics with a minimum of free
parameters, and in this way it bridges the gap between heavy ion and 
high energy physics phenomenology. It does not assume a hot thermalised medium, 
and the aim is to see how well such an extrapolation can reproduce experimental data,
thus exposing effects of collective behaviour. The model is a
generalisation of the model for \pA\ collisions in
\citeref{Bierlich:2016smv}, and is based on the following points:

\begin{itemize}

\item The basic \pp\ interaction is described by the \pytppp event
  generator, based on multiple partonic sub-collisions and string
  hadronisation.
\item The generalisation to nucleus collisions is inspired by the
  \fritiof model, and the notion of "wounded" or "participating"
  nucleons.
\item The number of wounded nucleons is calculated from the Glauber
  model in impact parameter space, including "Gribov corrections" due to
  diffractive excitation of individual nucleons.
\item The Glauber model is formulated in impact parameter space.
  Diffractive excitation is then most conveniently described by the
  Good--Walker formalism, as the result of fluctuations in the nucleon 
  substructure. We here for the fist time account for fluctuations in
  both the projectile and the target nucleons, in a Glauber calculation. 
  (As frequently in MC simulations, fluctuations in the position of 
  nucleons in the nuclei are also included.)
\end{itemize}
 
The model is implemented in an event generator, which generates
  exclusive final states. It is included in the \pytppp package, where
  the user simply specifies a nucleus instead of a hadron as
  projectile and/or target. The possibility to add a signal process
  (of electroweak or other origin) is also included, enabling the user
  to study every process one could normally study in a \pp\ collision.

We have shown that \angantyr gives a good description of general final 
state properties. This includes not only multiplicity and transverse 
momentum distributions both in \pPb\ and \PbPb\ collisions, but also 
its dependence on centrality. We note, however, that this dependence 
is very sensitive to the experimental definition of centrality. Thus 
we see that for low centrality the correlation between central 
multiplicity and "centrality" is more a correlation between central 
and forward activity, rather than between central activity and impact 
parameter. The model predictions for XeXe collisions are also in good 
agreement with \alice data published later. 

The model underestimates somewhat central particle production, 
when $p_\perp$ is integrated down to $p_\perp=0$. This may be not 
surprising, as it is an extrapolation of \pythia's description of pp 
dynamics, which is too low for small $p_\perp$ below 200 MeV. 
Future work is needed to improve the hadronisation models in this 
region, including their interface to the perturbative shower.

The description of data is quite sensitive to the handling of, in particular, 
secondary absorptive sub-events. We have investigated several different choices
relating to this treatment, relating to (i) the distributions in the covered rapidity
ranges, (ii) distributions of partons in the projectile nucleon, and (iii) 
energy--momentum conservation. For visualization we performed this
investigation in \pPb~collisions, noting that they will be significantly
smaller in \PbPb\ collisions. 
Although our final choices may not be based on completely solid 
theoretical grounds, the fact that alternatives investigated give a poorer 
description of data tells us, that the choices are reasonable. Certainly there
are other variations to investigate, but we postpone such studies
to a future publication.

In \pytppp all strings decay into hadrons independently. Thus it does 
not include a mechanism to reproduce the collective effects seen in pp 
collisions. Such effects are therefore also not reproduced by the 
present version of \angantyr, and the model should be thought of as a 
baseline for understanding the non-collective background to observables 
sensitive to collective behaviour. 

Also in high energy pp collisions the number of strings is quite large, 
in particular in events with high multiplicity.
In \citeref{Bierlich:2014xba} we showed that overlapping
strings forming "ropes" can qualitatively reproduce the increased 
strangeness in \pp\ \cite{ALICE:2017jyt}, as well as in \pPb\ and \PbPb\
\cite{Bierlich:2017sxk} collisions. In \citeref{Bierlich:2017vhg} we
further showed that the transverse pressure due to the increased
energy density provides a transverse expansion and a qualitative
description of the "ridge" observed in \pp\ collisions. An important
future direction will be to fully include these models in \angantyr,
and test to what degree they provide a description of the observed
collective effects in nucleus collisions. Besides the angular 
correlations, the transverse expansion may affect the $p_\perp$ 
distributions, which are less accurately reproduced in \pPb\ and 
\PbPb\ collisions.

To conclude we think that it is notable that a direct extrapolation of
\pp\ dynamics can reproduce general features of inclusive particle
production in \AA\ collisions to better than 10\%. This emphasises 
the importance of correlation studies, 
and in a future version of \angantyr we plan to include the collective
effects from string-string interactions in the description of
collisions with nuclei. In the future we also want to find
observables sensitive to the fluctuations related to diffractive
excitation and the internal substructure of nucleons. This is an
essential feature which distinguishes \angantyr from other event
generators available for nucleus-nucleus collisions.

\section*{Acknowledgments}

CB wants to thank C. H. Christensen for assistance in comparing to
ALICE data. This work was funded in part by the Swedish Research
Council, contracts number 2016-03291, 2016-05996 and 2017-0034, in
part by the European Research Council (ERC) under the European Union’s
Horizon 2020 research and innovation programme, grant agreement No
668679, and in part by the MCnetITN3 H2020 Marie Curie Initial
Training Network, contract 722104.

\appendix

\section{Generating absorptively and diffractively wounded nucleons}
\label{sec:gener-absorpt-diffr}

Here we will go through the technicalities of choosing the
interactions between projectile and target nucleons. In
\cite{Bierlich:2016smv} we showed that for a fixed nucleon--nucleon
impact parameter, $b$, and a fixed projectile state, the cross section
for the target nucleon to be wounded is given by the average of the
fluctuations in the target nucleon. Writing the imaginary part of the
scattering amplitude for given projectile and target states, $p$ and
$t$, in terms of the corresponding $S$-matrix,
$T_{pt}(b)\equiv 1- S_{pt}(b)$, we have
\begin{equation}
  \label{eq:basic-wounded-target}
  d\sigma\subwt =
  \left(1 - \llangle\llangle S_{pt}\rrangle_t^2\rrangle_p\right)d^2b.
\end{equation}
This works well for \pA\ collisions, but for \AA\ we also want to look
at the probability for the projectile nucleon being wounded, and on
top of this we want to be able to separate between absorptively and
diffractively wounded nucleons.

\subsection{Absorptively wounded nucleons}
\label{sec:absorpt-wound-nucl}

We expect the absorptively wounded nucleons will give the most
important contributions to the final state particle production, and we
therefore want to take special care to capture cross section
fluctuations in this case and at the same time make sure we correctly
reproduce the absorptive nucleon--nucleon cross section,
\begin{equation}
  \label{eq:abs-wounded}
  d\sigma\subabs = \left(1 - \llangle S_{pt}^2(b)\rrangle_{pt}\right)d^2b.
\end{equation}
The procedure will therefore be to generate one state for each nucleon
in the projectile and target nuclei and for each pair of nucleons
calculate
\begin{equation}
  \label{eq:abs-prob}
  P\subabs=1-S_{pt}^2(b),
\end{equation}
and declare the nucleon-nucleon interaction absorptive with this
probability. This will clearly give the correct absorptive
nucleon--nucleon cross section.

If we find the interaction is not absorptive we want to go on and
check if either the target or the projectile or both are diffractively
wounded, but this will then require us to consider averages over the
possible states of the projectile or target or both. In the following
we will consider a diffractively wounded target, but the corresponding
treatment of the projectile is completely analogous.

\subsection{Diffractively wounded nucleons}
\label{sec:diffr-wound-nucl}

In general it is not necessarily straight forward to analytically
calculate the average $\langle S_{pt}^2(b)\rangle_t$ needed to get 
the correct cross section for diffractive excitation. Instead we will
estimate the fluctuations by generating a secondary, or auxiliary
state for each projectile ($p'$) and target ($t'$) nucleon. We will
still calculate the probability for absorptive interaction using only
the primary states, but to get the probability of the target nucleon
to be wounded we note that the product $S_{pt}(b)S_{pt'}(b)$ will on
average yield the correct value for
$\langle\langle S_{pt}^2(b)\rangle_t\rangle_p$, so naively we could
use the probability $P\subwt=1-S_{pt}(b)S_{pt'}$. However, it is
clear that we will then have a negative probability for having a
diffractively wounded target $P\subwt-P\subabs<0$, for
$S_{pt}<S_{pt'}$. Therefore we also need to consider the statistically
equivalent situation where the absorptive interaction probability is
given by
\begin{equation}
  \label{eq:abs-prob-prime}
  P'\subabs=1-S_{pt'}^2(b),
\end{equation}
while the corresponding wounded probability is still
\begin{equation}
  \label{eq:naive-wounded-prime}
  P'\subwt=1-S_{pt}(b)S_{pt'}(b)=P\subwt,
\end{equation}
where the probability for a diffractively wounded target is then
positive.

The procedure we have chosen to handle this, is to shuffle
probabilities between the two situations so that we always get
non-negative probabilities for diffractively wounded nucleons
according to
\begin{eqnarray}
  \label{eq:shuffle}
  \tilde{P}\subwt&=&\left\{
             \begin{array}{ll}
               S_{tp}<S_{pt'}:&0\\
               S_{pt}>S_{pt'}:& P\subwt + P'\subwt-P'\subabs=1-2S_{pt}(b)S_{pt'}(b) +S_{pt'}^2(b)
             \end{array}\right.  \\
  \tilde{P}'\subwt&=&\left\{
             \begin{array}{ll}
               S_{tp}<S_{pt'}:& P'\subwt + P\subwt-P\subabs=1-2S_{pt}(b)S_{pt'}(b) +S_{pt}^2(b)\\
               S_{pt}>S_{pt'}:&0
             \end{array}\right.  
\end{eqnarray}
which will give the correct cross section for the target nucleon being
wounded.

By considering the auxiliary state for the projectile, $p'$, we can
then also find the probability for the projectile being diffractively
wounded. And if both are wounded we say that the interaction is a
double diffractive excitation\footnote{This will actually not give the
  correct double diffraction cross section, but as we are mainly
  concerned with the number of wounded nucleons, we postpone the
  detailed treatment of double diffraction to future improvements}.


\bibliographystyle{JHEP}
\bibliography{angRef}

\end{document}

%% file: figures/MPIladders.tex
\newsavebox{\ppMPIladdersproton}
\savebox{\ppMPIladdersproton}(0,0)[bl]{
  \SetWidth{2.0}
  \ArrowLine(-40,0)(0,0)
  \GOval(0,0)(5,5)(0){0}
  \SetWidth{1.0}
  \Line(0,0)(40,0)
  \Line(0,3)(40,10)
  \Line(0,-3)(40,-10)
}

\newsavebox{\ppMPIladdersprotonel}
\savebox{\ppMPIladdersprotonel}(0,0)[bl]{
  \SetWidth{2.0}
  \ArrowLine(-40,0)(0,0)
  \ArrowLine(40,0)(80,0)
  \GOval(0,0)(5,5)(0){0}
  \SetWidth{1.0}
  \Line(0,0)(40,0)
  \Line(0,3)(40,3)
  \Line(0,-3)(40,-3)
  \GOval(40,0)(5,5)(0){0}
}

\newsavebox{\ppMPIladdershardladder}
\savebox{\ppMPIladdershardladder}(0,0)[bl]{
  \SetColor{Red}
  \Gluon(0,0)(15,75){4}{7}
  \COval(15,75)(2,2)(0){Red}{Red}
  \Gluon(15,75)(15,125){4}{5}
  \Gluon(15,75)(50,75){4}{3}
  \COval(15,125)(2,2)(0){Red}{Red}
  \Gluon(15,125)(50,125){4}{3}
  \Gluon(15,125)(0,200){4}{7}
  \SetColor{Black}
}

\newsavebox{\ppMPIladdersfirst}
\savebox{\ppMPIladdersfirst}(0,0)[bl]{
  \put(50,50){\usebox{\ppMPIladdersproton}}
  \put(50,250){\usebox{\ppMPIladdersproton}}
  \put(50,50){\usebox{\ppMPIladdershardladder}}
  \Gluon(50,50)(85,100){4}{5}
  \GOval(85,100)(2,2)(0){0}
  \Gluon(85,100)(85,150){4}{5}
  \GOval(85,150)(2,2)(0){0}
  \Gluon(85,150)(50,250){4}{12}
  \Gluon(85,100)(115,100){4}{3}
  \Gluon(85,150)(115,150){4}{3}
  \Text(50,30)[c]{(a)}
}

\newsavebox{\ppMPIladder}
\savebox{\ppMPIladder}(0,0)[bl]{
  \put(50,50){\usebox{\ppMPIladdersproton}}
  \put(50,250){\usebox{\ppMPIladdersproton}}
  \put(50,50){\usebox{\ppMPIladdershardladder}}
}

\newsavebox{\ppMPIladdersecond}
\savebox{\ppMPIladdersecond}(0,0)[bl]{
  \put(50,50){\usebox{\ppMPIladdersproton}}
  \put(50,250){\usebox{\ppMPIladdersproton}}
  \Gluon(50,50)(50,250){4}{15}
  \GOval(50,75)(2,2)(0){0}
  \Gluon(50,75)(90,75){4}{3}
  \GOval(50,100)(2,2)(0){0}
  \Gluon(50,100)(90,100){4}{3}
  \GOval(50,125)(2,2)(0){0}
  \Gluon(50,125)(90,125){4}{3}
  \GOval(50,150)(2,2)(0){0}
  \Gluon(50,150)(90,150){4}{3}
  \GOval(50,175)(2,2)(0){0}
  \Gluon(50,175)(90,175){4}{3}
  \GOval(50,200)(2,2)(0){0}
  \Gluon(50,200)(90,200){4}{3}
  \GOval(50,225)(2,2)(0){0}
  \Gluon(50,225)(90,225){4}{3}
}

\newsavebox{\ppMPIladderthird}
\savebox{\ppMPIladderthird}(0,0)[bl]{
  \put(50,50){\usebox{\ppMPIladdersprotonel}}
  \put(50,250){\usebox{\ppMPIladdersprotonel}}
  \Gluon(50,50)(50,250){4}{15}
  \Gluon(90,50)(90,250){4}{15}
  \GOval(50,75)(2,2)(0){0}
  \GOval(90,75)(2,2)(0){0}
  \Gluon(50,75)(90,75){4}{3}
  \GOval(50,100)(2,2)(0){0}
  \GOval(90,100)(2,2)(0){0}
  \Gluon(50,100)(90,100){4}{3}
  \GOval(50,125)(2,2)(0){0}
  \GOval(90,125)(2,2)(0){0}
  \Gluon(50,125)(90,125){4}{3}
  \GOval(50,150)(2,2)(0){0}
  \GOval(90,150)(2,2)(0){0}
  \Gluon(50,150)(90,150){4}{3}
  \GOval(50,175)(2,2)(0){0}
  \GOval(90,175)(2,2)(0){0}
  \Gluon(50,175)(90,175){4}{3}
  \GOval(50,200)(2,2)(0){0}
  \GOval(90,200)(2,2)(0){0}
  \Gluon(50,200)(90,200){4}{3}
  \GOval(50,225)(2,2)(0){0}
  \GOval(90,225)(2,2)(0){0}
  \Gluon(50,225)(90,225){4}{3}
}

\newsavebox{\ppMPIladderforth}
\savebox{\ppMPIladderforth}(0,0)[bl]{
  \put(50,50){\usebox{\ppMPIladdersprotonel}}
  \put(50,250){\usebox{\ppMPIladdersprotonel}}
  \ZigZag(70,50)(70,250){4}{15}
}

\newsavebox{\ppMPIladderfifth}
\savebox{\ppMPIladderfifth}(0,0)[bl]{
  \put(50,50){\usebox{\ppMPIladdersprotonel}}
  \put(50,250){\usebox{\ppMPIladdersprotonel}}
  \ZigZag(60,50)(60,250){4}{15}
  \ZigZag(80,50)(80,250){4}{15}
}

\newsavebox{\ppMPIladdersixth}
\savebox{\ppMPIladdersixth}(0,0)[bl]{
  \put(50,50){\usebox{\ppMPIladdersprotonel}}
  \put(50,250){\usebox{\ppMPIladdersprotonel}}
  \ZigZag(70,50)(70,150){4}{8}
  \ZigZag(70,150)(50,250){4}{8}
  \ZigZag(70,150)(90,250){4}{8}
  \GOval(70,150)(2,2)(0){0}
}

\newsavebox{\ppMPIladdersevth}
\savebox{\ppMPIladdersevth}(0,0)[bl]{
  \put(50,50){\usebox{\ppMPIladdersprotonel}}
  \put(50,250){\usebox{\ppMPIladdersprotonel}}
  \ZigZag(70,120)(70,180){4}{6}
  \ZigZag(70,180)(50,250){4}{6}
  \ZigZag(70,180)(90,250){4}{6}
  \GOval(70,180)(2,2)(0){0}
  \GOval(70,120)(2,2)(0){0}
  \ZigZag(70,120)(50,50){4}{6}
  \ZigZag(70,120)(90,50){4}{6}
}



\newsavebox{\ppMPIladderssecond}
\savebox{\ppMPIladderssecond}(0,0)[bl]{
  \put(50,50){\usebox{\ppMPIladdersproton}}
  \put(50,250){\usebox{\ppMPIladdersproton}}
  \put(50,50){\usebox{\ppMPIladdershardladder}}
  \GOval(55,88)(2,2)(0){0}
  \Gluon(55,88)(85,100){4}{3}
  \GOval(85,100)(2,2)(0){0}
  \Gluon(85,100)(85,150){4}{5}
  \GOval(85,150)(2,2)(0){0}
  \Gluon(85,150)(56,202){4}{6}
  \GOval(56,202)(2,2)(0){0}
  \Gluon(85,100)(115,100){4}{3}
  \Gluon(85,150)(115,150){4}{3}
  \Text(50,30)[c]{(b)}
}

\newsavebox{\ppMPIladdersthird}
\savebox{\ppMPIladdersthird}(0,0)[bl]{
  \put(50,50){\usebox{\ppMPIladdersproton}}
  \put(50,250){\usebox{\ppMPIladdersproton}}
  \put(50,50){\usebox{\ppMPIladdershardladder}}
  \GOval(55,88)(2,2)(0){0}
  \Gluon(55,88)(85,100){4}{3}
  \GOval(85,100)(2,2)(0){0}
  \Gluon(85,100)(85,150){4}{5}
  \GOval(85,150)(2,2)(0){0}
  \Gluon(85,150)(63,162){4}{2}
  \GOval(63,162)(2,2)(0){0}
  \Gluon(85,100)(115,100){4}{3}
  \Gluon(85,150)(115,150){4}{3}
  \Text(50,30)[c]{(c)}
}

\newsavebox{\pAMPIladdersfourth}
\savebox{\pAMPIladdersfourth}(0,0)[bl]{
  \put(66,66){\usebox{\ppMPIladdersproton}}
  \put(50,50){\usebox{\ppMPIladdersproton}}
  \put(50,250){\usebox{\ppMPIladdersproton}}
  \put(50,50){\usebox{\ppMPIladdershardladder}}
  \Gluon(66,66)(85,100){4}{3}
  \GOval(85,100)(2,2)(0){0}
  \Gluon(85,100)(85,150){4}{5}
  \GOval(85,150)(2,2)(0){0}
  \Gluon(85,150)(56,202){4}{6}
  \GOval(56,202)(2,2)(0){0}
  \Gluon(85,100)(115,100){4}{3}
  \Gluon(85,150)(115,150){4}{3}
  \Text(50,30)[c]{(a)}
}

\newsavebox{\pAMPIladdersfifth}
\savebox{\pAMPIladdersfifth}(0,0)[bl]{
  \put(66,66){\usebox{\ppMPIladdersproton}}
  \put(50,50){\usebox{\ppMPIladdersproton}}
  \put(50,250){\usebox{\ppMPIladdersproton}}
  \put(50,50){\usebox{\ppMPIladdershardladder}}
  \Gluon(66,66)(85,100){4}{3}
  \GOval(85,100)(2,2)(0){0}
  \Gluon(85,100)(85,150){4}{5}
  \GOval(85,150)(2,2)(0){0}
  \Gluon(85,150)(75,202){4}{5}
  \GOval(75,202)(4,4)(0){0}
  \Line(75,202)(95,199)
  \Line(75,202)(95,205)
  \SetColor{Green}
  \ZigZag(50,250)(75,202){4}{7}
  \SetColor{Black}
  \Gluon(85,100)(115,100){4}{3}
  \Gluon(85,150)(115,150){4}{3}
  \Text(50,30)[c]{(b)}
}

\newsavebox{\AAMPIladdersfirst}
\savebox{\AAMPIladdersfirst}(0,0)[bl]{
  \put(66,66){\usebox{\ppMPIladdersproton}}
  \put(50,50){\usebox{\ppMPIladdersproton}}
  \put(50,250){\usebox{\ppMPIladdersproton}}
  \put(66,266){\usebox{\ppMPIladdersproton}}
  \put(50,50){\usebox{\ppMPIladdershardladder}}
  \Gluon(66,66)(85,100){4}{3}
  \GOval(85,100)(2,2)(0){0}
  \Gluon(85,100)(85,150){4}{5}
  \GOval(85,150)(2,2)(0){0}
  \Gluon(66,266)(85,150){4}{12}
  \Gluon(85,100)(115,100){4}{3}
  \Gluon(85,150)(115,150){4}{3}
  \Text(50,30)[c]{(a)}
}

\newsavebox{\AAMPIladderssecond}
\savebox{\AAMPIladderssecond}(0,0)[bl]{
  \put(66,66){\usebox{\ppMPIladdersproton}}
  \put(50,50){\usebox{\ppMPIladdersproton}}
  \put(50,250){\usebox{\ppMPIladdersproton}}
  \put(66,266){\usebox{\ppMPIladdersproton}}
  \put(50,50){\usebox{\ppMPIladdershardladder}}
  \Gluon(66,66)(85,100){4}{3}
  \GOval(85,100)(2,2)(0){0}
  \Gluon(85,100)(85,150){4}{5}
  \GOval(85,150)(2,2)(0){0}
  \Gluon(85,150)(75,202){4}{5}
  \GOval(75,202)(4,4)(0){0}
  \Line(75,202)(95,199)
  \Line(75,202)(95,205)
  \SetColor{Green}
  \ZigZag(50,250)(75,202){4}{7}
  \SetColor{Black}
  \Gluon(85,100)(115,100){4}{3}
  \Gluon(85,150)(115,150){4}{3}
  
  \SetColor{Cyan}
  \Gluon(66,266)(90,220){4}{5}
  \COval(90,220)(2,2)(0){Cyan}{Cyan}
  \Gluon(90,220)(120,220){4}{4}
  \Gluon(90,220)(90,163){4}{6}
  \COval(90,163)(2,2)(0){Cyan}{Cyan}
  \Gluon(90,163)(120,163){4}{4}
  \Gluon(90,163)(72,113){4}{6}
  \COval(72,113)(4,4)(0){Cyan}{Cyan}
  \Line(72,113)(91,110)
  \Line(72,113)(91,116)
  \SetColor{Black}
  \SetColor{Green}
  \ZigZag(72,113)(50,50){4}{7}
  \SetColor{Black}

  \Text(50,30)[c]{(b)}
}

\newsavebox{\AAMPIladdersthird}
\savebox{\AAMPIladdersthird}(0,0)[bl]{
  \put(66,66){\usebox{\ppMPIladdersproton}}
  \put(50,50){\usebox{\ppMPIladdersproton}}
  \put(50,250){\usebox{\ppMPIladdersproton}}
  \put(50,50){\usebox{\ppMPIladdershardladder}}
  \put(66,266){\usebox{\ppMPIladdersproton}}
  \Gluon(66,66)(85,100){4}{3}
  \GOval(85,100)(2,2)(0){0}
  \Gluon(85,100)(85,150){4}{5}
  \GOval(85,150)(2,2)(0){0}
  \Gluon(85,150)(75,202){4}{5}
  \GOval(75,202)(4,4)(0){0}
  \Line(75,202)(95,199)
  \Line(75,202)(95,205)
  \SetColor{Green}
  \ZigZag(50,250)(75,202){4}{7}
  \SetColor{Black}
  \Gluon(85,100)(115,100){4}{3}
  \Gluon(85,150)(115,150){4}{3}

  \SetColor{Cyan}
  \Gluon(66,266)(90,220){4}{5}
  \COval(90,220)(2,2)(0){Cyan}{Cyan}
  \Gluon(90,220)(120,220){4}{4}
  \Gluon(90,220)(90,163){4}{6}
  \COval(90,163)(2,2)(0){Cyan}{Cyan}
  \Gluon(90,163)(120,163){4}{4}
  \Gluon(90,163)(72,113){4}{6}
  \COval(72,113)(4,4)(0){Cyan}{Cyan}
  \Line(72,113)(91,110)
  \Line(72,113)(91,116)
  \SetColor{Black}
  \SetColor{Green}
  \ZigZag(72,113)(66,66){4}{5}
  \SetColor{Black}

  \Text(50,30)[c]{(c)}
}

\newsavebox{\ppMPIladders}
\savebox{\ppMPIladders}(0,0)[bl]{
  \put(0,0){\usebox{\ppMPIladdersfirst}}
  \put(160,0){\usebox{\ppMPIladderssecond}}
  \put(320,0){\usebox{\ppMPIladdersthird}}
}

\newsavebox{\pAMPIladders}
\savebox{\pAMPIladders}(0,0)[bl]{
  \put(80,0){\usebox{\pAMPIladdersfourth}}
  \put(240,0){\usebox{\pAMPIladdersfifth}}
}

\newsavebox{\AAMPIladders}
\savebox{\AAMPIladders}(0,0)[bl]{
  \put(0,0){\usebox{\AAMPIladdersfirst}}
  \put(160,0){\usebox{\AAMPIladderssecond}}
  \put(320,0){\usebox{\AAMPIladdersthird}}
}